\documentclass[twocolumn]{aastex63}
\usepackage{amsmath}
\usepackage{mathrsfs}
\usepackage{xcolor}
\usepackage[T1]{fontenc}
\usepackage{fourier}
\usepackage{hyperref}
\usepackage[normalem]{ulem}
\usepackage{relsize}
\usepackage{lineno}
\submitjournal{ApJ}

\shorttitle{Mass loss history of PS1-11aop}
\shortauthors{Ibik et al.}
\begin{document}
%\linenumbers

\title{PS1-11aop: Probing the Mass Loss History of a Luminous Interacting Supernova Prior to its Final Eruption with Multi-wavelength Observations}

\correspondingauthor{Adaeze Ibik}
\email{adaeze.ibik@mail.utoronto.ca}

\author[0000-0003-2405-2967]{Adaeze L.~Ibik}
  \affiliation{David A.~Dunlap Department of Astronomy \& Astrophysics, University of Toronto, 50 St.~George Street, Toronto, ON M5S 3H4, Canada}
  \affiliation{Dunlap Institute for Astronomy \& Astrophysics, University of Toronto, 50 St.~George Street, Toronto, ON M5S 3H4, Canada}

\author[0000-0001-7081-0082]{Maria R. Drout}
  \affiliation{David A.~Dunlap Department of Astronomy \& Astrophysics, University of Toronto, 50 St.~George Street, Toronto, ON M5S 3H4, Canada}
  \affiliation{Dunlap Institute for Astronomy \& Astrophysics, University of Toronto, 50 St.~George Street, Toronto, ON M5S 3H4, Canada}

\author[0000-0003-4768-7586]{Raffaella Margutti}
\affil{Department of Astronomy, University of California, Berkeley, CA 94720-3411, USA}
\affil{Department of Physics, University of California, 366 Physics North MC 7300, Berkeley, CA 94720, USA}

\author[0000-0002-4513-3849]{David Matthews}
\affiliation{Department of Astronomy, University of California, Berkeley, CA 94720-3411, USA}
\affiliation{Center for Interdisciplinary Exploration and Research in Astrophysics (CIERA), and Department of Physics and Astronomy, Northwestern University, Evanston, IL 60208, USA}

\author[0000-0002-5814-4061]{V.\ Ashley Villar}
\affiliation{Center for Astrophysics \textbar{} Harvard \& Smithsonian, 60 Garden Street, Cambridge, MA 02138-1516, USA}

\author[0000-0002-9392-9681]{Edo Berger}
\affiliation{Center for Astrophysics \textbar{} Harvard \& Smithsonian, 60 Garden Street, Cambridge, MA 02138-1516, USA}
\affiliation{The NSF AI Institute for Artificial Intelligence and Fundamental Interactions}

\author[0000-0002-7706-5668]{Ryan Chornock}
\affiliation{Department of Astronomy, University of California, Berkeley, CA 94720-3411, USA}

\author[0000-0002-8297-2473]{Kate D. Alexander}
\affiliation{Department of Astronomy and Steward Observatory, University of Arizona, 933 North Cherry Avenue, Tucson, AZ 85721-0065, USA}

\author[0000-0003-0307-9984]{Tarraneh Eftekhari}
\altaffiliation{NHFP Einstein Fellow}
\affiliation{Center for Interdisciplinary Exploration and Research in Astrophysics (CIERA), and Department of Physics and Astronomy, Northwestern University, Evanston, IL 60208, USA}

\author[0000-0003-1792-2338]{Tanmoy Laskar}
\affiliation{Department of Physics \& Astronomy, University of Utah, Salt Lake City, UT 84112, USA}

\author[0000-0001-9454-4639]{Ragnhild Lunnan}
\affiliation{The Oskar Klein Centre, Department of Astronomy, Stockholm University, AlbaNova, SE-10691 Stockholm, Sweden}

\author[0000-0002-2445-5275]{Ryan J. Foley}
\affiliation{Department of Astronomy and Astrophysics, University of California, Santa Cruz, CA 95064, USA}

\author[0000-0002-6230-0151]{David Jones}
\affiliation{Institute for Astronomy, University of Hawaii, 640 N. Aâohoku Pl., Hilo, HI 96720, USA}

\author[0000-0002-0763-3885]{Dan Milisavljevic}
\affiliation{Purdue University, Department of Physics and Astronomy, 525 Northwestern Ave, West Lafayette, IN 47907}

\author[0000-0002-4410-5387]{Armin Rest}
\affiliation{Space Telescope Science Institute, 3700 San Martin Drive, Baltimore, MD 21218, USA}
\affiliation{Department of Physics \& Astronomy, Bloomberg Center for Physics and Astronomy, Room 366 3400 N. Charles Street, Baltimore, MD 21218, USA}

\author[0000-0002-4934-5849]{Daniel	Scolnic}
\affiliation{Department of Physics, Duke University, Durham, NC 27708, USA}

\author[0000-0003-3734-3587]{Peter K. G. Williams}
\affiliation{Center for Astrophysics \textbar{} Harvard \& Smithsonian, 60 Garden Street, Cambridge, MA 02138-1516, USA}



\begin{abstract}
Luminous interacting supernovae are a class of stellar explosions whose progenitors underwent vigorous mass loss in the years prior to core-collapse. While the mechanism by which this material is ejected is still debated, obtaining the full density profile of the circumstellar medium (CSM) could reveal more about this process. Here, we present an extensive multi-wavelength study of PS1-11aop, a luminous and slowly declining Type IIn SN discovered by the PanSTARRS Medium Deep Survey. PS1-11aop had a peak r-band magnitude of $-$20.5\,mag, a total radiated energy $>$ 8$\times$10$^{50}$\,erg, and it exploded near the center of a star-forming galaxy with super-solar metallicity. We obtained multiple detections at the location of PS1-11aop in the radio and X-ray bands between 4 and 10\,years post-explosion, and if due to the SN, it is one of the most luminous radio supernovae identified to date.  Taken together, the multiwavelength properties of PS1-11aop are consistent with a CSM density profile with multiple zones. The early optical emission is consistent with the supernova blastwave interacting with a dense and confined CSM shell which contains multiple solar masses of material that was likely ejected in the final $<$10-100 years prior to the explosion,($\sim$0.05$-$1.0 M$_{\odot}$yr$^{-1}$ at radii of $\lesssim$10$^{16}$\,cm). The radio observations, on the other hand, are consistent with a sparser environment ($\lesssim$2$\times 10^{-3}$ M$_{\odot}$yr$^{-1}$ at radii of $\sim$0.5-1$\times$10$^{17}$\,cm)---thus probing the history of the progenitor star prior to its final mass loss episode. 
\end{abstract}



%\section{Introduction}\label{sec:intro}
\section{Introduction}\label{sec:intro}

Approximately 7$-$9\% of core-collapse supernovae (CCSNe) exhibit narrow and intermediate hydrogen emission lines in their spectra \citep{Smith2011,Schlegel1990}, which are indicative of interaction between the SN blastwave and a dense circumstellar medium (CSM; \citealt{Chugai1994,Chevalier1994}). The optical light curves and spectra of Type IIn SNe can be used to estimate the density of the CSM and hence a pre-SN mass-loss rate for the progenitor star. Inferred values from optical observations range from approximately $10^{-3}$ to $10^{-1}$ M$_\odot$ yr$^{-1}$ \citep[e.g.][]{Smith2014,Smith2017hsn,Villar2017,Taddia2013, Kiewe2012}. These values are higher than steady winds observed in evolved massive stars \citep{Crowther2007,Vink2011,Smith2014}, indicating that a subset of massive stars undergoes intense (possibly violent/eruptive) mass loss in the years immediately preceding core collapse.

The class of Type IIn SNe spans a wide range of observed properties (e.g.\ peak luminosities, timescales), which in turn map to a wide range of CSM properties \citep[e.g.][]{Nyholm2020,Ofek2014b,Ransome2021,Kiewe2012}. As a result, they are likely produced by multiple progenitor channels \citep[for a review, see][]{Smith2017hsn}. Of particular note was the discovery of a class of Type IIn SNe that achieves peak luminosities 10--100 times larger than those of normal CCSNe \citep[][]{Smith2007,Gal_Yam_2012}. Many of these events also have long timescales, leading to total radiated energies in excess of $10^{50}$ or even $10^{51}$ ergs \citep[e.g.][]{Smith2007,Smith_2008,Rest2011,Fransson_2014,Nicholl2020,Dickinson2024,Brennan2023,Tartaglia2020,Smith2023}. 
If powered entirely by interaction, these ``luminous'' Type IIn SNe require very massive progenitors that ejected several solar masses of material shortly before core-collapse \citep[e.g.][]{Smith2007B,vanMarle2010,Chatz2012}.

Such intense mass-loss episodes are not predicted by standard models of stellar evolution in the years immediately preceding core collapse. This has led to several proposed mass-loss mechanisms including (i) the transport of energy from the core to the surface of the star during the final nuclear burning stages via either turbulent mixing \citep{Arnett2011,Smith&Arnett2014} or convectively driven waves \citep{Quataert2012,Fuller2017,Wu2021}, (ii) strong dust or pulsation driven winds in red supergiants \citep{Hofner2018,Yoon2010}, (iii) the onset of episodes of pulsational pair-instability \citep{Woosley2007, Woosley2017,Nicholl2020}, or (iv) late stage binary interactions \citep{Sana2012,Chevalier_2012,Schroder2020, Zapartas2021,Cohen2024}. 

One means to constrain which mass loss mechanism(s) operate in luminous Type IIn SNe is to map the CSM over large physical scales, including the region of the CSM \emph{exterior} to the dense material that leads to the high optical luminosity. First, constraints on the outer radius of the dense material directly tie to the timescale over which the mass loss mechanism was operating prior to core-collapse \cite[e.g.][]{Smith2023,Renzo2020,Margutti2017}. Second, measurements of the outer density itself could be directly compared to expectations for winds from various classes of evolved massive stars \citep{Smith2014} as well as the inter-shell regions predicted for objects undergoing pulsational pair-instability \citep{Woosley2017}.

However, probing this region often requires late-time observations. 
While some nearby Type IIn SNe have been followed for years in the optical (e.g., SN1988Z, SN2005ip-- \citealt{Smith2017}; SN2010jl-- \citealt{Fransson_2014}; SN2015da-- \citealt{Smith2023}), many luminous events are located at larger distances. As a result, the optical flux (arising from thermal emission that is produced at the SN ejecta-CSM shock front when the optical depth is high; see e.g., \citealt{Chugai1994,Ofek2014,Chatz2012}) often falls below the detection threshold within 1--2 years post-explosion \citep{Nyholm2020,Villar2020}. Fortunately, radio and X-ray observations offer an additional means to probe the CSM at larger scales. 

Radio synchrotron emission is produced when the SN shock interacts with the surrounding medium \citep{Chevalier1998ApJ,ChevalierFransson2003, ChevalierFransson2017}. This emission is heavily absorbed at early times (due to synchrotron self-absorption and free-free absorption) but eventually enters the optically thin phase and brightens. Both the radio luminosity and the time of peak brightness at a given frequency are proportional to CSM density \citep{Chevalier_2006}. As a result, Type IIn SNe, with their high inferred mass-loss rates, are predicted to be bright radio sources on a timescale of years---if the region of dense mass-loss extends to large physical scales and the shock front is not too strongly decelerated. Indeed, some Type IIn SNe detected in the radio are among the brightest known radio SNe \citep{Chandra_2012,vanDyke1993ApJ,Weiler1990, Stroh2021}.  At the same time, the non-detection of other events at late times has been used to argue for dense mass-loss regions that are confined to small radii \citep{vanDyk1996}.

Thermal (free-free) X-ray emission is also produced at both the forward and reverse shock of interacting supernova, with a luminosity proportional to the square of the CSM density
\citep{Fransson1996,ChevalierFransson2003}. This emission is also heavily absorbed at early times (due to bound-free absorption), but as the column depth due to the external CSM decreases, X-rays can escape the interaction zone. Indeed, the few type IIn detected in X-rays are among the brightest X-ray supernova known, with luminosities reaching $\sim$10$^{41}$ ergs on timescales of $>$1000 days \citep[][]{Chandra2015,Dwarkadas2012,pooley2002ApJ}. However, while they are powerful probes of CSM interaction, interpretation of the origin of the emergent X-rays can be challenging. Higher densities at the reverse shock can lead to larger intrinsic X-ray luminosities than the forward shock \citep{Fransson1996}, but for high enough CSM densities the reverse shock will remain radiative even at late times \citep{Nymark2006,Nymark2009}. In this situation, absorption due to a ``cold dense shell'' formed between the forward and reverse shocks and any potential CSM asymmetries will impact the fraction of reverse shock X-ray photons that can escape. In addition, recent work has shown that in the case of a confined CSM, it is possible for the emergent X-ray light curve to be dominated by the ``post-interaction phase'' when the hot gas expands after the shock exists the dense shell \citep{Margalit2022}.

While several dozen luminous Type IIn SN have been discovered to date, few have late-time radio and X-ray follow-ups. 
The best studied are SN\,2010jl and SN\,2017hcc, both of which were located nearby ($z<0.02$), showed slowly declining optical emission, and were detected in the radio and X-ray bands at various times between $\sim$1 and 4 years post-discovery \citep{Chandra2015,Chandra2022}. 
While there are differences between these two objects (SN\,2010jl was one of the most X-ray luminous SN known, while SN2017hcc is relatively X-ray faint) both sets of observations have been used to argue for a sparser outer region of CSM surrounding the progenitor star \citep{Chandra2015,Chandra2022}.  
Similarly, while absorption could also play a role, the lack of detected radio, X-ray, and H$\alpha$ emission from SN2006gy at timescales $>$1 year was used to argue against continued CSM interaction as the late-time power source for that event \citep{Smith2008A}. \citet{Hatsukade2021} also present results from a VLA survey of luminous supernovae at more than 2 years post-explosion. They do not detect radio emission in excess of expectations for the host galaxy in any of the 8 luminous Type IIn SN in their sample---although these results have not yet been used to place physical constraints on the density of the CSM at large physical scales in the context of a non-relativistic explosion. 

In this paper, we present a multi-wavelength study of PS1-11aop, a luminous transient with time-variable intermediate width $H_{\rm{\alpha}}$ features, similar to many Type IIn SN. Located at redshift z$=$0.218, PS1-11aop was included in a survey program that aimed to search for radio and X-ray emission at the location of all luminous transients discovered during the PanSTARRS Medium-Deep Survey that showed spectroscopic features consistent with expectations for CSM interaction. In the case of PS1-11aop, luminous radio and X-ray emission was detected at the location of PS1-11aop between 4 and 10 years post-explosion. We, therefore, present a detailed case study of this object here, while results from the full radio and X-ray survey (which includes mainly upper limits) will be presented in an upcoming publication. In \S \ref{sec:obs}, we outline the discovery of PS1-11aop and the optical, radio, and X-ray observations obtained. In \S \ref{sec:opticalprops}, we describe the optical photometry and spectroscopic results. A description of the host galaxy properties and local environment of PS1-11aop can be found in \S \ref{sec:Host Galaxy} while a discussion of whether the detected optical, radio, and X-ray emission can be attributed to a SN is given in \S \ref{sec: argument for sn}. We then present the optical, radio, and X-ray modeling of the PS1-11aop in \S \ref{sec:optical modelling}, \S \ref{sec:radio modelling} and \S \ref{sec:x-ray modelling}. Finally, we discuss the implications of multi-wavelength analysis and modeling for the progenitor system in \S \ref{sec:Discussion and Conclusion} and present our conclusions in \S \ref{sec:conclusion}. Throughout this paper, we assumed a flat $\Lambda_{\rm{CDM}}$ cosmology with H$_{0}$ $=$ 71\,km\,s$^{-1}$ Mpc$^{-1}$, $\Omega_{m}$ $=$ 0.27, and $\Omega _{\Lambda}$ $=$ 0.73 \citep{Komatsu2011}. 

\begin{figure}[t]
\centering
\includegraphics[width=\columnwidth]{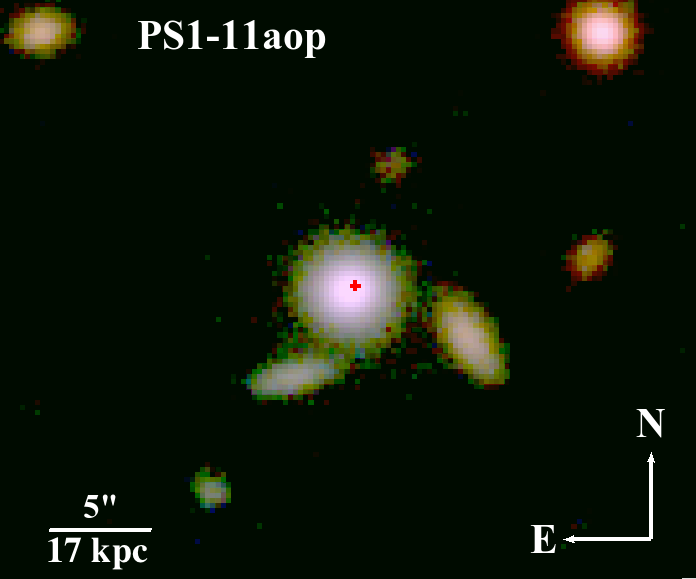}
\caption {Color image of the host galaxy of PS1-11aop, constructed from $g_{P1}$ ,$r_{P1}$, and $z_{P1}$ images. The location of PS1-11aop near the galaxy core is marked as a red cross.} 
\label{fig:hostimg}
\end{figure}

\begin{figure*}[ht!]
\centering
\includegraphics[width=.6\textwidth]{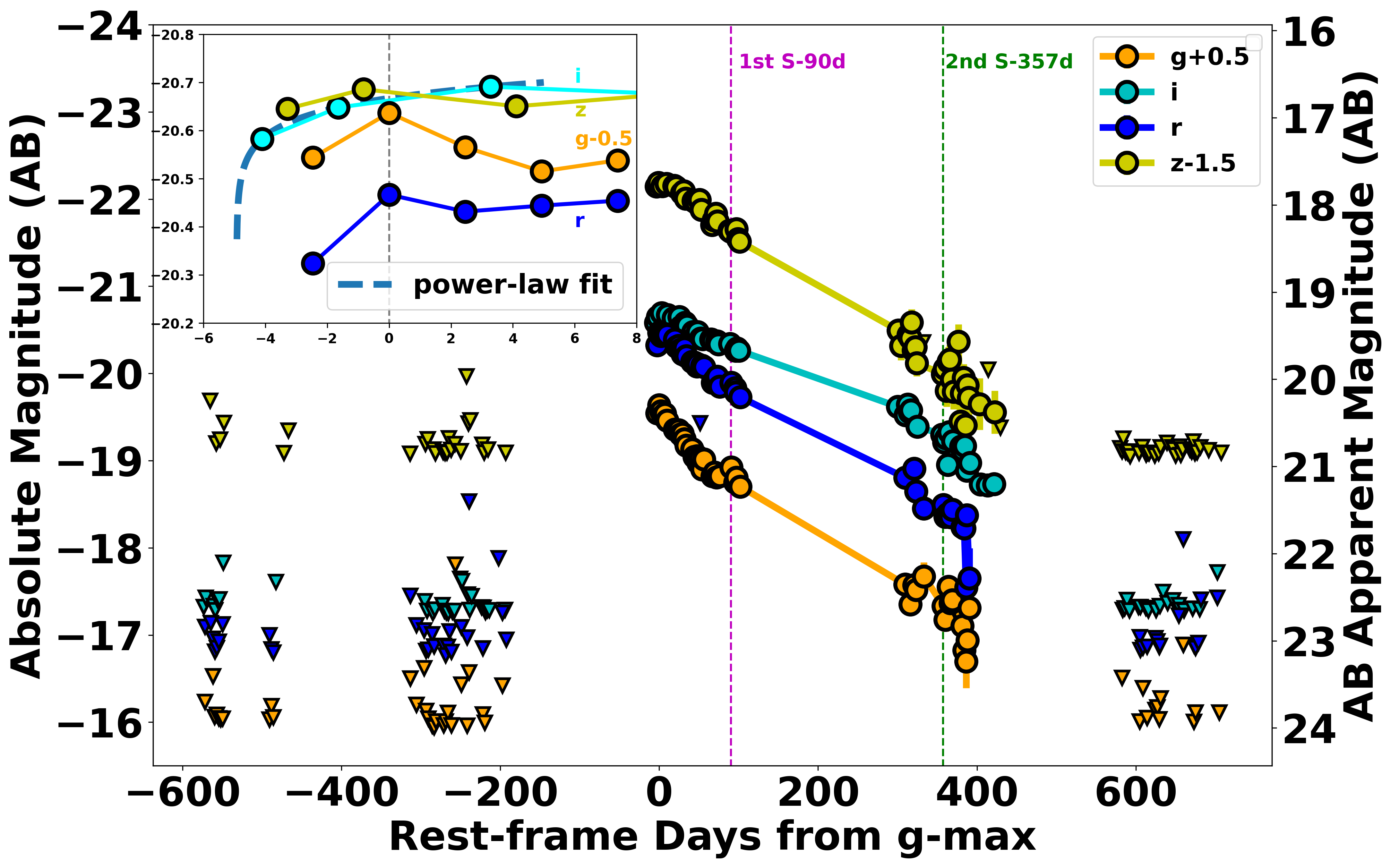}
\includegraphics[width=.3\textwidth]{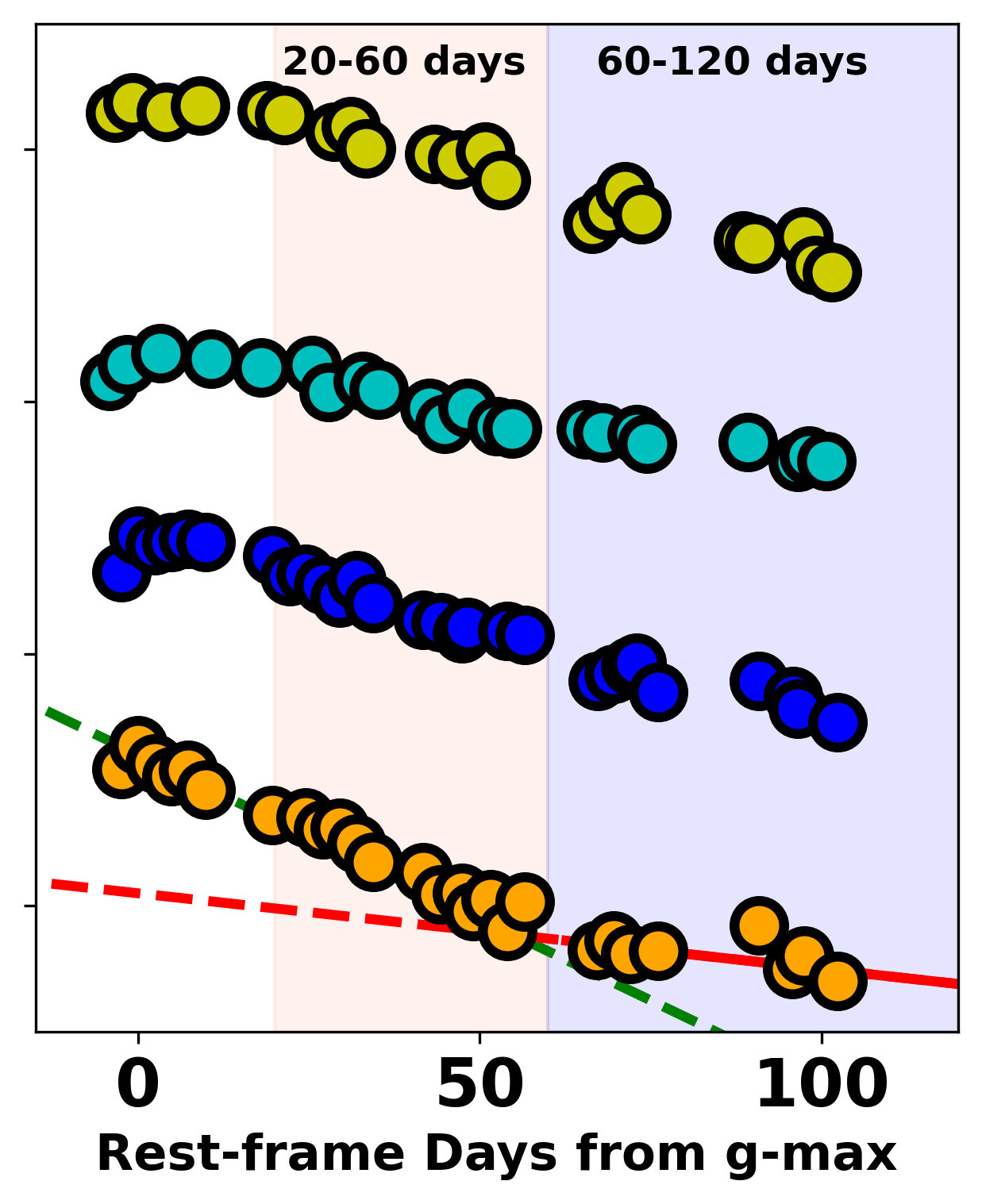}
\caption {\emph{Left:} Absolute magnitude rest-frame light curve for PS1-11aop. Circles are $g_{P1}r_{P1}i_{P1}z_{P1}$ detections while triangles represent 3$\sigma$ upper limits at early times before the explosion and later times when the explosion faded below the PS1-MDS detection threshold in the optical. The vertical lines are rest-frame times when the spectra were taken relative to the time of g-band maximum.  The inner panel is zoomed in into the early rise of the light curve. While we have relatively few points on the rise, a power-law fit to the i-band would imply a time of explosion only shortly before discovery. The right y-axis showing the corresponding apparent magnitudes is not corrected for band-specific extinction. While a few very extreme points in the second season (e.g. the two very low points in the g- and r-band) have very low signal-to-noise and are likely outliers, a change in decline rate is observed in some bands near the end of the first season, which subsequently extends into the second.
\emph{Right:} A detailed view of the first season's light curve which highlights the aforementioned change in decline rate. This is mainly evident in the g-band, while similar effects are weaker or absent in r, i, and z. In \S \ref{subsec:basic LC props} we fit a linear decline rates to each band in two regimes, which are highlighted here: 20 to 60 days and 60 to 120 days. The green and red lines show results from these fits for the g-band (see also Table~\ref{tab:SN-props}).
}
\label{fig:LC-zoom}
\end{figure*}

\section{Observations and Data Reduction} \label{sec:obs}

We present multi-wavelength observations of PS1-11aop, which was discovered by the Panoramic Survey Telescope and Rapid Response System (Pan-STARRS) Medium-Deep Survey (PS1-MDS). 

\subsection{Discovery with PS1-MDS} \label{subsec:discovery}

PS1 is a wide-field imaging system designed for dedicated survey observations. It has a 1.8\,m diameter primary mirror and an $f/4.4$ Cassegrain focus which, when coupled with a set of 4800$\times$4800 pixel detectors, produces images covering a 3.3\,$\deg$ diameter field with low distortion \citep{Kaiser2010}.
Observations are obtained through a set of five broad filters: $g_{P1}r_{P1}i_{P1}z_{P1}y_{P1}$. A full description of the filters and photometry system used by PS1 are given in \cite{Stubbe2010} and \cite{Tonry2012}.

The PS1-MDS was a component of the PanSTARRS1 Science Consortium observational program that ran between 2009 and 2014. Approximately 25\% of observing time on PS1 was dedicated to revisiting 10 PS1-MDS fields (each consisting of a single PS1 footprint) on a nearly nightly basis. Initial reduction of the PS1-MDS images was performed through the  PS1 image processing pipeline \citep{Maigner2008}. Difference images were then produced from nightly stacked images and potential transients identified using the \textsc{photpipe} pipeline \citep{Rest2014}. These potential transients were visually inspected by humans for possible promotion to the status of transient alerts and flagged for potential spectroscopic follow-up. 

PS1-11aop was first detected in a PS1-MDS i-band image obtained on 23 July 2011 UTC (MJD\,55765.93). It was located at $\alpha(J2000) = 23^{h}30^{m}48^{s}.917$ and $\delta(J2000) = -01\arcdeg12\arcmin26\arcsec.59$, which is close to the center of its host galaxy (see Figure~\ref{fig:hostimg}) and had an apparent magnitude of $m_{\rm{i}} = 19.41 \pm 0.02$ AB mag. This was the first image obtained of its field in the 2011 season, after a roughly $\sim$195 day (observer frame) gap.

\subsection{Optical Photometry from PS1-MDS} \label{subsec:optical data}

During the normal operations of the PS1-MDS, the field of PS1-11aop was observed with an approximately nightly cadence during five $\sim$4 month seasons between 2009 and 2014. In optimal observing  conditions, $g_{P1}$ and $r_{P1}$ were observed  on  the  same night with $i_{P1}$ and $z_{P1}$ observations following on consecutive evenings to a depth of $\sim$23.3\,mag. Thus, densely sampled light curves were produced, with each band having a roughly three-day cadence.

To produce the final $g_{P1}$,$r_{P1}$,$i_{P1}$ and $z_{P1}$ photometry of PS1-11aop, deep template images were first constructed by stacking pre-explosion survey images. On average, 120--130 high-quality nightly stacked images were combined, weighted by the product of the inverse variance and the inverse area of the point-spread function (PSF). The typical depth achieved in the template is $\sim$25.0\,mag. With these in hand, \textsc{photpipe} was used to produce forced PSF photometry on a set of difference images, as described in \citet{Rest2014} and \citet{Villar2020}. An appropriate PSF was chosen for each difference image while applying a common centroid from the high signal-to-noise detections. 

\begin{deluxetable}{lccccccc}
\tabletypesize{\small}
\centering
\tablecaption{Optical Photometry \label{tab:photometryresults}}
\tablehead{
\colhead{MJD} & \colhead{Phase\tablenotemark{a}} & \colhead{Band} & \colhead{mag} & \colhead{$\sigma_{mag}$} \\
\colhead{} & \colhead{(days)} & \colhead{} & \colhead{(AB)} & \colhead{(AB)}
  }
\startdata
55768.62 & -1.94 & g$_{\rm{P1}}$ & 20.00 & 0.02 \\
55770.56 & 0.00 & g$_{\rm{P1}}$ & 19.91 & 0.02 \\
55773.02 & 2.46 & g$_{\rm{P1}}$ & 19.98 & 0.02 \\
55775.49 & 4.92 & g$_{\rm{P1}}$ & 20.03 & 0.02
\enddata
\tablenotetext{a}{This is the rest-frame days since observed g-band maximum with reference epoch MJD 55770.56}
\tablecomments{Table \ref{tab:photometryresults} is published in its entirety in the electronic edition. A portion is shown here for guidance regarding its form and content.}
\end{deluxetable}

PS1-11aop remained bright above the detection threshold of the PS1-MDS for $\gtrsim$1\,year (2 seasons) in all bands. Deep upper limits were also obtained starting from $\sim$600\,days prior to the g-band maximum to more than 600\,days after. The $g_{P1}$,$r_{P1}$,$i_{P1}$,$z_{P1}$ lightcurves of PS1-11aop were released as part of the PS1-MDS supernova sample in \citet{Villar2020}. The transient is named PSc300140 in the above article. For ease, we provide the measured apparent magnitudes and 3$\sigma$ upper limits in the AB system in Table \ref{tab:photometryresults}. This data is plotted in Figure~\ref{fig:LC-zoom}. 

Throughout this manuscript, unless stated otherwise, we will give the rest-frame phase relative to the time of observed g-band maximum, which occurred on MJD\,55770.562.

\subsection{Optical Spectroscopy} \label{subsec:optical spec}

We obtained two optical spectra of PS1-11aop while the transient was active. The first was obtained with the Hectospec multi-fiber spectrograph \citep{Fabricant2005} on the Multiple Mirror Telescope (MMT) on 27 October 2011 (90 days after g-band maximum). The second was obtained with the Blue Channel spectrograph \citep{Schmidt1989} on the MMT on 20 July 2012 (357 days after g-band maximum). Subsequently, after the transient had faded, we obtained a spectrum of the host galaxy with the Low Dispersion Survey Spectrograph-3 (LDSS3; \citealt{Allington-Smith1994}) on the Magellan-Baade telescope on 23 September 2014. Initial reduction (overscan correction, flat fielding, extraction, wavelength calibration) for the long-slit spectra was carried out using the standard packages in IRAF\footnote{IRAF was distributed by the National Optical Astronomy Observatory, which was managed by the Association of Universities for Research in Astronomy (AURA) under a cooperative agreement with the National Science Foundation}, while the Hectospec spectrum was reduced using the CfA pipeline for this instrument\footnote{\url{https://www.mmto.org/hsred-reduction-pipeline/}}.  Flux calibration and telluric correction were performed using spectrophotometric standard stars observed on the same night as the science observations and a set of IDL scripts described in \citealt{Blondin2012}.

\begin{figure}[t!]
\centering
\includegraphics[trim=1.5cm 0cm 2.5cm .10cm,clip,width=\columnwidth]{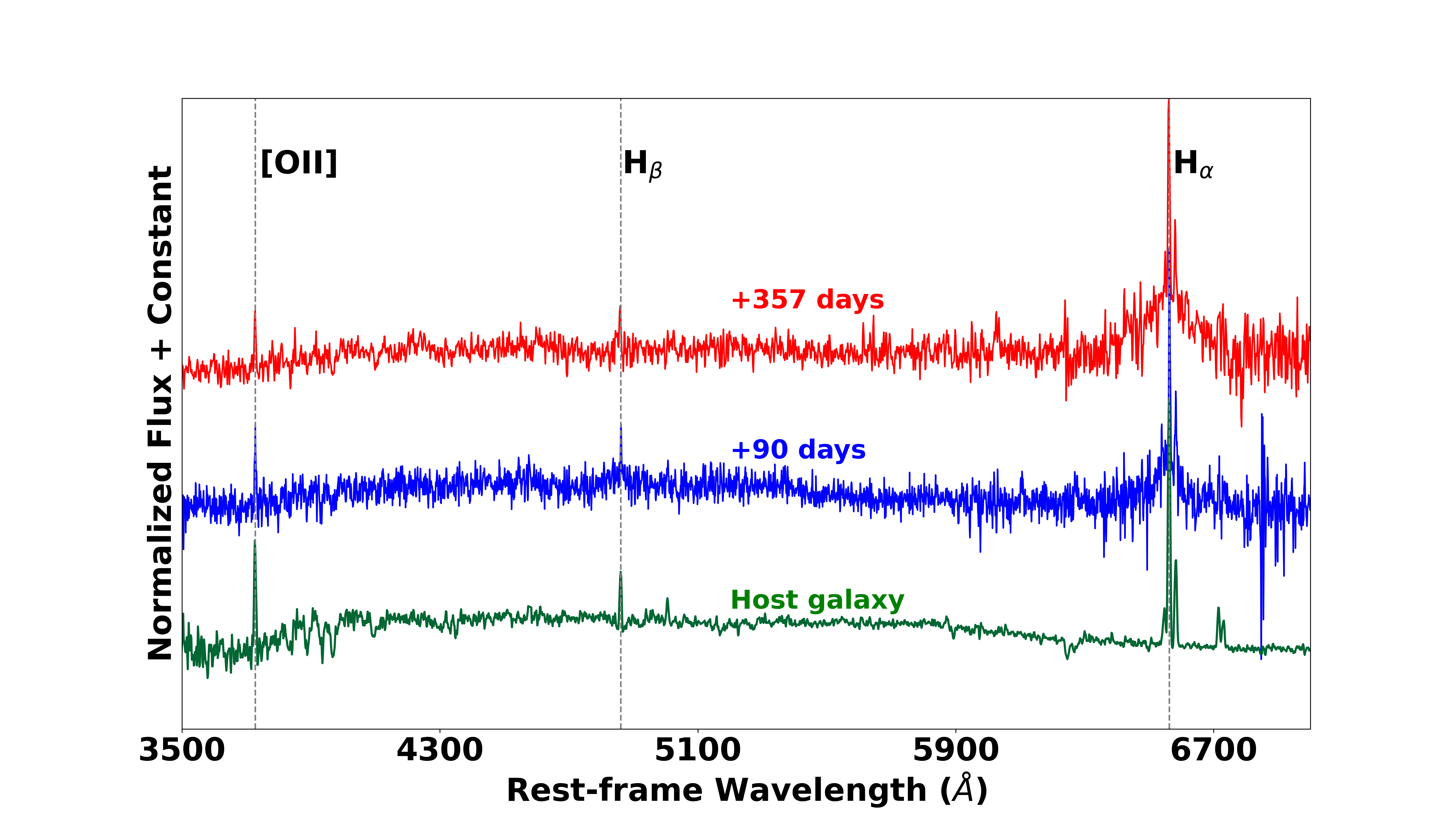}
\caption {Spectra of PS1-11aop at two epochs relative to the time of g-band maximum in the rest frame as seen on the legend. The green spectrum is for the host galaxy. Vertical lines mark the dominant strong spectral features, identified as $H_{\alpha}$, $H_{\beta}$, and [OII].}
\label{fig:spectra}
\end{figure}

\begin{figure*}
\centering
\includegraphics[width=0.9\textwidth]{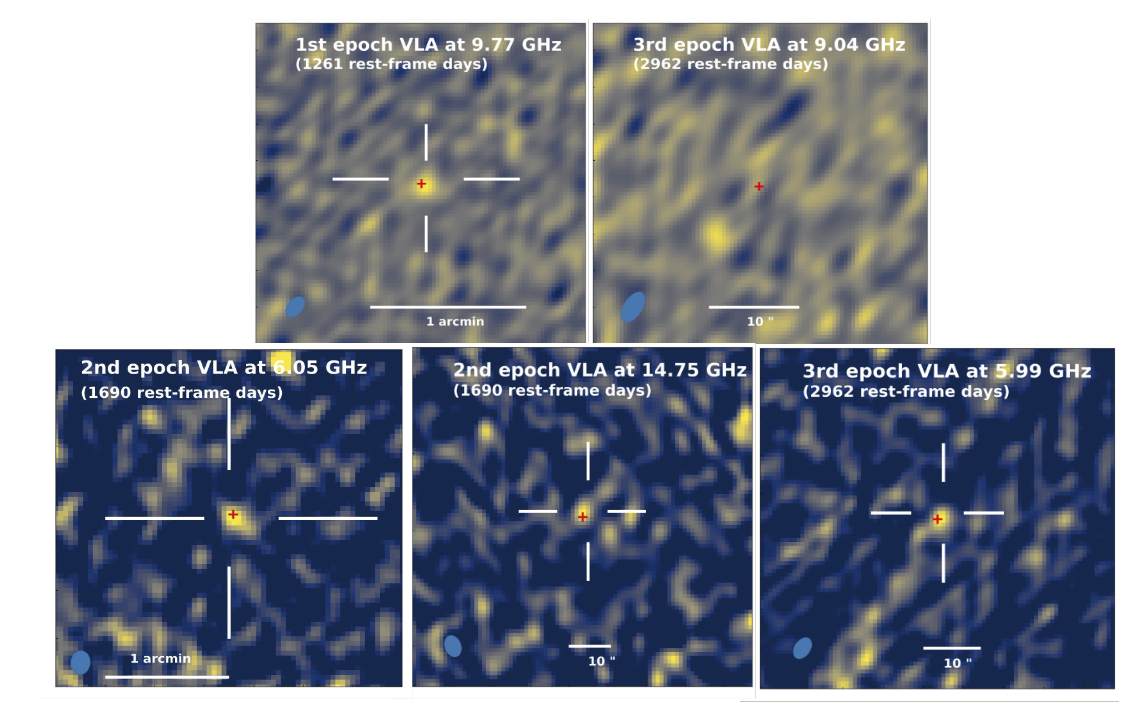}
\caption{VLA observations of PS1-11aop at multiple epochs and frequencies.\emph{Top:} X-band images at the first and third epochs showing fading of the radio emission at 9 GHz. \emph{Bottom:} Observations at 6 GHz and 14 GHz from the 2nd epoch along with observations at 6 GHz from the 3rd epoch. }
\label{fig:radio-images}
\end{figure*}

Broadly, we observed intermediate-width hydrogen emission features.  If due to a SN, these lines most closely resemble those of Type IIn SN (e.g., SN2005ip-- \citealt{Smith2017}; SN2010jl-- \citealt{Fransson_2014}; SN2015da-- \citealt{Smith2023}) as opposed to the PCygni lines observed in other Type II SN \citep{Filippenko1992}. However, intermediate width emission features have also been observed in other classes of transients such as tidal disruption events (TDEs; \citealt{Charalampopoulos2022}). In addition, the presence of narrow emission lines of nebular features such as [OIII] indicate a non-negligible contribution from the host galaxy, even in our explosion spectra. Properties of the intermediate-width lines will be further addressed in \S \ref{sec:spec props} and a discussion of the classification of this object as a supernova is provided in \S \ref{sec: argument for sn}.

We use narrow emission lines of H$\alpha$ (6563\AA),
H$\beta$ (4861\AA), [OII] (3727\AA) and the doublets of
[NII] (6548,6582\AA) and [SII] ( 
6716,6731\AA) in the host galaxy spectrum to measure a redshift of 
0.218$\pm$0.001. This corresponds to a luminosity distance of 
1070$\pm$4 Mpc, which we adopt for PS1-11aop throughout this 
manuscript. Both the explosion and host galaxy spectra are shown in Figure~\ref{fig:spectra}.

\subsection{Radio Observations} \label{subsec:radio obs}
We carried out observations of PS1-11aop with the Karl G. Jansky Very Large Array (VLA) interferometer on three epochs between 2015 and 2021 ($\sim$4--10 years post-discovery). The first epoch was obtained on the 12th of October 2015 (MJD 57307.08) as part of a late-time X-band ($10$ GHz) survey of PS1$-$MDS Type IIn supernovae (program number 15B-237; PI: Drout). Observations were carried out in the D configuration of the VLA, which is where the array is most compact (and hence is when the beam size is largest), and a radio source coincident with the optical transient position was identified. As a result of the detection, a second epoch of multi-frequency observations was obtained on the 17th of March 2017 (MJD 57829.76), also in the D configuration, in the S-(3\,GHz), C-(6\,GHz),  X-(10\,GHz), and K$_U$-(15 GHz) bands (program number 17A-226; PI: Drout). Finally, a third epoch was obtained on the 13th of June, 2021 (MJD 59378.45) tracking the continued emission, using C configuration, in the S-, C-, and X-bands (program number 21A-317; PI: Drout). All observations were taken with the 8-bit VLA samplers.

The Common Astronomy Software Applications package (CASA) \citep{casa2007} was used to perform the data reduction, flux measurement calibration, and imaging of the data. Specifically, we used the python-based CASA pipeline tool \texttt{pwkit} released by \cite{peterwilliams2017}. We flagged radio frequency interference (RFI) using the automatic AOFlagger. The bandpass calibration was done using 3C147 which is the primary calibrator while the secondary calibrator is J0145$-$2733. After data reduction, we imaged the total intensity component (Stokes I) of the source visibilities, setting the cell size so there would be 4$-$5 pixels across the width of the synthesized beam. All calibrated data were imaged using the CASA \textsc{tclean} task and Briggs weighting with a robust factor of 0.5. 

The task \texttt{fitsrc} was used to measure fluxes and associated uncertainties within the image. In cases where no radio source was identified at the location of the optical transient, 
we compute an upper limit based on the root-mean-square (rms) noise in the final image at the location of PS1-11aop. At each epoch, a detection was found at greater than 3$\sigma$ in one of the observed bands (X-band during epoch 1 and C-band at epochs 2 and 3). In addition, during epoch 2, we report a marginally significant (2.7$\sigma$) source at the location of the SN in the Ku-band. 
The summary of all measured radio flux densities and upper limits for the three epochs as well as their corresponding luminosities are presented in Table \ref{tab:radioresults}. 

\begin{figure*}[ht]
\centering
\includegraphics[width=0.45\textwidth]{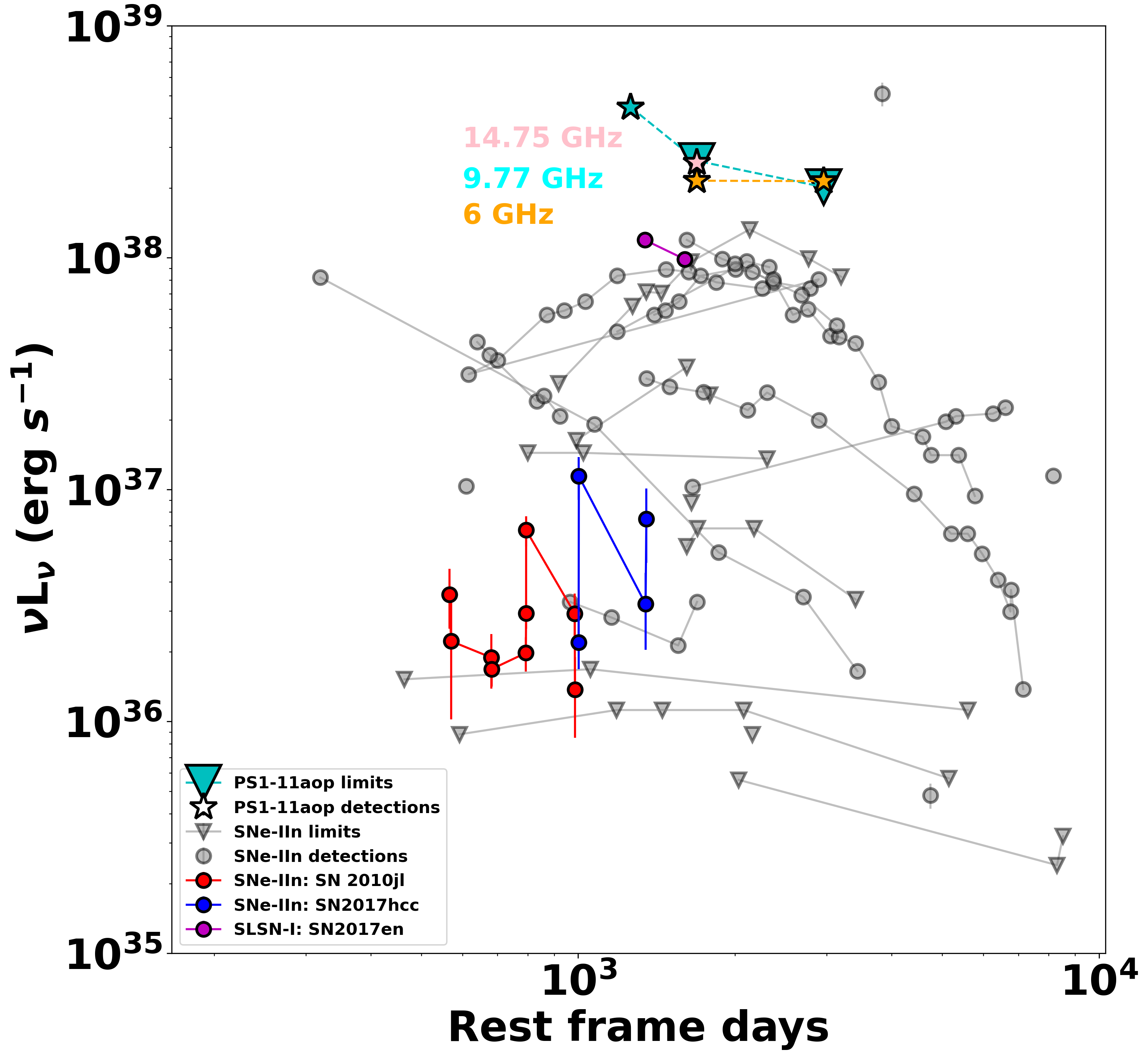}
\includegraphics[width=0.45\textwidth]{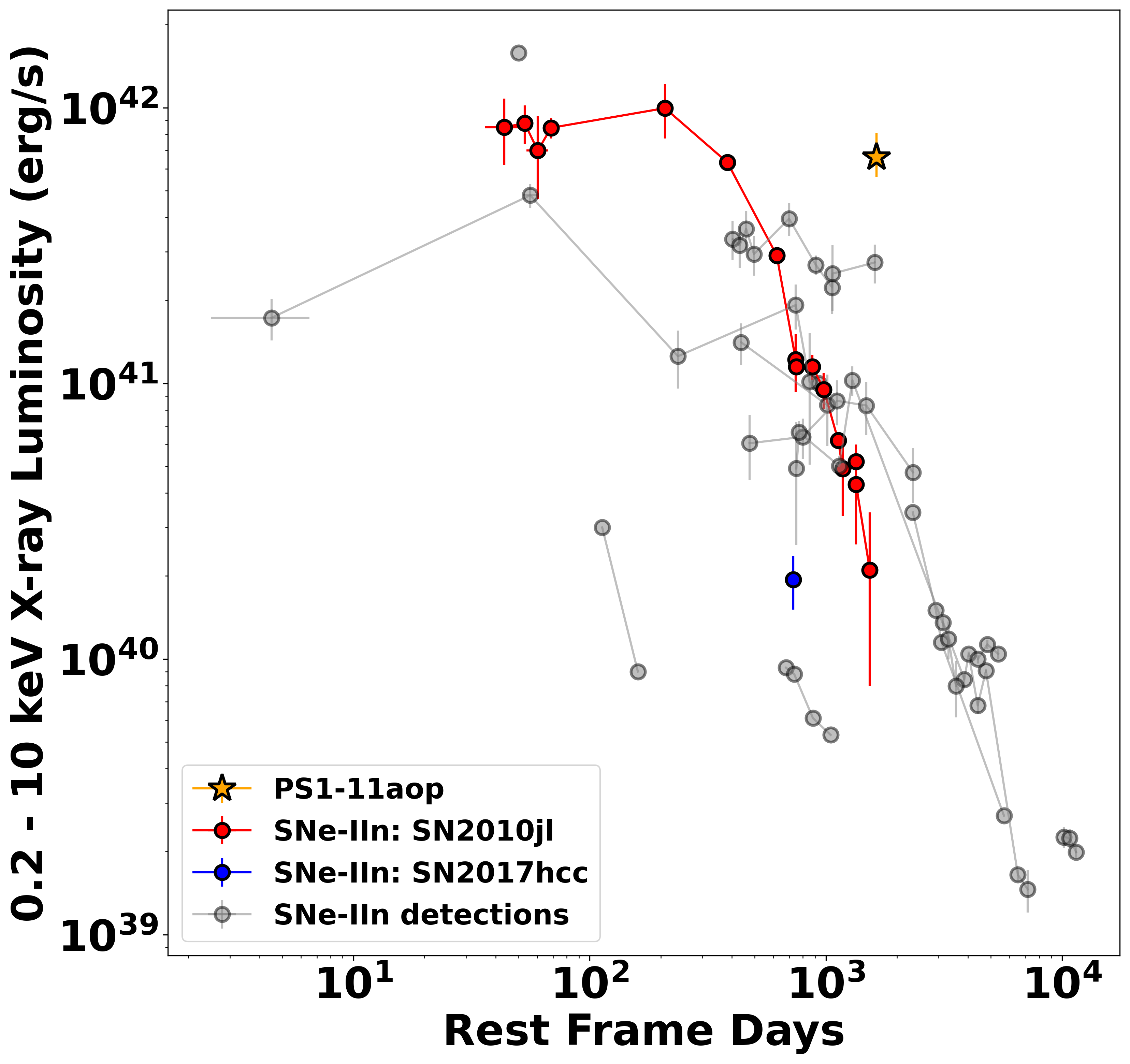}
\caption{\emph{Left:} Comparison radio emission detected at the location of PS1-11aop to other SN$-$IIn in the literature. Circles represent detections while triangles represent 3$\sigma$ upper limits. Colored stars are observed points for PS1-11aop, while other events from the literature are shown as gray markers: \citep{vandyk1996A, Chandra2009, Chandra2012, pooley2002ApJ, Shle1999AJ, SmithI2007, PonaamandAlicia2007, Chevaliar1987Natur, vanDyke1993ApJ, Hoffman_2008, schinelz2009ApJ, Prez_Torres_2009, Bauer2008ApJ,Chandra2015, Stroh2021, Chandra2022}. The red light curve of the long-lived SN2010jl and the blue light curve of SN2017hcc are plotted for context. We also highlight SN2017en, which was initially observed as a hydrogen-poor superluminous SN, in magenta. This event was which was detected in radio at multiple frequencies and was found to be interacting with H-rich CSM at late times \citep{Margutti2023}.  SN\,2002hi (single gray point with the highest luminosity) is the brightest radio emission ever detected and was reported by \cite{Stroh2021}. \emph{Right:} Unabsorbed X-ray comparison plot for PS1-11aop and other SN$-$IIn in the literature. The orange star is the single epoch detection of PS1-11aop while the gray circles are detections of other SN$-$IIn from \citealt{Dwarkadas2012, Ofek2014, Temple2005, Schlegel2006, pooley2002ApJ, Brennan2023, Chandra2015,Chandra2022}. We also show the detections of SN2010jl and SN2017hcc for context. The time of PS1-11aop on both plots is in rest-frame days since the observed g-band maximum which is MJD 55770.562. The times plotted for other SN$-$IIn were taken from their respective published articles.}
\label{fig:LCradio/x-ray}
\end{figure*}

The two 6-GHz detections in epochs 2 and 3 are consistent within the errors, while we find a fading of a factor $\gtrsim$2 at 9 GHz between epochs 1 and 3. Based simply on the flux density error of the first epoch, this fading would be significant at a level of 2.64$\sigma$. However, given the importance of understanding whether the observed radio emission is variable, we perform an additional test. We assume that the true flux densities for epochs 1 and 3 at 9 GHz are equal (with  F$_{\nu} = 40$\,$\mu$Jy). We then generated 1000 data realizations based on this mean flux density and the rms noise of each observation (7.8 and 6.6\,$\mu$Jy, respectively). The results show that only 2.5\% of the realizations had a flux difference as large as the observed upper limit ($>$20$\mu$Jy), indicating that the flux difference is statistically significant (p-value $=$ 0.025).

In Figure~\ref{fig:radio-images} we show a series of five VLA image cutouts showing the detection and subsequent non-detection at 9 GHz, the two 6 GHz detections, and the marginal Ku-band source. In Figure~\ref{fig:LCradio/x-ray}, we plot $\nu$L$_{\rm{\nu}}$
for our observations of PS1-11aop compared to (5$-$10\,GHz) radio observations of Type IIn SN from the literature.
If the radio emission from the location of PS1-11aop is truly from a SN, then it would be one of the most luminous radio SN discovered to date.

\begin{deluxetable}{lcccccc}
\tabletypesize{\small}
\tablecaption{Radio Observations of PS1-11aop\label{tab:radioresults}}
\tablehead{\colhead{MJD } & \colhead{Phase\tablenotemark{a}} & \colhead{Central} & \colhead{Flux density$^{b}$}  & \colhead{L$_{\nu}$} & \colhead{$\nu$ L$_{\nu}$}\\
\colhead{} & \colhead{} & \colhead{Freq.} & \colhead{} & \colhead{$\times$10$^{28}$} & \colhead{ $\times$10$^{38}$} \\
\colhead{} & \colhead{(days)} & \colhead{(GHz)} & \colhead{($\mu$Jy)} & \colhead{(ergs/s/Hz)} & \colhead{ (ergs/s)} }
\startdata
          57307.08 & 1261.51 & 9.77 & 40.4$\pm$7.8 & 4.55$\pm$0.88  & 5.87$\pm$1.13 \\ 
\hline
          57829.76 & 1690.64 & 3.00 &  $<$ 210.5 & $<$ 23.7 & $<$9.39 \\
          57829.76 & 1690.64 & 6.05 & 31.8$\pm$9.5 & 3.57$\pm$1.07  & 2.86$\pm$0.85\\
          57829.76 & 1690.64 & 9.02 & $<$ 25.8 & $<$ 2.90 & $<$3.46\\
          57829.76 & 1690.64 & 14.75 &  15.5$\pm$5.6$^{c}$ & 1.74$\pm$0.63  & 3.39$\pm$1.23\\
\hline
          59378.45 & 2962.14 & 2.99 &  $<$ 56.7 &  $<$6.38 & $<$2.52\\
          59378.45 & 2962.14 & 5.99 & 31.7$\pm$6.3 & 3.57$\pm$0.71  & 2.82$\pm$0.56 \\
          59378.45 & 2962.14 & 9.04 &  $<$ 19.8 &  $<$2.23 & $<$2.67  \\
\hline	
\enddata
\tablenotetext{a}{Rest-frame days since observed g-band maximum which is MJD 55770.562}
\tablenotetext{b}{Upper limits presented are 3$\sigma$}
\tablenotetext{c}{This is a marginal detection at 2.7$\sigma$}
\end{deluxetable}

\subsection{X-ray Observations} \label{subsec:x-ray obs}
X-ray follow-up observations of PS1-11aop were taken on the 16th January 2017 (MJD 57769.35; which is 1641 rest-frame days after g-band maximum) with an exposure time of 9.92\,ks with the Chandra X-ray Observatory (CXO) as part of a late-time X-ray survey of PS1-MDS luminous interacting SN through the program GO7-18045A (PI: Maria Drout). After applying standard ACIS data filtering with the CIAO software package (version 4.11 \citealt{Fruscione2006}), we measure counts within a 1$\arcsec$ radius source region (chosen to encompass 90\% of the PSF at the source location) and a 35$\arcsec$ background region. Three counts are present in the source region, which is significant above the background at a 4.2$\sigma$ level (assuming Poisson statistics are appropriate for the low count region). Using the CIAO function \texttt{srcflux}, we measure a count rate at the location of PS1-11aop of 3.26$^{+2.45}_{-1.64} \times 10^{-4}$ counts/s, where the upper and lower bounds denote the 68\% confidence interval.  

\begin{deluxetable}{l|cccc}
\tabletypesize{\small}
\tablecaption{X-ray Flux Measurements (0.2$-$10\,keV)\label{tab:xrayflux}}
\tablehead{\colhead{kT } & \colhead{Flux$_{abs}$} & \colhead{Flux$_{unabs}$} & \colhead{Luminosity} & \colhead{EM} \\
\colhead{(keV)} & \multicolumn{2}{c}{(10$^{-15}$ erg/s/cm$^{2}$)}  & \colhead{(10$^{41}$ erg/s)} & \colhead{(10$^{64}$ cm$^{-3}$)} }
\startdata
80.0 & 5.2$_{-2.6}^{+3.9}$ & 5.4$_{-2.7}^{+4.1}$ & 7.4$_{-3.7}^{+5.6}$  & 6.7$_{-3.4}^{+5.0}$ \\ 
20.0 & 4.6$_{-2.3}^{+3.4}$ & 4.8$_{-2.4}^{+3.6}$ & 6.6$_{-3.3}^{+5.0}$ & 5.2$_{-2.6}^{+3.9}$ \\ 
8.0 & 3.9$_{-2.0}^{+2.9}$ & 4.2$_{-2.1}^{+3.1}$ & 5.7$_{-2.9}^{+4.3}$  & 5.1$_{-2.6}^{+3.9}$ \\ 
2.0 & 2.9$_{-1.4}^{+2.2}$ & 3.4$_{-1.7}^{+2.6}$ & 4.7$_{-2.4}^{+3.5}$ & 8.8$_{-4.4}^{+6.6}$\\ 
0.8 & 3.4$_{-1.7}^{+2.6}$ & 4.5$_{-2.2}^{+3.3}$ & 6.1$_{-3.1}^{+4.6}$ & 25.0$_{-12}^{+19}$\\ 
\enddata
\end{deluxetable}

We convert this count rate to flux assuming a thermal bremsstrahlung spectrum at a redshift of $z = 0.218$ and a range of temperatures between $kT = 0.8$ and 80 keV using the \texttt{xszbremss} model within CIAO/Sherpa. This range was chosen to encompass a wide range of possible temperatures for both the forward and reverse shock in interacting SNe. A shock temperature of $\sim$20 keV was measured for SN\,2010jl at two years post-explosion \citep{Ofek2014,Chandra2015}, while 3 keV was adopted for SN\,2017hcc \cite{Chandra2022} and higher temperatures have been predicted for the forward shock in some CSM configurations \citep[e.g.][]{Fransson1996}. 
We summarize our results in Table~\ref{tab:xrayflux} where we provide the measured absorbed fluxes, unabsorbed fluxes, and luminosities. All quoted values are in the range of $0.2-10$ keV. For the unabsorbed fluxes and luminosities, we have corrected only for the galactic $N_H$ of 4.09$\times$10$^{20}$ cm$^{-2}$ \citep{HI4PI2016}. The possibility of additional absorption local to the explosion site of PS1-11aop will be discussed further in Section~\ref{sec:x-ray modelling}. Finally, we calculate the required emission measure using the normalization of each of these spectral models ($EM = \int n_e n_I dV$). These values are provided in the final column of Table~\ref{tab:xrayflux}.

For the temperatures considered, the resulting fluxes and luminosities span less than a factor of two, indicating that our adopted flux is relatively insensitive to the specific temperature of the emission within this range. 
In the right panel of Figure~\ref{fig:LCradio/x-ray}, we plot the luminosity inferred for a temperature of kT$=$20 keV. The X-ray emission from the location of PS1-11aop, if caused by a SN, would be more luminous than any observation of a Type IIn SN at $>1000$ days post-explosion, rivaled only by SN\,2010jl at times $\lesssim$400 days and SN\,2002hi at 50 days post-explosion.

\section{Photometric and Spectroscopic Properties} \label{sec:opticalprops}
Here, we present the basic properties of PS1-11aop, based on an analysis of its optical light curves and spectra.

\subsection{Basic light curve properties} \label{subsec:basic LC props}

In Figure~\ref{fig:LC-zoom}, we plot the light curves for PS1-11aop. This plot and all calculations below are given in rest-frame days. Absolute magnitudes are shown on the left axis and were calculated after correcting for both galactic foreground extinction and the distance to the transient. We adopt a value of $E(\bv)_{\rm{MW}} = 0.0347(9)$\,mag from \citep{Schlafly2011} and correct the PS1 photometry using a standard Milky Way extinction curve with R$_{V}$ $=$ 3.1. We do not correct for any extinction intrinsic to the host galaxy of PS1-11aop. To correct for cosmological expansion, we do not perform full k-corrections to shift the wavelengths covered by each band to the rest frame, but instead use the equation: M $=$ m $- 5 log_{10}(\rm{d}_{\rm{L}}/ 10 \rm{pc}) + 2.5\,log_{10}(1+z)$ where d$_{\rm{L}}$ is the luminosity distance and $z$ is the redshift obtained from the host galaxy spectrum (See \S \ref{subsec:optical spec}).  Basic properties of the light curve are given in Table~\ref{tab:SN-props}.

The light curves have peak absolute magnitudes which range from $M = -20.14$ mag (g-band) to $M = -20.69$ mag (i- and z-bands). This would place PS1-11aop on the luminous end of Type IIn SNe, although it is fainter than the energetic transients with narrow/intermediate width emission lines that were identified in the cores of galaxies by \citet{Kankare2017}. In addition, PS1-11aop displays a remarkably slow decline, with magnitudes brighter than $-18.5$ mag for more than 1 year post-explosion.

\begin{deluxetable}{llcc}
\tabletypesize{\small}
\tablecaption{PS1-11aop optical properties \label{tab:SN-props}}
\tablehead{\colhead{Properties} & \colhead{values} }
\startdata    
          Peak absolute g-band magnitude (AB) & $-$20.14 $\pm$ 0.02  \\
          Peak absolute r-band magnitude (AB) & $-$20.47 $\pm$ 0.02  \\
          Peak absolute i-band magnitude (AB) & $-$20.69 $\pm$ 0.03  \\
          Peak absolute z-band magnitude (AB) & $-$20.69 $\pm$ 0.05  \\
          g-band decline rates (10$^{-2}$ mag day$^{-1}$)\tablenotemark{a} & 1.3(0.1), 0.3(0.1), 0.4(0.2) \\
          r-band decline rates (10$^{-2}$ mag day$^{-1}$)\tablenotemark{a} & 0.7(0.1), 0.5(0.1), 0.7(0.1)\\
          i-band decline rates (10$^{-2}$ mag day$^{-1}$)\tablenotemark{a} & 0.7(0.1), 0.3(0.1), 0.7(0.1)\\
          z-band decline rates (10$^{-2}$ mag day$^{-1}$)\tablenotemark{a} & 0.6(0.1), 0.7(0.1), 0.8(0.2) \\
          Peak bolometric luminosity, L$_{bol}$ (erg $s^{-1}$) & $5.42 \pm 0.25 \times 10^{43}$ \\
          Total radiated energy, $E_{rad}$ (ergs)  & $>$ (8.42 $\pm$ 0.17) $\times 10^{50}$ \\  
\enddata
\tablenotetext{a}{Measured between 20--60, 60--120, and 300--380 days post g-band maximum, respectively.}
\end{deluxetable} 

In terms of light curve morphology, we observe 1$-$2 rising light curve points in each observed band, and the $riz-$band light curves all appear to be rising or relatively flat for the first $\sim$20 days, indicating we may have observed the true light curve peak (the g-band begins to decline more steeply immediately after maximum). 

We then measure a linear decline rate for all four bands in three-time windows (all relative to g-band maximum): 20--60 days, 60--120 days, and 300--380 days. The breaks between these windows are set by an apparent change in slope observed in the g-band light curve (see Figure~\ref{fig:LC-zoom}, right panel) and the observing gap between seasons. Indeed, we find that the g-band initially shows a decline rate of $\sim$0.013 mag\,day$^{-1}$, before shallowing to $\sim$0.003 mag\,day$^{-1}$ after 60 days. The i-band light curve (which contains H$\alpha$ at the redshift of PS1-11aop) also shows a slight shallowing in decline rate after 60 days, although less significant ($\sim$0.007 to $\sim$0.003 mag\,day$^{-1}$) while there is no evidence for a change in decline rate during the first season in either the r$-$ or z$-$bands (See Table \ref{tab:SN-props} for associated uncertainties). In general, the decline rate remained similar in all bands during the second season of observations, although we caution that our data during this phase has significantly lower signal-to-noise\footnote{In particular, we note that the two low points near the end of the r-band light curve in Figure~\ref{fig:LC-zoom} have errors of $\sim$0.35 mag and are only separated from other points by $<$2$\sigma$. Hence, they are likely outliers.}. 

Finally, we also measure deep pre-explosion limits in all four bands. These limits were obtained over two distinct time ranges spanning $\sim$190--310 and $\sim$485--570 days prior to the g-band maximum (with a gap in between when the field of PS1-11aop was near solar conjunction). We find no evidence for pre-explosion outbursts to median depths of $\sim$ $-$16.57 mag for g-band, $-$16.96 mag for r-band, $-$17.31 mag for i-band, and $-$17.68 mag for z-band.

\subsection{Constraints on the explosion epoch} \label{subsec:estimate of exp epoch}

As described above, PS1-11aop was present in the first observations taken of its field in the 2011 season. The last pre-explosion upper limit was an r-band observation at 193 rest-frame days prior to the g-band maximum. This places an upper bound on the explosion epoch. However, we also observe a rising light curve in the first 2$-$3 epochs of all four PS1 bands, indicating that the explosion may have been more recent (see inset of Figure~\ref{fig:LC-zoom}). 

While we caution that there are large uncertainties, we investigate the explosion epoch that would be inferred from fitting a power law to the rising portion of the observed light curve. Specifically, we fit a power-law to the fluxes of the rising portion of the i-band light curve of the form $F_{\nu} \propto (t - t_0)^{\alpha}$ and find a best-fit power law index of $\alpha = 0.043 \pm 0.026$. We chose i-band because it has three points rising consecutively. Extrapolating the best-fit power law, we would infer an explosion epoch ($t_0$ in the equation above) only $\sim$5\,days prior to the g-band maximum, indicating that PS1-11aop would have risen quite rapidly. However, due to the small number of points available on the rise, we consider this a lower limit since the explosion may have occurred at any point between $\sim$5 and 190 days prior to the g-band maximum. For the rest of this section, we will continue to quote phase relative to g-band maximum.

\begin{deluxetable*}{lccccccccccccccc}
\tabletypesize{\small}
\tablecaption{Pseudo-Bolometric Luminosity and Blackbody properties of PS1-11aop\label{tab:bol-BB}}
\tablehead{ \colhead{MJD} & \colhead{Phase\tablenotemark{a}} & \colhead{g-band\tablenotemark{b}} & \colhead{g$_{err}$} & \colhead{r-band} &  \colhead{r$_{err}$} & \colhead{i-band} &  \colhead{i$_{err}$} & \colhead{z-band} &  \colhead{z$_{err}$} & \colhead{T$_{BB}$} & \colhead{T$_{BB}err$} & \colhead{R$_{BB}$} & \colhead{R$_{BB}err$} & \colhead{L$_{bol}$} & \colhead{L$_{bol}err$}\\
\colhead{} & \colhead{} & \colhead{} &  \colhead{} & \colhead{} &  \colhead{} & \colhead{} &  \colhead{} & \colhead{} &  \colhead{} & \colhead{}  & \colhead{} & \colhead{10$^{14}$} & \colhead{10$^{14}$} & \colhead{10$^{43}$} & \colhead{ 10$^{42}$} \\
\colhead{} & \colhead{(days)} & \colhead{(Jy)} &  \colhead{(mag)} & \colhead{(mag)} &  \colhead{(mag)} & \colhead{(mag)} &  \colhead{(mag)} & \colhead{(mag)} &  \colhead{(mag)} & \colhead{(K)}  & \colhead{(K)} & \colhead{(cm)} & \colhead{(cm)} & \colhead{(erg/s)} & \colhead{ (erg/s)} }
\startdata
55767.67 & $-$2.892 & 19.817	& 0.033	 & 19.553 & 	0.029	& 19.314	& 0.026	& 19.271	& 0.025 &	7126.41	& 126.99 &	30.26	& 1.023	& 5.36	& 2.37 \\ 
55769.93 & $-$0.633 & 19.846 & 0.034 &	19.519 & 0.029 &	19.289 &	0.025 & 19.266 &	0.024 &	7040.46 & 122.45 &	31.13	& 1.037	& 5.40	& 2.33 \\
55772.94 &	2.379 &	19.884 & 0.034 & 	19.495 &	0.028 &	19.270 &	0.025 &	19.263 &	0.025 &	6934.97 & 117.829 &	32.13 &	1.06 &	5.42 & 2.54 \\        
55774.45 &	3.89 &	19.903 &	0.035 & 19.489 &	0.028 &	19.265 &	0.0251 &	19.263 &	0.025 &	6886.05	& 116.002 &	32.57 &	1.07	& 5.42	& 2.32 \\
55777.46 & 	6.89 &	19.939 &	0.035 &	19.488 &	0.028 &	19.260 &	0.025 &	19.265 &	0.025 &	6794.44	& 113.081	& 33.345 &	1.09 &	5.39	& 2.10
\enddata
\tablenotetext{a}{This is the rest-frame days since observed g-band maximum with reference epoch 55770.56}
\tablenotetext{b}{These are AB magnitudes interpolated to a common set of epoches and used to produce the color evolution plot Figure \ref{fig: tempradcolor}.}
\tablecomments{Table \ref{tab:bol-BB} is published in its entirety in the electronic edition. A portion is shown here for guidance regarding its form and content.}
\end{deluxetable*}

\subsection{Color, Temperature, and Radius evolution} \label{subsec:temprad}

To probe the color, photospheric temperature, and photospheric radius evolution of PS1-11aop, we interpolate the multi-band data to a set of common epochs (since the $g_{P1}r_{P1}i_{P1}z_{P1}$ bands are not observed at the same time). We use low-order polynomials to fit the data to carry out the interpolation and chose epochs by requiring that all bands were observed within a time period of $\sim$3\,days. Uncertainties are propagated using a Monte Carlo technique where we create 1000 realizations of our light curves, drawing from the mean and error appropriate for each data point.

\begin{figure}[t]
\includegraphics[width=.45\textwidth,trim=0.9cm 1cm 2.25cm 2.50cm,clip]{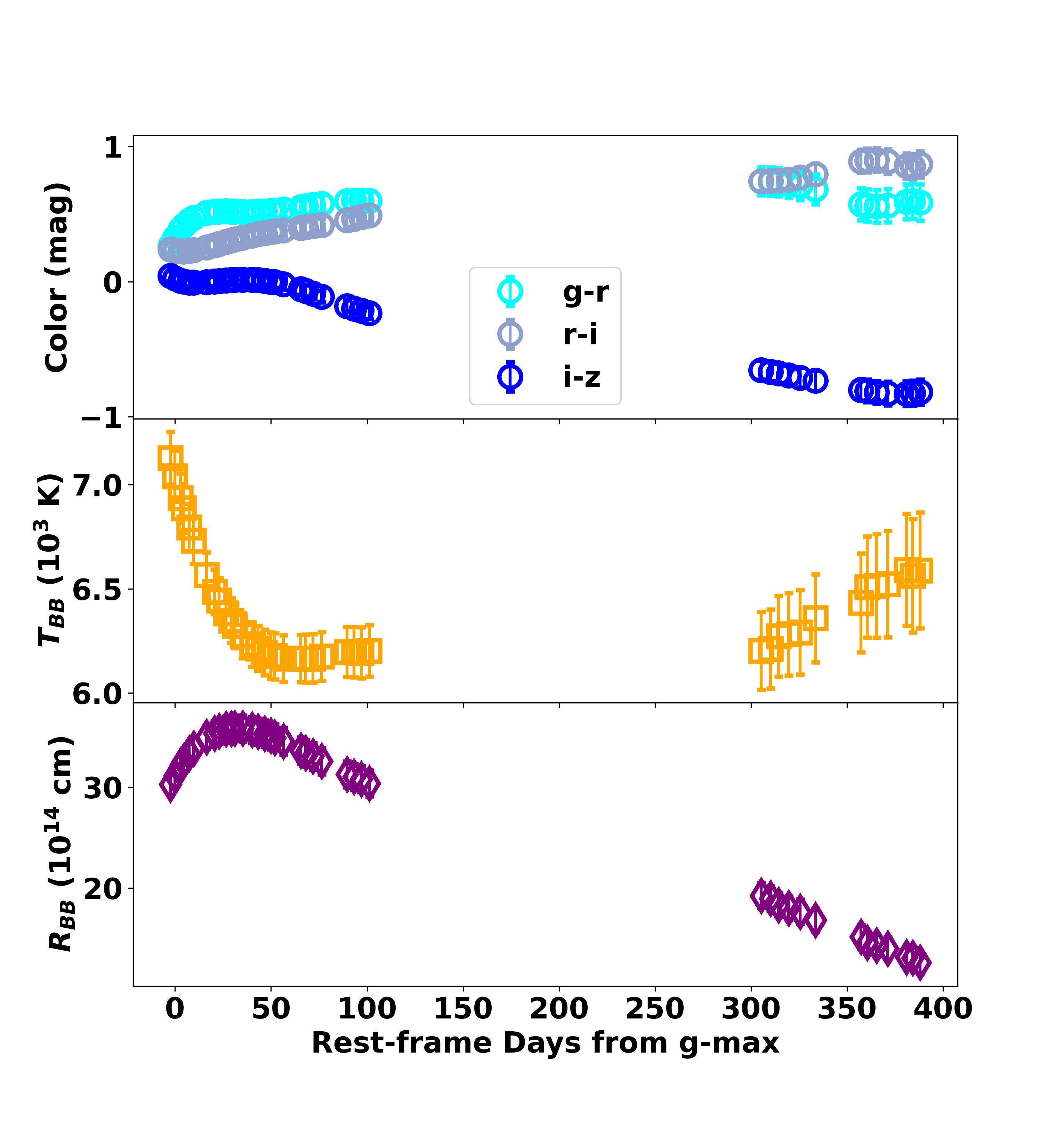}
\caption {\emph{Top:} Inferred Color evolution of PS1-11aop. $g-r$ and $r-i$ evolved to redder colors with time while $i-z$ evolves blueward possibly caused by increased contribution to the $i-$band flux from $H_{\alpha}$ emission. \emph{Middle:} Blackbody temperature evolution. At early times the ejecta cools before plateauing around $\sim$6000 K. During the second observing season temperature is poorly constrained and the apparent rise in temperature may be due to contamination from $H_{\alpha}$ emission. \emph{Bottom:} Blackbody radius evolution. At early times the radius increases, before receding at later times.}
\label{fig: tempradcolor}
\end{figure}

The top panel of Figure \ref{fig: tempradcolor} shows the color evolution of PS1-11aop. We see that $g-r$ and $r-i$ evolve to redder colors with time, while $i-z$ is initially flat before evolving to bluer colors between $\sim$50$-$400 days after g-band maximum. This $i-z$ evolution may be due to an increase in the relative contribution of the $H_{\alpha}$ line from the transient, which falls within the i-band at the redshift of PS1-11aop.

Using the interpolated light curves, we then fit blackbodies to the spectral energy distributions (SEDs) formed by the $g_{P1}r_{P1}i_{P1}z_{P1}$ magnitudes. The inferred photospheric temperature (T$_{BB}$) and radius (R$_{BB}$) evolution are plotted in the middle and lower panels of Figure \ref{fig: tempradcolor}, respectively, and provided in Table~\ref{tab:bol-BB}. During the first $\sim$40--50 days after the g-band maximum, we observe a decrease in the inferred color temperature coupled with an increase in the blackbody radius, consistent with an expanding and cooling photosphere. The velocity that would be inferred from this early radius evolution is only $\sim$4000\,km\,s$^{-1}$ and decreases gradually over time. By approximately 12\,days after discovery, the velocity that would be inferred reduces to $\sim$3000\,km\,s$^{-1}$, although we caution that these values inherently assume that the blackbody radius describes the true radius of the optical emitting material. 

Once it decreases to $\sim$6300 K, the observed color temperature plateaus, and the blackbody radius begins to recede. This behavior is analogous to that observed in SNe of all types, but is distinct from what is found in other classes of transients such as tidal disruption events (TDEs; \citealt{Chornock2014}). While in normal Type IIP SN, this type of behavior is attributed to the recombination front moving back through the expanding ejecta, in Type IIn SN, the interaction region is expected to be geometrically thin. As a result, it has been argued that this behavior can instead be explained in terms of the clumpiness of the cold dense shell formed by the post-shock gas \citep[e.g.\ as in SN2006tf and SN2006gy;][]{Smith2008,SmithRyan2010}. In this case, R$_{BB}$ is different from the true radius of the optical emitting shell. The true radius continues to increase with time, while the observed R$_{BB}$ is modified by a dilution factor related to the covering fraction of the denser clumps of material that remain optically thick. In this case, the velocities inferred above from the blackbody radii near their maximum would be lower limits, and this change in photospheric behavior would be linked with the onset of a transition to a phase where the material becomes less dense and more transparent to light.

Finally, while a slight increase in best-fit color temperature is found at late times ($>$ 1 year), we caution that the errors are large and the SED at these times suffers from increased contamination from the $H_{\alpha}$ feature. The interpolated griz-band light curves and inferred blackbody temperatures and radii are provided in Table~\ref{tab:bol-BB}.

\subsection{Bolometric Light Curve and Total Radiated Energy} \label{subsec:bolometric LC}

\begin{figure}[t!]
\includegraphics[width=.45\textwidth,trim=.5cm 0.5cm 2.cm 2.50cm,clip]{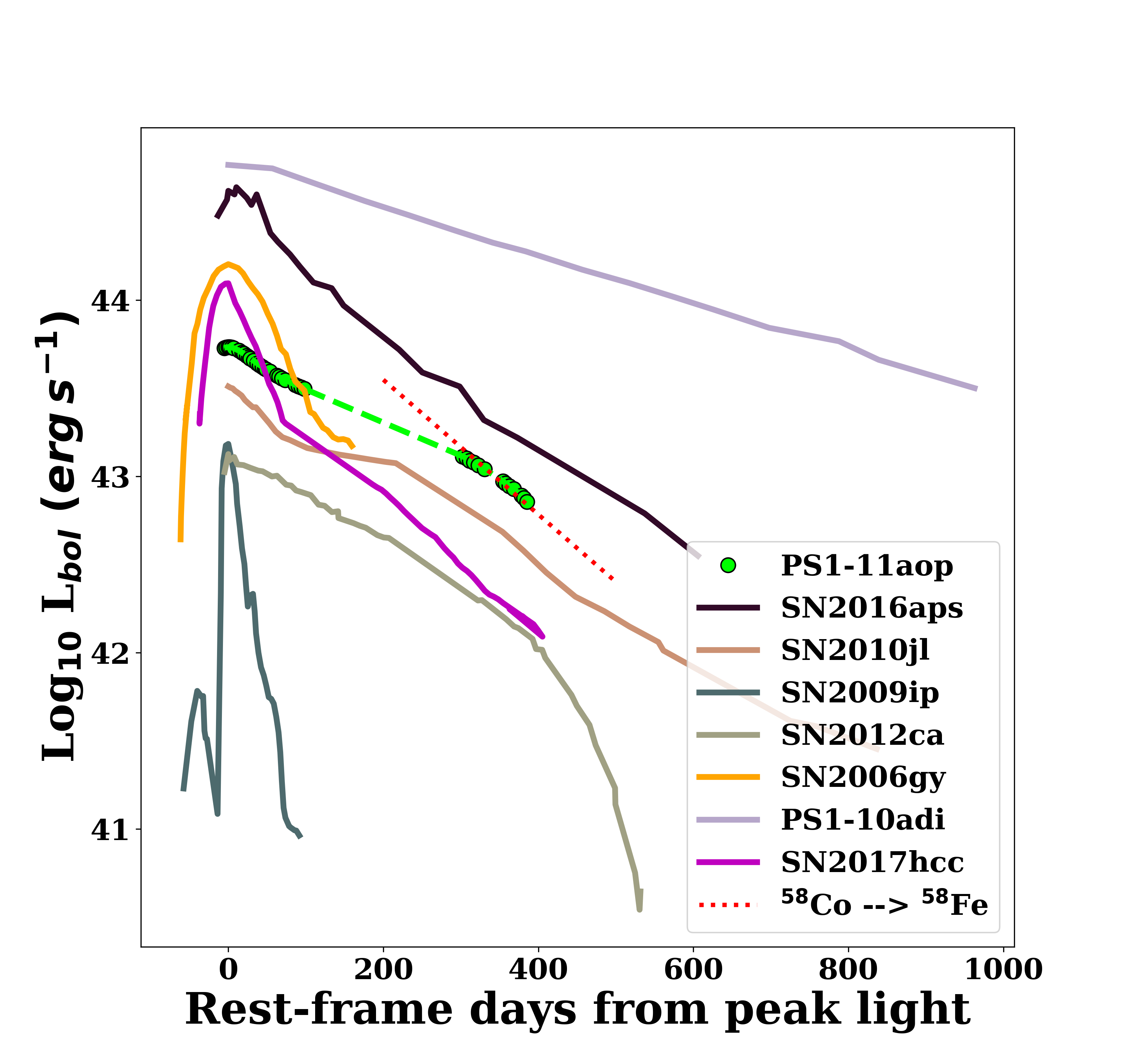}
\caption {Pseudo-bolometric light curve of PS1-11aop shown in cyan color alongside other Type IIn SN published in the literature: SN2012ca [Ugriz: \cite{Inserra2013}], SN2010jl [UBVRI: \cite{Fransson_2014}], SN2017hcc [BB: \citep{Moran2023}] PS1-10adi [BB: \cite{Kankare2017}],  SN2006gy [BB: \citep{Smith2007}],  SN2009ip [BB: \cite{Margutti2014}], SN2016aps [BB: \cite{Nicholl2020}]. The red dotted line represents the decay rate associated with $^{56}$Ni $\rightarrow$ $^{56}$Co $\rightarrow$ $^{56}$Fe radioactive decay chain, assuming full trapping of gamma-rays.
}
\label{fig: bolsne}
\end{figure}

We construct a pseudo-bolometric light curve for PS1-11aop using the interpolated $g_{P1}r_{P1}i_{P1}z_{P1}$ light curves and best-fit blackbodies described in Section~\ref{subsec:temprad}. Specifically, we sum the flux in the observed light curves via a trapezoidal integration and then integrate the best-fit blackbody for each epoch to account for the unobserved flux that is both (i) redwards of the z-band, and (ii) bluewards of the g-band extending back to 2000 \AA. As mentioned above, errors are propagated using a Monte Carlo technique based on 1000 realizations of our data. The resultant pseudo-bolometric light curve, which spans $\sim$400 days is shown in Figure~\ref{fig: bolsne} and provided in Table~\ref{tab:bol-BB}.

We find the PS1-11aop has a peak bolometric luminosity of $\sim$ ($5.42 \pm 0.25$) $\times 10^{43}$ \,erg $s^{-1}$. As described above, the light curve evolves slowly: at the end of our first season of observations ($\sim$120 days post-discovery), PS1-11aop still has a luminosity of ($3.14 \pm 0.17$) $\times 10^{43}$\,erg $s^{-1}$, similar to a typical Type Ia SN at peak. As a result of its slow decline, the total radiated energy is also large. Integrating the pseudo-bolometric light curve, we find a lower limit on the total radiated energy of $E_{\rm{rad}} \approx$ (8.42 $\pm$ 0.17) $\times 10^{50}$\,erg. These bolometric properties are summarized in Table~\ref{tab:SN-props}. 

In Figure~\ref{fig: bolsne}, we also show bolometric light curves for other luminous transients with narrow/intermediate width emission lines in their spectra. PS1-11aop is less extreme than some very luminous events such as SN\,2006gy \citep{Smith2007}, SN2016aps \citep{Nicholl2020}, SN2017hcc \citep{Moran2023}, and the nuclear transient PS1$-$10adi \citep{Kankare2017}, but shows a similar overall timescale and slightly higher luminosities than the well-studied SN2010jl \citep{Fransson_2014}.

\subsection{Spectroscopic Properties} \label{sec:spec props}

\begin{figure}[t!]
\centering
\includegraphics[trim=4.5cm 0cm 2.75cm .15cm,clip,width=\columnwidth]{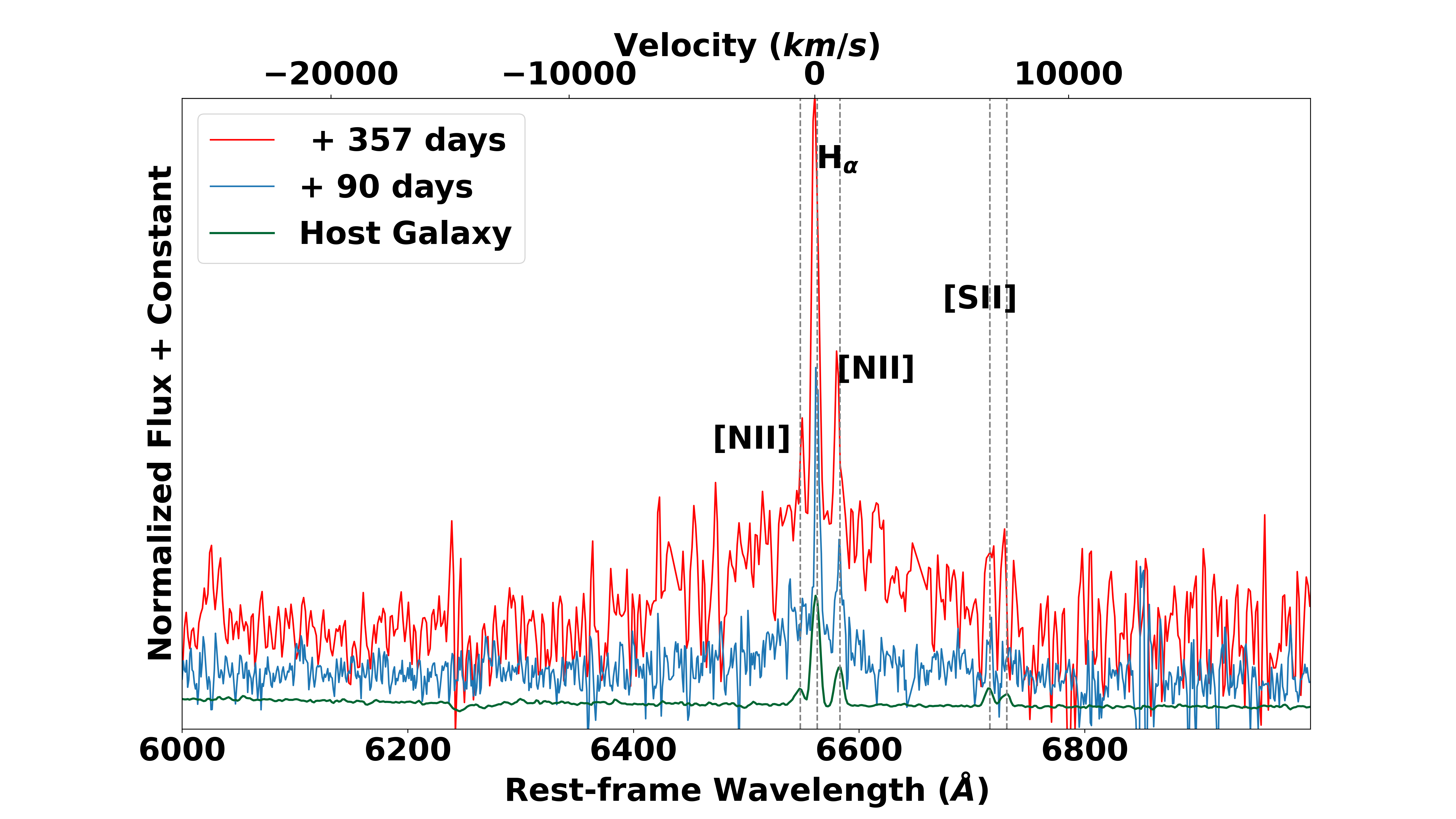}
\caption{
Zoom-in of the spectra around $H_{\alpha}$ showing lines of the doublets [NII] and [SII]. 
}
\label{fig:spectra2}
\end{figure}

In Figure~\ref{fig:spectra2}, we show the region centered around H$\alpha$ ($\lambda$6562.85 $\AA$) for the two spectra of PS1-11aop that were obtained while the transient was active. For comparison, the spectrum of the host galaxy (obtained $\sim$3 years later) is also shown. All spectra have been shifted to rest-frame wavelengths and corrected for $E(\bv)_{\rm{MW}} = 0.0347(9)$ mag of Milky Way reddening \citep{Schlafly2011} assuming a \citet{Cardelli1989} extinction law. In addition to narrow lines of H$\alpha$
and [NII] (which are likely attributed to contamination from the host galaxy), both spectra also exhibit a broader component of H$\alpha$. 
Given the low resolution and moderate signal-to-noise of these spectra, we cannot comment on the possible presence of a narrow P-Cygni component, which could be produced by unshocked CSM material local to the environment of the transient. Similarly, while there is tentative evidence for a broad component of H$_{\rm{\beta}}$ in these spectra, the low signal-to-noise does not allow a quantitative assessment of its properties.

We fit a Gaussian to the broader emission component of H$\alpha$ in both explosion spectra in order to estimate the velocity of the emitting material. For the spectrum obtained 90 days after the g-band maximum, the best-fit Gaussian has a full-width half maximum (FWHM) of $\sim$3200$\pm$270 km s$^{-1}$. This velocity is broadly consistent with that inferred from the expansion of the blackbody radius at early times (see Section~\ref{subsec:temprad}). It is also typical for the ``intermediate'' width component in many Type IIn SN \citep[e.g.,][]{Fransson_2014, smith2017RS, Brennan2023}, but relatively narrow compared to the $\sim$5000-20,000\,km s$^{-1}$ typically found for with TDEs at similar epochs \citep[e.g.,][]{vanVelzen2021, Charalampopoulos2022}. While the ``Bowen'' TDEs have been found to have systematically lower H$_{\alpha}$ widths (including some with $<$5000 km s$^{-1}$; \citealt{Charalampopoulos2022}), these events are identified by the presence of strong [NIII] in their spectra which is not evident in PS1-11aop.

In contrast, for the spectrum obtained $\sim$1 year after discovery, we find that the apparent velocity of the emitting material has increased: the best-fit Gaussian to this epoch has a FWHM of $\sim$7500$\pm$400 km s$^{-1}$.
Similar broad components of the H$_{\rm{\alpha}}$ feature have been observed at late times in some luminous interacting SN. These are typically attributed to either electron scattering \citep[e.g., SN2010jl;][]{Fransson_2014} or to fast-moving material \citep[e.g., SN1988Z and SN2006tf;][]{Smith2008}. In the latter case, we note that the wings of the H$_{\rm{\alpha}}$ feature in the $+$357 day spectrum of PS1-11aop are located beyond 7,500 km s$^{-1}$. If we were to assume that PS1-11aop is an interacting SN and that the earliest blackbody radius evolution (prior to its peak, see discussion in Section~\ref{subsec:temprad}) traces the interaction shell, then the velocities observed in the second spectra would be higher than the inferred expansion speed of the interaction shell at early times ($\sim$4000 km s$^{-1}$). In this case, the fast-moving material seen from this late-time spectrum may represent inner freely-expanding ejecta that has not yet reached the reverse shock (as assumed in \citealt{Smith2008}).
Observation of this material could indicate the presence of a clumpy or asymmetric CSM (such emission from the inner ejecta is not fully blocked by an optically thick interaction shell).

\section{Host Galaxy Properties} \label{sec:Host Galaxy}

Here we provide constraints on both the global properties of the host galaxy of PS1-11aop and on the offset of the transient from the nucleus of its host. 

\subsection{Absolute Magnitude and Total Stellar Mass}

In order to estimate the B-band absolute magnitude and total stellar mass of the host galaxy of PS1-11aop, we use \texttt{Prospector} \citep{Johnson2021} to fit a combination of multi-band photometry and optical spectroscopy available for the host galaxy. As shown in Figure~\ref{fig:hostimg}, the host of PS1-11aop is located near two other galaxies in the field. We, therefore, rely on the SDSS DR10 $ugriz-$band model magnitudes of galaxy \citep{Ahn2012}, which have been deblended to obtain the host magnitudes. The galaxy is also detected in the WISE bands, but by visual inspection, it appears to be blended with another nearby source and we therefore do not include them in our galaxy fits\footnote{Although we note that its WISE ratios are consistent with star-forming galaxies in the diagnostic plot of \citep{Wright2010}}. We supplement this photometry with our optical host galaxy spectrum from LDSS described in \S~\ref{subsec:optical spec}. Prior to inclusion in the \texttt{Prospector} fits we adjusted the flux calibration by matching to the SDSS photometry. This was done in order to correct for any errors in the absolute flux calibration as well as the fact that the 1-arcsecond width of the LDSS slit does not encompass the entire galaxy.

\texttt{Prospector} uses the stellar population synthesis library provided in the Flexible Stellar Populations Synthesis (FSPS) stellar population code \texttt{python-fsps} \citep{Conroy_2009} to fit the observed data. We assumed a Chabrier initial mass function \citeyearpar{Chabrier2003} and allowed a delayed-$\tau$ star formation history (SFH) model \citep{Carnall_2019} that has 5 free parameters (stellar mass, stellar metallicity, galaxy age, dust attenuation optical depth for a foreground screen, and star formation timescale for an exponentially declining SFH). We run \texttt{Prospector} in the absorption-only mode (i.e. not including nebular emission lines).
The likelihood given the free parameters was then sampled using a nested sampling approach through the \texttt{dynesty} code \citep{Speagle2020} and the posterior distributions were calculated.

We use the resulting best-fit model from \texttt{Prospector} to calculate the absolute B-band magnitude of the host galaxy. The model was shifted to the rest frame and synthetic B-band photometry was performed, yielding an absolute magnitude of M$_B = -20.6$~mag. In addition, \texttt{Prospector} provides the total stellar mass of the best-fit model along with 1$\sigma$ confidence intervals based on the simulations that take uncertainties in both the models and measured fluxes into account. We find the stellar mass of the galaxy to be $\log(M_{gal}/M_{\odot})  = 10.74^{+0.03}_{-0.05}$.

\subsection{Emission Line Diagnostics}\label{subsec:EmissionLine}

\begin{figure}[t!]
\centering
\includegraphics[width=.45\textwidth,height = 0.3 \textheight,trim=.05cm 0cm 0cm .10cm,clip]{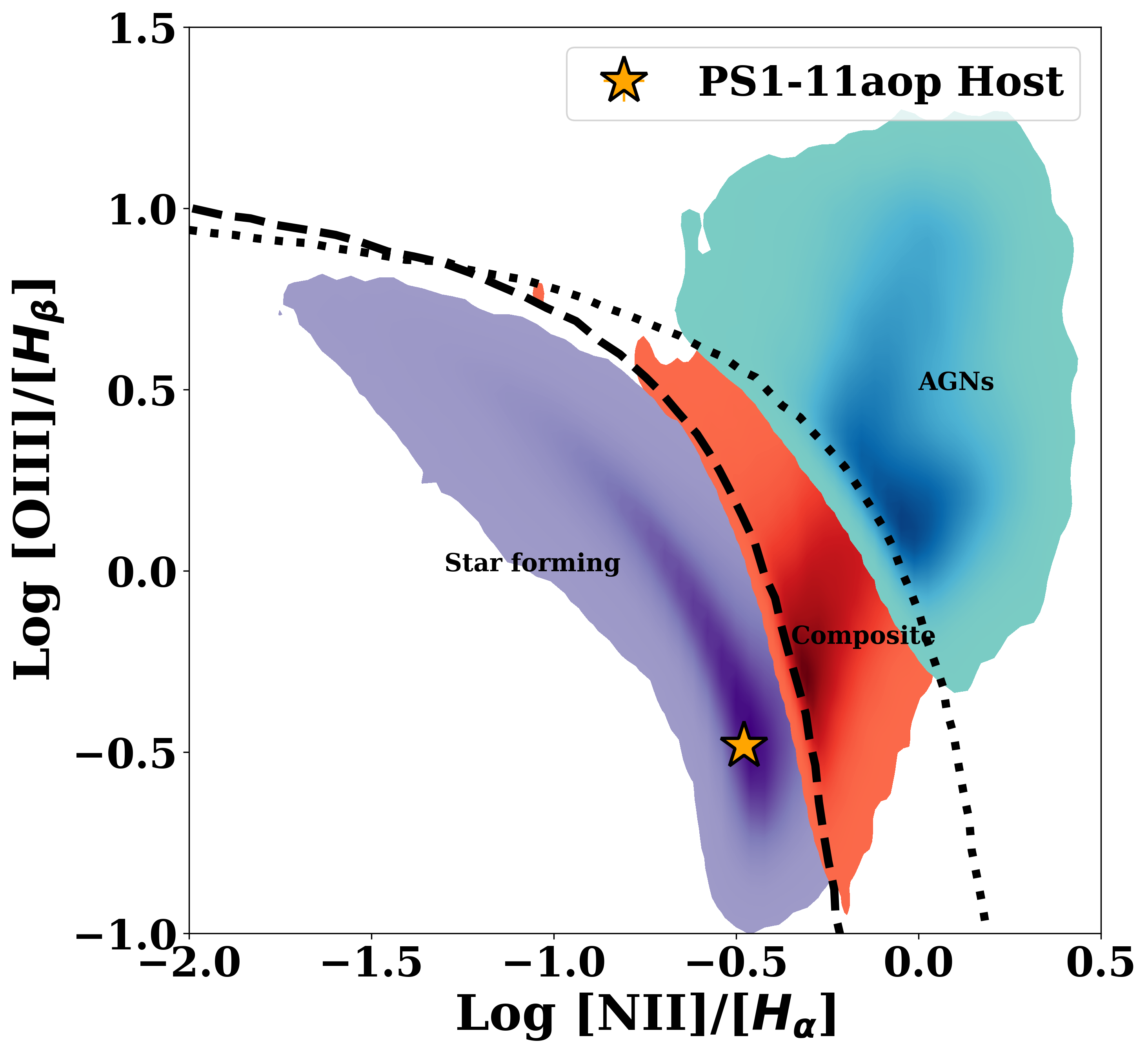}
\caption {BPT diagram showing the dominant source of photo-ionization using the nebular emission line ratios from PS1-11aop as an orange star. The background colored kernel distribution function is a representation of the SDSS DR8 data used \citep{Aihara2011}. The dashed line \citep{Kauffmann2003} indicates the empirical division between star-forming galaxies and AGNs while the dotted line \citep{Kewley2001} shows the theoretical demarcation. Regions in purple, red, and blue, broadly designate star-forming, composite, and AGN-dominated hosts (slight blurring between categories is caused by the density plotting function). The host of PS1-11aop host is a star-forming galaxy.} 
\label{fig:BPT}
\end{figure}

We use the line fluxes of the narrow emission lines in the host galaxy spectrum to assess several properties of the host galaxy. We first scale the host galaxy spectrum to the SDSS photometry, correct for galactic extinction using the \cite{Cardelli1989} extinction law, and subtract off the best-fit model absorption spectrum obtained from \texttt{Prospector} fitting (described above). We then fit Gaussians to the H$_{\alpha}$, H$_{\beta}$, [OIII]$\lambda 5007$ and [NII]$\lambda 6584$ emission lines to measure their fluxes. The local continuum was determined from the mean value in the region surrounding each line. To estimate our uncertainties, we approximate the error on each data point as the standard deviation of the fluxes within the continuum region and then create 100 realizations of our data allowing each point to vary. The fitting process is then repeated on each realization of our spectrum. We take our flux uncertainties to be the standard deviation of these results.

Using these fluxes, we construct and place the host galaxy of PS1-11aop on the updated ``Baldwin, Phillips \& Terlevich'' (BPT) \citep{BPT1-1981} diagram according to \cite{BPT22011}. This enables us to probe the major source of photoionization and distinguish between star-forming galaxies and active galactic nuclei (AGN). PS1-11aop is represented by an orange star while in the background, we plot the density distribution of nearby (0.02 $<$ z $<$ 0.35) emission line galaxy ratios with S/N $>$5 from the SDSS spectroscopic data from Data Release 8 \citep[DR8:][]{Aihara2011}. The result shows that the PS1-11aop host galaxy is consistent with a star-forming galaxy (See Fig \ref{fig:BPT}). Integrating our H$_\alpha$ line flux, we calculate an H$_\alpha$ luminosity of 4.25(1) $\times$ 10$^{41}$\,ergs\,s$^{-1}$. This leads to an inferred star formation rate (SFR) of  3.35$\pm$0.01\,M$_\sun$ yr$^{-1}$ using the equation (SFR$_{H_{\alpha}}$ = 7.9$\times$10$^{-42}$L$_{H_{\alpha}}$) from \citet{kennicutt1998}. We use a Monte Carlo method to propagate the uncertainty in the H$_\alpha$ flux density to SFR$_{H_{\alpha}}$. We simulate 1000 flux density values, drawing from a Gaussian probability distribution with the same 1-sigma uncertainty as for the emission. We then use these to calculate 1000 SFR$_{H_{\alpha}}$ values and record their mean and standard deviation.

We also estimate a metallicity of $12 + \log(O/H) = 9.09(1)$ using Equation 1 from \citet{Pettini2004}: $12 + \log(O/H)= 8.90 + 0.57 \times$ N2, where N2 $=$ [NII]$\lambda$6583/H$_{\alpha}$. This value is approximately 2 times solar metallicity assuming $12 + \log(O/H)_{\rm{solar}} = 8.69$, and should be relatively insensitive to uncertain extinction due to the proximity of the H$_{\alpha}$ and [NII]$\lambda$6583 lines. Coupled with the stellar mass measurement described above, the host of PS1-11aop is consistent with the mass-metallicity relationship for SDSS galaxies from \citet{Tremonti2004}. Specifically, among galaxies with masses approximately $\log(M_{\text{gal}}/M_{\odot}) \sim 10.75$ in the sample from \citet{Tremonti2004}, 68\% exhibit metallicities falling between $12 + \log(O/H)$ 8.8 and 9.2. The host galaxy of PS1-11aop is situated within this specified range. A summary of the global host galaxy properties can be found in Table \ref{tab:galaxy-props}.

\subsection{Transient-Galaxy Offset}\label{subsec:Local environment}

\begin{deluxetable}{llcc}
\tabletypesize{\small}
\tablecaption{PS1-11aop Host galaxy properties \label{tab:galaxy-props}}
\tablehead{\colhead{Properties} & \colhead{values}   }
%\multicolumn{2}{c|}{\multirow{2}{*}{offset}} 
\startdata
Observational properties & \\
\hline
Apparent r-band magnitude, AB (mag) & 19.08$\pm$0.01  \\
Redshift & 0.218$\pm$0.001 \\
Luminosity Distance (Mpc) & 1070$\pm$4\\
%E(B-V) (mag) & 0.0347(9) \\
H$_{\alpha}$ flux (ergs\,s$^{-1}$\,cm$^{-2}$\,\AA$^{-1}$) & 3.77(1) $
                \times$10$^{-15}$\\
\hline
Inferred properties & \\
\hline
Absolute B-band magnitude, AB (mag) & $-$20.6  \\
Metallicity [12+log(O/H)] & 9.09(1) \\ %9.089(4) \\
H$_{\alpha}$ SFR (M$_\odot$\,yr$^{-1}$) & 3.35$\pm$0.01 \\%1.193(2)\\
Log Stellar Mass (M$_\odot$) &  $10.74^{+0.03}_{-0.05}$ \\
\enddata
\end{deluxetable} 

PS1-11aop exploded very close to the center of its galaxy (see Figure~\ref{fig:hostimg}). We therefore utilize the PS1-MDS individual images and deep-stack templates described in Section~\ref{subsec:optical data} to quantify the offset and its uncertainty between the SN and its host galaxy. First, we determine the centroid of PS1-11aop within the PS1-MDS template images using the astrometry framework within the \texttt{photpipe} pipeline \citep{Rest2014}. In brief, the transient centroid and uncertainty are first obtained in each individual difference image. These measurements are then combined to calculate a weighted, 3$\sigma$ clipped, SN centroid in each band. Through this process, we achieve an uncertainty on the location of PS1-11aop ($\sigma_{SN}$) of $\sim$0.01$\arcsec$.
Second, we determine the flux-weighted centroid of the host galaxy (and associated uncertainty, $\sigma_{\rm{gal}}$) within the PS1-MDS template images using \texttt{SOURCE EXTRACTOR} \citep{Bertin1996}. For this purpose, we use all pixels associated with the galaxy with signal-to-noise greater than 3 (\texttt{DETECT\_THRESH} $=$ 3) and visually inspect to ensure that there is minimal contamination from the nearby galaxies (Figure~\ref{fig:hostimg}). This yields uncertainties of $\sim$0.01--0.03$\arcsec$ (see Table~\ref{tab:galaxy-offset}).

Using these positions, we then calculate the transient-galaxy offset and associated uncertainties. In Table~\ref{tab:galaxy-offset}, we list the resulting offsets in both arcseconds and kiloparsecs. From this, we see that while PS1-11aop is located within the inner $\sim$1 kpc of the center of its host---similar to other luminous transients such as SN2006gy \citep{Smith2007} and CSS100217 \citep{Benetti2013}---it is not coincident with the nucleus itself.

\begin{deluxetable}{lcc|cc}
\tabletypesize{\small}
\tablecaption{PS1-11aop Host galaxy Offset \label{tab:galaxy-offset}}
\tablehead{
\colhead{band} & 
\colhead{$\sigma_{sn}$} & 
\colhead{$\sigma_{gal}$} & 
\multicolumn{2}{c}{offset} \\
\colhead{}  & \colhead{($\arcsec$)} & \colhead{($\arcsec$)} & \colhead{($\arcsec$)} & \colhead{(kpc)}} 
\startdata
g & 0.01 & 0.01  & 0.35 $\pm$ 0.01 & 1.22 $\pm$ 0.04 \\
r & 0.01 & 0.02  & 0.35 $\pm$ 0.02 & 1.22 $\pm$ 0.07 \\
i & 0.01 & 0.03  & 0.34 $\pm$ 0.03 & 1.19 $\pm$ 0.10 \\
z & 0.01 & 0.02  & 0.32 $\pm$ 0.02 & 1.12 $\pm$ 0.07 
\enddata
\end{deluxetable}

\section{Constraints on the Origin of the Optical, Radio, and X-ray Emission} \label{sec: argument for sn}
As shown above, PS1-11aop was a luminous and long-lived optical transient that displayed time-variable intermediate-width hydrogen emission lines in its optical spectrum. In addition, radio and X-ray emission was detected at a position coincident with the transient between $\sim$4 and 10 years post-explosion. These properties are all consistent with a luminous supernova that is interacting with a dense circumstellar medium. However, the relative proximity of PS1-11aop to the nucleus of its galaxy ($<0.5\arcsec$), coupled with the varying spatial resolution of instruments used to gather the data described in Section~\ref{sec:obs} imply that other origins should also be considered. Here we discuss the optical, X-ray, and radio emission in turn.

\emph{Optical Emission:} The intermediate-width hydrogen lines observed in the optical spectrum of PS1-11aop are a defining feature of Type IIn SNe \citep{Schlegel1990} but also show similarities to some tidal disruption events (TDEs; \citealt{Arcavi2014}) and Seyfert 1 galaxies \citep{Osterbrock1985}. We, therefore, consider whether PS1-11aop could be due to either a TDE or a ``changing-look quasar'', wherein an AGN undergoes a rapid change in both continuum emission and spectroscopic state, likely due to a change in line-of-sight absorption or accretion rate. A major argument against both interpretations is that, as described above, the optical emission from PS1-11aop is \emph{not} coincident with the host galaxy nucleus. However, for completeness, we also highlight 
that, while the full diversity of changing look AGN is still being uncovered, most discovered to date involve a transition between two classes of Seyfert galaxies \citep{Ruan2016}. In contrast, the host spectrum of PS1-11aop shows no evidence of an active nucleus and we find no evidence of pre-explosion optical variability of the host. Finally, concerning TDEs, we note that the photosphere of PS1-11aop shows clear cooling and expansion over the first $\sim$year post-discovery. This is distinct from most optically-discovered TDEs (which have relatively constant blue colors, \citealt{Chornock2014}).

Thus, we find that the optical emission from PS1-11aop is mostly consistent with a SN origin. We note that due to host galaxy contamination in our explosions spectra, we are unable to confirm the presence of a time-variable narrow ($<$1000 km s$^{-1}$) line component. However, based on the presence of the intermediate width emission lines (and lack of broad P-Cygni lines typically observed Type IIP SN, which are powered mainly by recombination of the shock heated stellar envelope), we associate PS1-11aop with the Type IIn SN class. 
We therefore model the optical emission in the context of a luminous interacting transient in Section~\ref{sec:optical modelling}.

\emph{Radio Emission:} Using the second epoch of radio observations, we estimate a spectral index (F$_{\nu}$ $\propto$ $\nu^\alpha$) of $\alpha \lesssim -0.8$, consistent with optically-thin synchrotron emission. However, this could be consistent with an interacting SN, non-thermal star formation, or an AGN.  We also observe a factor of $\sim$2 fading in the X-band between epochs 1 and 3, while the C-band observations in epochs 2 and 3 are consistent. While such slow evolution is expected for old radio transients (e.g., SN1995N in \citealt{Chandra2009}) this may also be indicative of a steady underlying source.

The detected radio emission is unresolved, but observations were carried out in the D- and C-array configurations of the VLA. The smallest synthesized beam width amonst our observations was of 2.1 arcseconds (X-band; C-array). This is comparable to the SDSS Petrosian radius of the host galaxy ($\sim 2 - 2.5 \arcsec$) and we therefore consider if the emission could instead be due to star formation in the host galaxy. 

We use the equations from \cite{Tabatabaei2017} to assess what star formation rate would be inferred from our observed radio fluxes.  If due to star formation, the 6 GHz from from epoch 2 and 9.77 GHz flux from epoch 1 would yield 
mid-radio continuum star formation rates (SFR$_\mathrm{MRC}$) of $12.85\pm3.80$\,M$_\mathrm{\odot}\,yr^{-1}$ and  $20.92\pm3.98$ M$_\mathrm{\odot}\,yr^{-1}$ respectively. 
The equations from \cite{Tabatabaei2017} include contributions to the radio flux from both thermal and non-thermal sources. For these calculations, we assume an electron temperature of $10^{4}$ K, a non-thermal spectral index of $\alpha = -0.8$, and the ratio of thermal to total star formation corresponding to the frequency of the observation (see Table 6 of \citealt{Tabatabaei2017}).
We use a Monte Carlo method to propagate the uncertainty in SFR$_\mathrm{MRC}$ similar to that of SFR$_{H_{\alpha}}$ as described in \S \ref{subsec:EmissionLine}. Our analysis shows that the SFR$_\mathrm{MRC}$ that would be implied by our radio observations is a factor of $4-6$ higher the optical star formation rate based on the H$_{\alpha}$ line emission (SFR$_\mathrm{H_\alpha} = 3.35\pm0.01$ M$_\mathrm{\odot}\,yr^{-1}$). This discrepancy implies that either there is obscured star formation which is not visible in the optical, or that another source of emission contributes to our observed radio flux.

Thus, while it is likely that at least the X-band detection in epoch 1 (which subsequently faded) has non-negligible contribution from an emission mechanism other than star formation, we cannot rule out a non-negligible contribution at later epochs. As a result, in Section~\ref{sec:radio modelling}, below, we will consider two scenarios: (i) the radio emission observed at the location is entirely due to the expansion of the SN blastwave into the ambient medium and (ii) all of the observed radio observations are upper limits on the true flux due to the SN. We will discuss the implications of both scenarios.

\emph{X-ray Emission:} The X-ray detection at the location of PS1-11aop is relatively luminous in the context of interacting SNe if truly associated with the SN (SN\,2010jl showed similar luminosities, but at earlier times). We therefore also consider other origins. First, a baseline level of X-ray emission is expected from all galaxies due to a combination of X-ray binaries and the warm interstellar medium, both of which correlate with star formation in the host \citep[e.g.][]{Mineo2012a,Mineo2012b}. To understand the contribution from these sources, we use the relationships found in  \cite{Lehmer2016}, which provide the X-ray luminosity as a function of the redshift, stellar mass, and star formation rate of the host galaxy. These predict a 0.5-10 keV X-ray luminosity from the host of PS1-11aop of $\sim$5$\times$10$^{40}$ erg s$^{-1}$. This is approximately a factor of 10 lower than detected (see Table~\ref{tab:xrayflux}), indicating that there is likely another source of X-ray emission.

Next, we note that the radius that contains 90\% of the \emph{Chandra} PSF ($\sim 1 \arcsec$) also encompasses the galaxy core. We therefore consider whether the detected X-ray emission could be due to an AGN. First, as described above, the host galaxy spectrum obtained 3 years post-explosion shows no evidence of AGN-activity, either in the line ratios of the narrow emission lines or in the presence of a broad base to H$_{\alpha}$. In addition, both the X-ray luminosity (L$_{X,0.3-10 keV} \sim  5 \times 10^{41}$ erg/s) and X-ray to optical flux ratio ($X/O = \log (F_{X,0.5-2kev}) + 0.4 m_r + 5.71 \sim -1.5$; where $m_r$ is the apparent r-band magnitude of the host galaxy; see Table~\ref{tab:galaxy-props}) are slightly below the thresholds typically adopted for AGN (L$_{X} > 10^{42}$ and $X/O < -1$, respectively; \citealt{Ranalli2012}). Thus, if the emission is associated with an AGN it would need to be an X-ray faint AGN that shows no signatures in the optical. However, we have only a single epoch of observation and thus no constraints on the time variable behavior. Thus, in the sections below we will comment on the implications if the X-ray emission is due to the SN, and whether this is consistent with inferences from the combination of the optical and radio emission. The bulk of this modeling is presented in Appendix~\ref{apsec:x-ray model} while we provide a summary of our modeling framework and conclusions in Section~\ref{sec:x-ray modelling}.

\section{Optical Light Curve Models} \label{sec:optical modelling}
 
We have shown that the intermediate width hydrogen features observed in the explosion spectra of PS1-11aop are similar to those observed in Type IIn SN. Here we model the optical light from PS1-11aop in the context of interacting transients. We first place limits on the amount of radioactive $^{56}$Ni that can contribute to the light curve. We then examine several models to constrain the structure of the CSM probed by the optical emission over the first $\sim$1 year post-discovery.

\subsection{Upper limit on Contribution from Radioactive Nickel}\label{subsec:Ni-56 content}

In normal core-collapse SNe, a significant fraction of the bolometric luminosity is powered by the radioactive decay of $^{56}$Ni. In PS1-11aop, the initial decline rate is \emph{shallower} than would be expected from the $^{56}$Ni $\rightarrow$ $^{56}$Co $\rightarrow$ $^{56}$Fe decay chain---indicating that another power source must be present. However, as shown in 
Figure \ref{fig: bolsne}, between approximately 300 and 400 days post-discovery, the slope of the bolometric light curve increases slightly. At these epochs, the decay rate is broadly comparable to expectations for the $^{56}$Co $\rightarrow$ $^{56}$Fe decay, assuming full trapping of gamma rays. Using the equations of \cite{Valenti2008} for the late-time light curve of a $^{56}$Ni-powered transient (when the luminosity is directly related to the rate of energy deposition), we find that we would require $\approx15-37$ M$_{\odot}$ of $^{56}$Ni to match the observed luminosity (this range corresponds to assumed explosion epochs between $-5$ and $-100$\,days).

However, such large nickel masses cannot be accommodated by the early-time light curve. For example, running a large grid of models for the photospheric phase of a transient powered by radioactive decay \cite{Valenti2008, Arnett1982}, we find upper limits on the amount of radioactive Nickel that range from M$_{Ni} \lesssim 5$ M$_{\odot}$ for $t_{exp} = -5$\,days to  M$_{Ni} \lesssim 10$ M$_{\odot}$ for $t_{exp} = -100$\,days. The predicted nebular phase luminosities associated with the radioactive decay of 10 M$_{\odot}$ of $^{56}$Ni are a factor of $\sim$10 fainter than the luminosity of PS1-11aop at similar epochs. Thus, while we cannot place strict limits on the amount of $^{56}$Ni (with values of 5$-$10\,$M_\sun$ accommodated by the observed light curve), we find that the late-time light curve ($\sim$ 1\,year post-explosion) must be dominated ($>$90\% of the observed flux) by another power source. Within the context of SN models, some options for this additional emission include CSM interaction and cooling/recombination of shocked material. Given the wide range of decline rates that are possible due to power sources such as CSM interaction (due to varying densities, geometries, etc.), it's not unreasonable to have a decline rate broadly similar to that of the radioactive decay of $^{56}$Ni by chance.

\begin{figure*}[t]
\centering
\includegraphics[width=\textwidth]{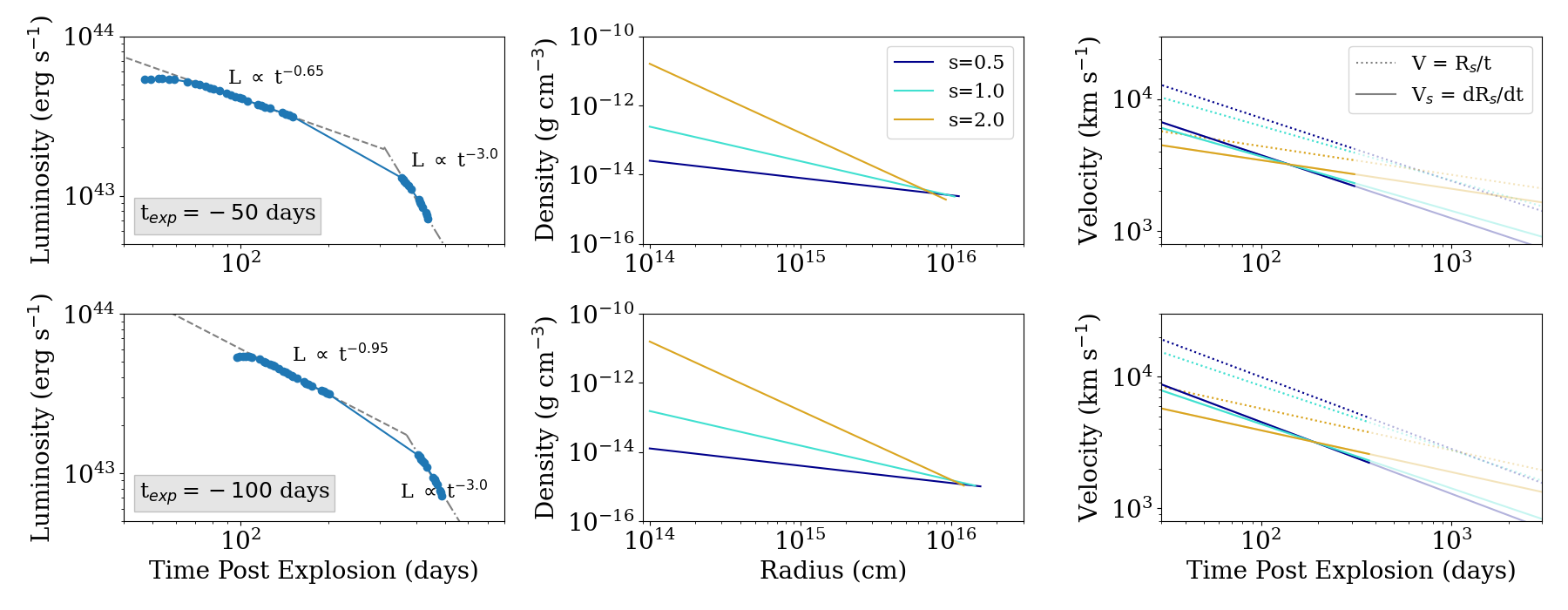}
\caption {Example, results from simple CSM models for PS1-11aop, adopting fiducial explosion times 50 days prior to g-band maximum (top row) and 150 days prior to g-band maximum (lower row). \emph{Left:} Bolometric light curve of PS1-11aop along with a broken power-law fit to the early and late-time data. \emph{Center:} Inferred CSM density distributions based on the power-law index and normalization of the early light curve of PS1-11aop for different assumed CSM slopes ($\rho \propto r^{-s}$ with s$=$0.5,1.0, and 2.0). Lines are truncated at the radius inferred at the time of the break in the power-law index in the bolometric light curve. \emph{Right:} Evolution of both the true shock velocity ($v_s = dR_s/dt$) and the velocity of the homologously expanding ejecta that would have reached the interaction zone ($V = R/t$) as a function of time.} 
\label{fig:SimpleCSM}
\end{figure*}

\vspace{0.2in}
\subsection{Simple CSM Models} \label{subsec:CSM model}

To probe the CSM densities that would be required to reproduce the high luminosity of PS1-11aop for more than a year post-explosion, we first consider a simple CSM model where the observed luminosity is directly proportional to and tracks in real time the rate of energy injection by the ejecta-CSM interaction \citep[e.g., as in][]{Smith_2008,Ofek2014,Fransson_2014}. Mathematically, this can be written as: 
\begin{equation}\label{eq:simpleCSM}
    L = \left( \frac{\eta}{2} \right) 4\pi R_{s}^{2} \rho_{CSM}  v^{3}_{s} 
\end{equation}
where $\eta$ is the efficiency of the conversion of kinetic energy to radiation, $v_{s}$ and $R_s$ are the velocity and radius of the interaction region, respectively, and $\rho_{CSM}$ is the density of the CSM at radius $R_s$. 

In order to use this expression to estimate $\rho_{CSM}$ as a function of radius based on the bolometric light curve of PS1-11aop, we require an expression for how the shock velocity evolves with time. Using the self-similar solutions of \citet{Chevalier1982}, \citet{Fransson1996} consider an ejecta profile of the form $\rho_{\rm{ejecta}} \propto r^{-n}$ expanding into a general CSM density profile of: 

\begin{equation}
    \rho_{CSM} = \left( \frac{\dot{M}}{4\pi v_w r_0^2} \right) \left( \frac{r_0}{r} \right)^s
\end{equation}
where $\dot{M}$ is the implied pre-explosion mass-loss rate of the progenitor star at the specific reference radius, $r_0$, and $s$ is the power law index of the density distribution. $s=2$ corresponds to a wind-like medium. They find that the shock velocity is expected to evolve as $v_s \propto t^{-(3-s)/(n-s)}$ and that the shock radius at time $t$ is $R_s = [(n-s)/(n-3)] v_s t$. In this case, equation~\ref{eq:simpleCSM} becomes \citep[see also][]{Fransson_2014}:
\begin{equation}\label{eq:simpleCSM2}
\begin{split}
    L = \left( \frac{\eta}{2} \right) \left(\frac{\dot{M}}{v_w}\right) r_0^{s-2} \left(\frac{n-s}{n-3}\right)^{2-s} v_s^{5-s} t^{2-s} \\
    \propto t^{-[15+s(n-6)-2n]/(n-s)} \quad \quad \quad \quad \quad \quad
\end{split}
\end{equation}
where the final proportionality follows from the time dependence of the shock velocity (and assuming a constant efficiency factor, $\eta$).
Thus, with assumptions for (i) the efficiency factor, $\eta$, (ii) either $n$ or $s$ and (iii) the shock velocity at any specific time, the normalization and index of a power-law fit to the bolometric light curve of PS1-11aop can be used to infer a density distribution for the CSM. We also note that due to the assumption that the SN ejecta can be described by a single power-law, this model will eventually break down at late times once the reverse shock propagates into the inner regions of the ejecta which are typically characterized by a second, shallower, powerlaw \citep{Chevalier1994}. This will be discussed below.

The power-law index found by fitting the bolometric light curve of PS1-11aop depends on the assumed epoch of explosion, and we consider dates ranging from 10 to 150 days prior to maximum light. For all explosion dates considered, we find that it is not possible to fit the entire light curve with a single power-law as observations during the second observing season decline more rapidly\footnote{We also do not consider data within the first $\sim$ 20 days of observations as the light curve is still rising/near peak and is likely to be impacted by diffusion through the dense CSM, which is not treated in this model.}. The timing of this apparent break ranges from $\sim$250 to 460 days, depending on the adopted epoch of the explosion. We show examples of a broken power-law fit for explosions 50 and 150 days prior to the maximum in the left panels of Figure~\ref{fig:SimpleCSM}. A similar break was observed in the light curve of SN\,2010jl, and was interpreted to be due to the shock emerging from the dense shell \citep{Fransson_2014, Chandra2015} or by a transition to a momentum-conserving ``snow-plow'' phase (\citealt{Ofek2014}; though see \citealt{Moriya2014} for a critique of this interpretation).

The power-law index found for the first portion of the PS1-11aop light curve ranged from $L \propto t^{-0.4}$ to $L \propto t^{-1.4}$ for explosion dates between 10 and 150 days prior to the maximum, respectively. We consider CSM density distributions with slopes ranging from $s=2$ to $s=0.5$ and in Table~\ref{tab:simpleCSM} list the normalization required in each case to reproduce the observed luminosity of PS1-11aop.  We note that for the assumed explosion dates more than 100 days prior to g-band maximum, the inferred value for $n$ (the slope of the ejecta density distribution) falls below 5, which was the limit where the self-similar solution of \cite{Chevalier1982} are valid. We therefore do not report values beyond this assumed explosion date.

To derive CSM properties from equation~\ref{eq:simpleCSM2}, it is necessary to adopt a shock velocity at some reference time. The values listed in Table~\ref{tab:simpleCSM} are scaled to $v_s = 3200$ km s$^{-1}$ at 90 days after g-band maximum\footnote{i.e. we take $v_s = 3200 \times (t/t_{\rm{ref}})^{-(3-s)/(n-s)}$ km s$^{-1}$, where $t_{\rm{ref}}$ is the time post-explosion that corresponds to the epoch of our first spectrum.}, based on the observed FWHM of the H$\alpha$ feature in the spectrum obtained at this epoch (see Section~\ref{sec:spec props}). This is broadly comparable to value adopted for SN\,2010jl (3000 km s$^{-1}$ at 320 days post-explosion; \citealt{Ofek2014}) and slightly higher than that for SN\,2006tf (2000 km s$^{-1}$; \citealt{Smith_2008}). We also adopt an efficiency factor of $\eta = 1$.

\begin{deluxetable}{lllc|cc|ccc}
\tabletypesize{\small}
\centering
\tablecaption{Results from Simple CSM Modelling \label{tab:simpleCSM}}
\tablehead{
\colhead{t$_{\rm{exp}}$} & \colhead{PL1} & \colhead{PL2} & \colhead{t$_{\rm{break}}$} & \colhead{$s$} & \colhead{$n$} & \colhead{$\dot{M}$} & \colhead{$R_{\rm{break}}$} & \colhead{M$_{\rm{tot}}$} \\
\colhead{(days)} & \colhead{} & \colhead{} & \colhead{(days)} & \colhead{} & \colhead{} & \colhead{(M$_{\odot}$ yr$^{-1}$)} & \colhead{(cm)} & \colhead{(M$_{\odot}$)}
  }
\startdata
-10 & -0.4 & -2.5 & 250 & 2.0 & 9.5 & 0.32 & 7.1$\times$10$^{15}$ & 7.0 \\
%& & & & 1.5 &  7.3 &  0.37 & 7.4e15 &   6.9 \\
& & & & 1.0 &  6.7 &  0.37 & 7.7$\times$10$^{15}$ &   7.1 \\
& & & & 0.5 &  6.4 &  0.34 & 8.1$\times$10$^{15}$ &   7.3 \\
\hline
-50 & -0.65 & -3.0 & 310 & 2.0 & 6.6 & 0.32& 9.2$\times$10$^{15}$ & 9.2 \\
%&  &  &  & 1.5 & 6.1 & 0.30 & 9.8e15& 8.6\\
&  &  &  & 1.0 & 5.8 & 0.24 & 1.0$\times$10$^{16}$ & 8.5\\
&  &  &  & 0.5 & 5.7 & 0.18 & 1.1$\times$10$^{16}$ & 8.5\\
\hline
-75 & -0.80 & -3.0 & 342& 5.8 & 2.0 & 0.31  & 1.1$\times$10$^{16}$ & 10.5 \\ 
 &  &  & & 1.0 & 5.5 & 0.19 & 1.2$\times$10$^{16}$ & 9.2 \\ 
 &  &  & & 0.5 & 5.4 & 0.12 & 1.3$\times$10$^{16}$ & 9.1\\ 
\hline
-100 & -0.95 & -3.0 & 370 & 2.0 & 5.2 & 0.32 & 1.2$\times$10$^{16}$ & 12.0 \\
&  &  &  & 1.0 & 5.1 & 0.15 & 1.4$\times$10$^{16}$ & 10.1  \\
&  &  &  & 0.5 & 5.1 & 0.09 & 1.6$\times$10$^{16}$ & 9.9 
\enddata
\tablecomments{Description of Table Columns. t$_{\rm{exp}}$ is the adopted explosion epochs in rest-frame days relative to the observed g-band maximum. PL1 and PL2 are the power-law indices that fit both the early and late-time bolometric light curve of PS1-11aop, while t$_{\rm{break}}$ is the time post-explosion inferred for the break between the two components. $s$ is the assumed power-law index for the CSM density distribution and $n$ is the inferred value for the power-law index of the SN ejecta based on the values of PL1 and $s$ (see text). $\dot{M}$ is the inferred mass loss rate at a reference radius of 5.5$\times$10$^{15}$ cm for an assumed pre-SN wind speed of 100 km s$^{-1}$. R$_{\rm{break}}$ is the shock radius inferred at the time of the break-in power-law behavior of the bolometric light curve, and M$_{\rm{tot}}$ is the integral of the density distribution between 1$\times$10$^{14}$ cm and R$_{\rm{break}}$. Values quoted for $\dot{M}$, R$_{\rm{break}}$, and M$_{\rm{tot}}$ are valid for an assumed shock velocity of 3200 km s$^{-1}$ at 90 days after g-band maximum. See text for the scalings to other assumed shock velocities. }
\end{deluxetable}

The mass-loss rates derived from this set of assumptions range from $\dot{M} = 0.09-0.37 \times$($v_w$/100 km/s)$^{-1}$ M$_\odot$ yr$^{-1}$. These numbers were calculated at a reference radius of 5.5$\times10^{15}$ cm (necessary because we consider a range of CSM density distributions) and are scaled to a pre-explosion wind speed of 100 km s$^{-1}$.  Integrating these density distributions out to the radius of the shock at the time of the break in the power-law evolution of the bolometric light curve (which range from $\sim$0.7--2 $\times10^{16}$ cm) we infer total CSM masses between $\sim$7 and 12 M$_\odot$. 

In the middle and right panels of Figure~\ref{fig:SimpleCSM}, we show examples of the implied CSM density distributions as well as the evolution of the shock velocity, $v_s$, and the velocity of the homologously expanding ejecta that would have reached the interaction zone $V = R_s/t$. Using the equations in \cite{Chevalier1994} we also calculate the velocity expected for the transition between the inner and outer portions of the ejecta assuming a flat inner ejecta profile and the values of $n$ listed in Table~\ref{tab:simpleCSM}. For an explosion energy of 10$^{51}$ ergs and an ejecta mass of 10 M$_\odot$ the transitions velocities range of from $\sim$900--3500 km s$^{-1}$. These are all smaller than the values found for $V$ during the time of the optical emission, indicating that the assumption made by the model that the shock is still in the outer portion of the ejecta is not unreasonable.

\begin{deluxetable*}{llccccccc}
\tabletypesize{\small}
\tablecaption{Optical MOSFiT modelling priors and posteriors for PS1-11aop \label{tab:MOSFiTresults}}
\tablehead{\colhead{Model} & \colhead{log $f_{Ni}$} & \colhead{$M_{ej}$} & \colhead{$M_{CSM}$} & \colhead{$\rho_{CSM, inner}$}& \colhead{$t_{exp}$\tablenotemark{a}} & \colhead{$v_{ej}$}& \colhead{$s$} \\
\colhead{Type} & \colhead{} & \colhead{\,(M$_\sun$)} & \colhead{\,(M$_\sun$)} & \colhead{\,(g\,cm$\rm{^{-3}}$)}& \colhead{\,(days)} & \colhead{\,(km\,s$\mathrm{^{-1}}$) } & \colhead{} 
  }
\startdata
& & & &Priors & & & \\
\hline
           griz (CSM + Ni) & 1$\times 10^{-3}$ $-$ 1 & 0.1 $-$ 30 & 0.1 $-$ 30 & 1$\times$10$^{-15}$ $-$ 1$\times$10$^{-11}$ & -200 $-$ 0 & 1$\times$10$^{3}$ $-$ 1$\times$10$^{5}$ & 0$-$2 \\
           griz (CSM) & 0 & 0.1 $-$ 30 & 0.1 $-$ 30 & 1$\times$10$^{-15}$ $-$ 1$\times$10$^{-11}$ & -200 $-$ 0 & 1$\times$10$^{3}$ $-$ 1$\times$10$^{5}$ & 0$-$2 \\
\hline
& & & &Posteriors & & & \\
\hline
           griz (CSM + Ni) & $-2.41$ $_{-0.48}^{+0.58}$ & 10.96$_{-7.32}^{+7.32}$ & 26.92$_{-4.34}^{+1.86}$ & 2.57$_{-1.89}^{+2.48}$ $\times$ 10$^{-12}$ & -151.34$_{-14.02}^{+14.62}$ & 4671$_{-315.75}^{+526.24}$ & 0.64$_{-0.26}^{+0.34}$ \\
           griz (CSM) & 0 & 11.48$_{-10.84}^{+6.61}$ & 19.49$_{-3.14}^{+2.24}$ & 5.49$_{-4.68}^{+1.77}$ $\times$ 10$^{-12}$ & -125.50$_{-19.74}^{+12.36}$ & 5754$_{-1060}^{+530}$ & 0.76$_{-0.27}^{+0.11}$ \\ 
\enddata
\tablenotetext{a}{This is the rest-frame days since observed g-band maximum with reference epoch MJD 55770.562.}
\end{deluxetable*}

The results shown in Table~\ref{tab:simpleCSM} are highly dependent on the adopted shock velocity. If the shock velocity at the time of our first spectrum differs from 3200 km s$^{-1}$, then the inferred reference mass-loss rates would scale as $\dot{M} \propto (v_{s}/3200 ~\rm{km/s})^{s-5}$, the inferred shock radius at the time of the power-law break as $R_s \propto (v_s/3200~\rm{km/s})$, and the total CSM mass out to the radius of the power-law break as $M_{\rm{tot}} \propto (v_s/3200~\rm{km/s})^{-2}$. Lower shock velocities (for instance, if part of the line width in the observed spectrum is due to electron scattering) would therefore correspond to slightly higher inferred mass-loss rates, smaller radii, and higher total CSM masses.

We can also consider a more agnostic approach to the CSM density distribution (and therefore shock evolution), rewriting Equation~\ref{eq:simpleCSM} as $L = \left( \frac{\eta}{2} \right) \left( \frac{\dot{M}}{v_{w}} \right) v^{3}_{s}$. In this case, we have simply substituted $\rho_{CSM} = \dot{M}/(4 \pi v_w R_s^2)$ such that $\dot{M}$ now represents the inferred pre-explosion mass-loss rate at whatever radius the shock is at during the time of the luminosity being considered. We apply this simplified model to the bolometric light curve of PS1-11aop, adopting a wide range of possible shock velocities between 2000 and 7000 km s$^{-1}$ (consistent with the range of velocities inferred from the observed expansion of the blackbody radii and FWHM of the observed spectra in Sections ~\ref{subsec:temprad} and ~\ref{sec:spec props}). For an efficiency of $\eta = 1$, we infer pre-explosion mass loss rates that range from 
$\sim$0.05--2 $\times$($v_w$/100 km/s)$^{-1}$ M$_{\odot}$ yr$^{-1}$ (from the luminosity during the bolometric peak) to $\sim$0.01--0.5 $\times$($v_w$/100 km/s)$^{-1}$ M$_{\odot}$ yr$^{-1}$ (from the luminosity during the second observing season). These values span a wider range but are broadly consistent with the results above.

\subsection{MOSFiT modelling} \label{subsec:MOSFiT modelling}

\begin{figure*}[ht]
\centering
\vspace{-0.5in}
\includegraphics[width=.8\textwidth,trim=.05cm 0cm 0cm .10cm,clip]{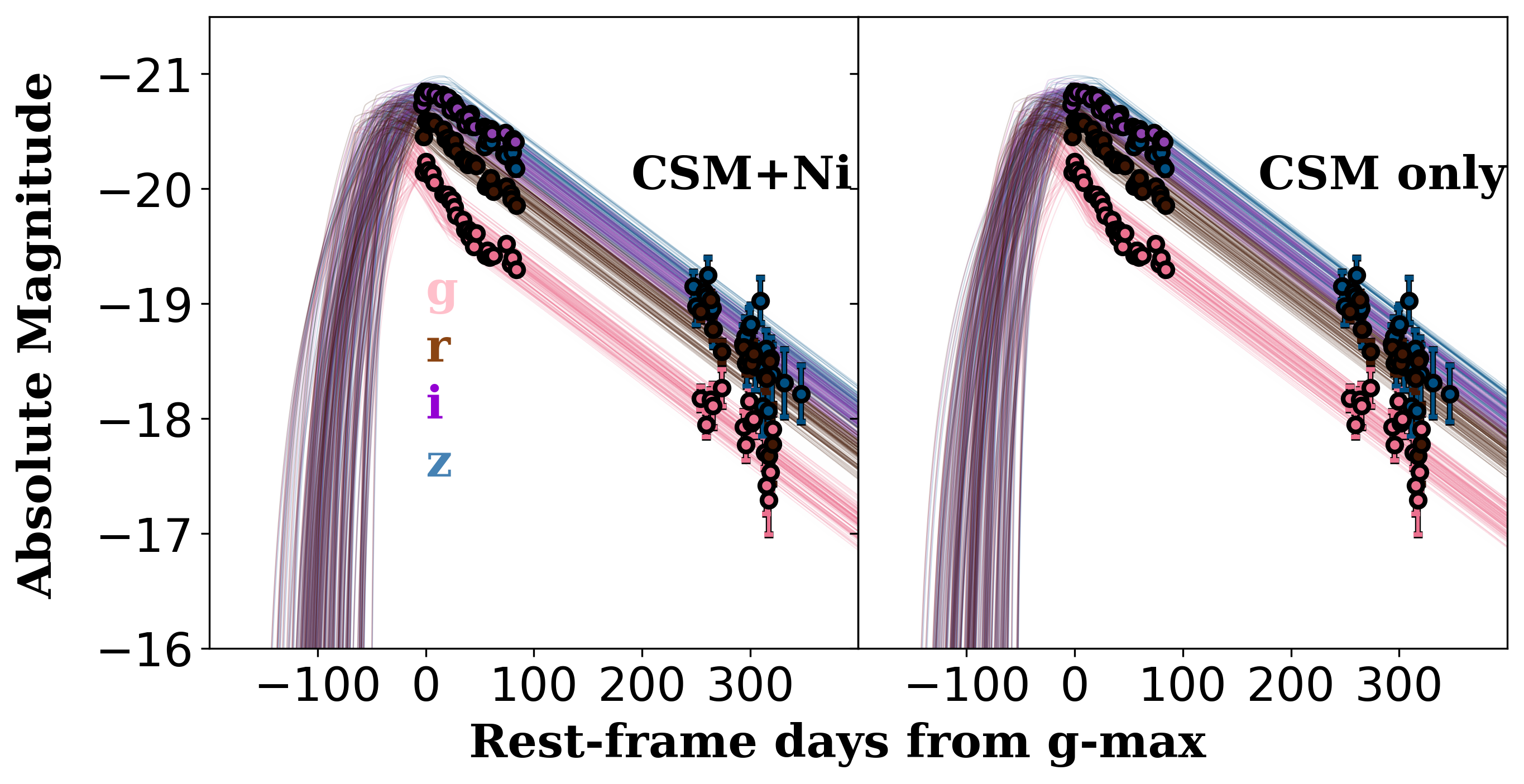}
\caption {MOSFiT realizations of a full MCMC fit to the $g_{P1}r_{P1}i_{P1}z_{P1}$ light curves for PS1-11aop. The error bars show 3$\sigma$ uncertainty. The left panel represents the model with radioactive decay of Nickel and circumstellar interaction with explosion as a power source while the right panel is for circumstellar interaction only. Fixed parameters based on the physics of the models are presented in \S \ref{subsec:MOSFiT modelling} while posteriors of the model parameters are presented in Table \ref{tab:MOSFiTresults}.  
} 
\label{fig:MOSFiTLC-bol}
\end{figure*}

In addition, we model the optical light curve of PS1-11aop using MOSFiT: the Modular Open Source Fitter for Transients \citep{Guillochon_2018}. In broad terms, MOSFiT takes as input a model for the power source of a given SN and, under the assumption of a blackbody spectrum, will perform a Markov Chain Monte-Carlo (MCMC) fit to the observed photometry to produce a posterior distribution for the free parameters of the input model. We consider two models when fitting the light curve of PS1-11aop: (i) pure circumstellar interaction and (ii) circumstellar interaction plus the radioactive decay of $^{56}$Ni. The model for circumstellar interaction implemented within MOSFiT is described in \cite{Villar2017} and based on that of \cite{Chatz2012}. It includes luminosity contributions from both the forward and reverse shocks which then undergo diffusion through the CSM assuming a stationary photosphere.

Throughout this modeling, we assume a flat core ejecta density profile ($\delta = 0$), a steep ejecta envelope density profile ($ n = 12$), an inner CSM radius $R_{0}$ of 6.685\,AU, and the default MOSFiT value of 50\% efficiency in radiating the deposited energy.  The important free parameters in the modeling are the characteristic ejecta velocity, ($v_{ej}$), time of the explosion ($t_{exp}$), the inner CSM density ($\rho_{CSM, inner}$) which is evaluated at $R_{0}$, the CSM density profile ($s$\footnote{The inner CSM density is $\rho_{CSM} =$ constant for a slope of $s = 0$ which is a shell-like density structure and $\rho_{CSM}$ $=$ $\rho_{CSM, inner}$ ($R_{0}$/r)$^{s}$  at a given radius, $r$ for a slope of $s = 0-2$ which is a wind-like CSM structure, where the inner CSM radius $R_{0}$ is 6.685\,AU.}), the total CSM mass ($M_{CSM}$), the total ejecta mass ($M_{ej}$) and the mass fraction of Nickel ($f_{\rm{Ni}}$) which is multiplied by the $M_{ej}$ to obtain the mass of radioactive Nickel ($M_{Ni}$). We set priors for these free parameters as shown in Table \ref{tab:MOSFiTresults}.

With this setup, we perform a fit to the absolute magnitude $g_{P1}r_{P1}i_{P1}z_{P1}$ light curves of PS1-11aop. To optimize the fit, we removed the i-band points from the data for the second season since this portion of the spectrum corresponds to the frequency range of the $H_{\alpha}$ line, which is thought to dominate the emission in that region (See \S \ref{subsec:temprad}). Throughout the run, we set the Gelman-Rubin statistics to 1.2 \citep{Gelman1998}, used 100 MCMC walkers, and 1000 points in each filter for each walker.

Table \ref{tab:MOSFiTresults} gives the results of our MOSFiT runs, while Figure \ref{fig:MOSFiTLC-bol} shows multiple realizations of the MCMC fit to the light curve for both models. We find that both models can broadly reproduce the observed photometry. Notable aspects of the best-fit solutions include: 
\begin{itemize}
    \setlength\itemsep{-0.15em}
    \item The total CSM mass is large ($\sim$20--27 M$_{\odot}$) and inner CSM density is high ($\sim3-5\times10^{-12}$ g cm$^{-3}$). These inner densities and radii correspond to equivalent progenitor mass-loss rates of 0.05$-$0.1 \,M$_\odot$ yr$^{-1}$, for an assumed wind speed of $v_{w}=100$\,km\,s$^{-1}$.
    \item Even when allowing contributions from radioactive decay, the model preferred only very small amounts ($\sim$0.04 M$_{\odot}$) of $^{56}$Ni such that it would contribute negligibly to the overall luminosity. As a result, the CSM properties inferred from the two models are broadly similar.
    \item MOSFiT strongly prefers an explosion date of $\sim$125$-$150\,days prior to the first PS1 detection. This is because the high luminosity requires a dense CSM, with a correspondingly long diffusion time. Such a long rise time is allowed by our data (See \ref{subsec:estimate of exp epoch}).
    \item MOSFiT favors a CSM density profile that is flatter than expected for a wind-like medium (s$\sim$0.6--0.8). This could be possibly caused either by changing mass-loss rates from the progenitor system over time or by a complex geometry \citep{Fransson1996}.
    \item Combining the inner CSM density ($\rho_{CSM, inner}$), slope (s), and total CSM mass ($M_{CSM}$) from MOSFiT, the inferred outer edge of the density CSM is $\sim$3.5$\times$10$^{15}$ cm. This is smaller than the radii inferred due to the break in the power-law behavior of the bolometric light curve in Section~\ref{subsec:CSM model}.
    \item This small outer CSM radius implies that the shock breaks out of the densest portion of the CSM during the middle of the first season of PS1 observations. This is evident as the change in decline rate in the light curve $\sim$60 days after the g-band maximum described in Section~\ref{subsec:basic LC props}. Within the MOSFiT CSM model, unless other power sources (such as $^{56}$Ni) are included, once the shock exists the dense CSM the light curve will be powered solely by adiabatic cooling and will follow a power-law decline \cite{Villar2017}.
    \item Although there are large uncertainties on the ejecta mass ($M_{ej}$), the best-fit models imply that the total CSM mass ($M_{CSM}$) is greater than that of the $M_{ej}$. This will lead to substantial deceleration.
\end{itemize}

In summary, MOSFit is broadly consistent with the simple CSM model described in Section~\ref{subsec:CSM model}, in requiring a large CSM mass to produce the optical light curve of PS1-11aop\footnote{In particular, if we had adopted an efficiency factor of 50\% in Section~\ref{subsec:CSM model}, which is closer to MOSFit's assumptions, then the total CSM masses in Table~\ref{tab:simpleCSM} would increase to $\sim$14--30 M$_\odot$}. However, MOSFiT prefers a CSM that is somewhat denser and confined to a smaller radius than what was found using the simple CSM model, above. 

We note that while MOSFit includes the effects of diffusion, it makes other simplifying assumptions in order to calculate the emission due to CSM interaction. In particular, MOSFit is a one-zone model that considers the diffusion of a deeply embedded power source through a fixed photospheric radius. Recent comparisons have identified some discrepancies between results from similar semi-analytic approaches and those from radiation hydrodynamic codes that both resolve the interaction layer, tracking its outward movement with time and employ a more sophisticated approach to solving the radiation transport \citep{Suzuki2021,Chatzopoulos2013}. Most notable is the example of the super-luminous SN\,2016aps, for which \citet{Nicholl2020} estimated a CSM mass of $M_{CSM} = 40-150\,$M$_{\odot}$ using MOSFit, while \citet{Suzuki2021} later found a more moderate amount of $M_{CSM} \sim 8\,$M$_{\odot}$, using a 1D radiation hydrodynamic simulation. 

It is therefore possible that the densities/masses inferred from MOSFit are overestimated. However, we note that light curve properties of PS1-11aop (peak luminosity, total radiated energy, shallow light curve evolution) are similar to both: (i) the model from \cite{Dessart2015} in which 9.8 M$_{\odot}$ of ejecta collides with a 17 M$_{\odot}$ CSM shell located within 1$\sim$10$^{16}$ cm and (ii) the models from \citet{Suzuki2020} in which a 10\,M$_{\odot}$ ejecta collides with a 10$-$20\, M$_{\odot}$ CSM shell located within 5$\times$10$^{15}$ cm of the progenitor star. Each of these models was based on radiation hydrodynamic simulations with an explosion kinetic energy of 10$^{51}$ ergs and are broadly comparable to the results found here and in Section~\ref{subsec:CSM model} based on semi-analytic models. When discussing implications for the overall CSM density distribution around PS1-11aop in Section~\ref{subsec:Overall CSM structure}, we will consider the broad range of solutions found by both methods.

\section{Radio Modelling} \label{sec:radio modelling}

In order to understand the structure of the CSM at larger physical scales, we present modeling of radio observations obtained between $\sim$4$-$10 years post-explosion. Our baseline model and assumptions are described in Section~\ref{subsec:Radio SSA+FFA Model} and the application to PS1-11aop in Section~\ref{subsec:Results from Radio Modelling}. We consider two cases: (i) the luminous radio emission detected is due to PS1-11aop and (ii) all radio observations represent upper limits to the true flux from the transient.

\subsection{Radio SSA and FFA Models} \label{subsec:Radio SSA+FFA Model}

Radio emission from interacting SNe primarily comes from synchrotron emission at the forward shock created by the fastest-moving ejecta as the SN shock expands into the CSM. We model the radio SEDs of PS1-11aop using a modified version of the framework described by \cite{Chevalier1998ApJ} and \cite{Chevalier_2006} for synchrotron emission (SE) and self-absorbed synchrotron emission (SSA) from interacting SNe. In particular, we include the effects of external free-free absorption (FFA) and describe the spectral energy distribution as: 
\begin{equation}\label{eq1}
    F_{\nu(SE+ SSA + FFA)} = F_{\nu(SSA)} \times e^{-\tau_{ff}}
\end{equation}
where F$_{\nu(SSA)}$ is a synchroton self-absorbed spectrum and $\tau_{ff}$ is the optical depth to free-free absorption. 

We represent the synchrotron self-absorbed spectrum as a function of frequency:
\begin{equation}
\begin{split}
\label{eq2}
F_{\nu(SSA)} = 1.582 F_{\nu_p} \left(\frac{\nu}{\nu_{p}}\right)^{5/2} 
 \left\{1-\exp\left[-\left(\frac{\nu}{\nu_p}\right)^{-(p+4)/2}\right]\right\} \quad
\end{split}
\end{equation}

\noindent where $F_{\nu_p}$ is the peak flux at frequency $\nu_{p}$, and $\nu_{p}$ is the synchrotron self-absorption frequency. In this equation, $p$ is the spectral index of the relativistic electrons \citep{Alexander2015}. We adopt $p=3$ \citep{Chevalier_2006}.

In order to estimate the free-free optical depth due to CSM material \emph{exterior} to the radio-emitting material at any point in time, it is necessary to combine constraints on the current radius and CSM density at the forward shock with an assumption about the external CSM density profile. For any combination of $F_{\nu p}$ and $\nu_{p}$, the radius of the radio-emitting material is \citep{Alexander2015,Chevalier1998ApJ,Chevalier_2006}:
\begin{equation}
\begin{split}
\label{eq3}
 R = 4.0 \times 10^{14} \alpha^{-1/19} \left(\frac{f}{0.5}\right)^{-1/19} \left(\frac{F_{\nu_p}}{\rm{mJy}}\right)^{9/19} \\  \times \left(\frac{D}{\rm{Mpc}}\right)^{-18/19}\left(\frac{\nu_{p}}{5~\rm{GHz}}\right)^{-1} \rm{cm} \quad \quad \quad \quad
\end{split}
\end{equation}

\noindent where $\alpha = \epsilon_E/\epsilon_B$ is the ratio of the fraction of post-shock energy contained in the relativistic electrons and amplified magnetic fields, $f$ is the radio filling factor, and $D$ is the (angular diameter) distance to the supernova. We adopt baseline values for these parameters of $\alpha = 1$, $f = 0.5$, and $\epsilon_B = 0.1$. The impact of varying these values will be discussed in \S \ref{subsec:radio-param-impact}. This expression comes from relating the equations for the optically thin and thick portions of the synchrotron-self absorbed spectrum \cite[see][for details]{Chevalier1998ApJ}. 

When coupled with the time post-explosion, $t$, the same parameters can also be used to calculate the density of the circumstellar material at the forward shock as described in \citep{Alexander2015,Chevalier_2006,Chevalier1998ApJ}:

\vspace{-0.2in}
\begin{equation}
\begin{split}
\label{eq4}
\rho_{CSM} = 4.64 \times 10^{11}\alpha^{\frac{-8}{19}}  \left(\frac{\epsilon_{B}}{0.1}\right)^{-1} \left(\frac{f}{0.5}\right)^{-8/19} \quad \\ \times \left(\frac{F_{\nu p}}{\rm{mJy}}\right)^{-8/19} \left(\frac{D}{\rm{Mpc}}\right)^{-4/19}\left(\frac{\nu_{p}}{\rm{5~GHz}}\right)^{2} \quad \quad \\ \times \left(\frac{t}{10~\rm{days}}\right)^{2} \left(\frac{R}{\rm{cm}}\right)^{-2} \, \rm{g} \, \rm{cm}^{-3} \quad \quad 
\end{split}
\end{equation}

\noindent To calculate the free-free optical depth, we then assume that the external CSM has a power-law profile of the form $\rho_{CSM} = \rho_0 (r/R_0)^{-s}$ where $\rho_0$ is the density at radius $R_0$. In this case, the free-free optical depth can be expressed in cgs units as:

\begin{equation}
\label{eq5}
      \tau_{ff} =  \frac{2.39 \times 10^{27} Z_{ave}}{\mu^{2}T^{1.35}} \frac{1}{\nu^{2.1}} \frac{\rho_0^2 R_0}{(2s-1)}
\end{equation}

\begin{figure*}[ht]
\centering
\includegraphics[width=.99\textwidth,height = 0.27 \textheight,trim=.05cm 0cm 0cm .10cm,clip]{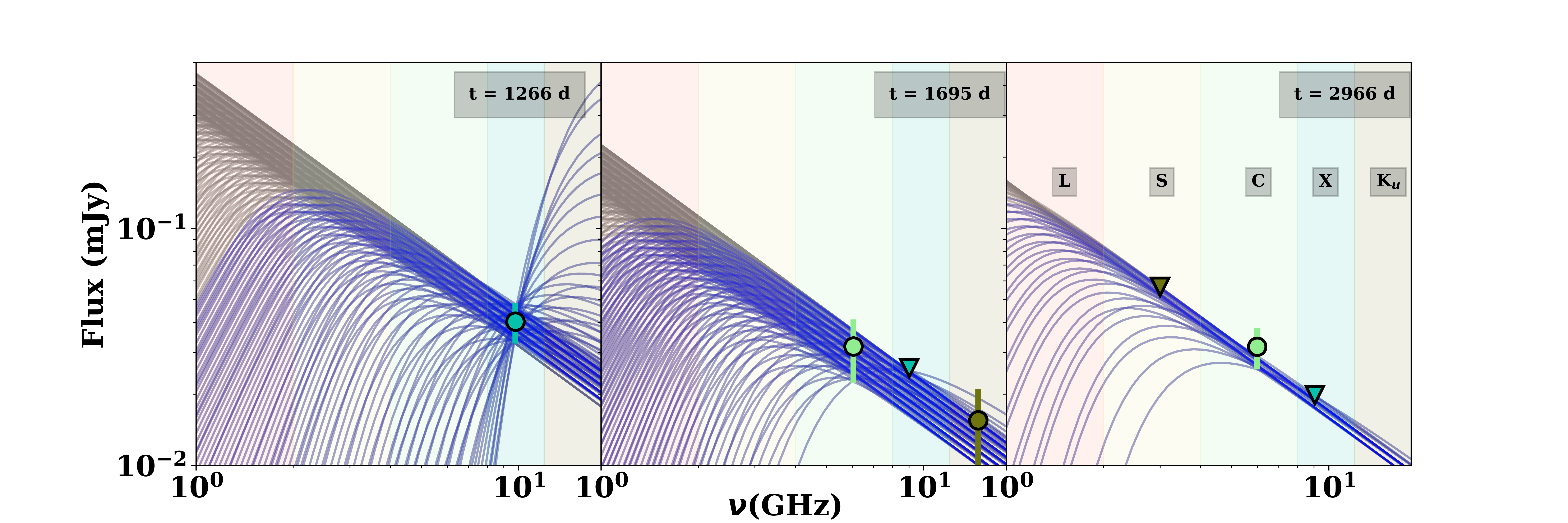}
\caption {Model SEDs that are consistent with the observed radio fluxes and upper limits for PS1-11aop. Models include the effects of both synchrotron self-absorption and free-free absorption and are shown in comparison to our 2015 X-band observations (left), the 2017 multi-frequency observations (middle), and the 2021 multi-frequency observation (right). 
All times are rest-frame days with respect to the estimated explosion epoch (g-band maximum $-$5\,days). Models in blue have inferred characteristic velocities ($v=R/t$) that are less than 20,000 km s$^{-1}$.} 
\label{fig:sedradio}
\end{figure*}

\noindent where $Z_{ave}$, $\mu$, and $T$ are the average metallicity, mean molecular weights, and temperature of the electrons. Throughout our analysis, we assume $\mu = 1$ and $Z_{ave} = 1$ for pure hydrogen, and $T= 10^4$ K. In the case of a wind-like medium ($s=2$), this reduces to the expression from \citet{Chevalier1998ApJ}, who also chose to parameterize the density in terms of a progenitor mass-loss rate, $\dot{M}$, and wind-speed, $v_w$, as $\rho_{0} = \dot{M}/(4\pi R_0^{2}v_{w})$. 
Using this value of $\tau_{ff}$, we can then modify the SSA spectrum given in Equation \ref{eq2} to account for the effects of free-free absorption.

\subsection{Application of Models to PS1-11aop } \label{subsec:Results from Radio Modelling}

In order to interpret the radio detections and/or upper limits of PS1-11aop, we generated a dense grid of SSA+FFA SEDs for a wide range of peak fluxes ($10^{-5}$ mJy $<$ $F_{\nu_p}$ $<$ 10 mJy) and frequencies ($10^{-2}$ GHz $<$ $\nu_p$ $<$ $10^2$ GHz) using equations \ref{eq1}--\ref{eq5}. When constructing our SED grid, we initially apply values of $\tau_{ff}$ assuming a wind-like medium for the external CSM ($s=2$). Given that the explosion epoch is relatively unconstrained within the $\gtrsim$100 days prior to first detection, we adopt an explosion epoch of 5\,days before g-band maximum when calculating physical parameters implied for each SED. The impact of varying both of these assumptions will be examined in \S \ref{subsec:radio-param-impact}. 

We then determine which combinations of $F_{\nu_p}$ and $\nu_p$ are allowed or ruled out by each epoch of our data. We require that a given SED does not violate any upper limits and (when modeling detected radio emission as originating from that SN) that it falls within the 1$\sigma$ error bars of any detections. We initially consider all three epochs independently and remain agnostic to the physical parameters implied by each SED. As such, the parameter space of SEDs allowed by the first epoch (with only 1 observed frequency) was significantly larger than that of either the second or third epochs (with 3-4 observed frequencies each). We subsequently apply a set of physically motivated cuts to further restrict the parameter space of allowed SEDs, as described below.

\subsubsection{Case 1 Results: Radio Emission Due to SNe}\label{subsec:radio-model-result}

Results of the radio modeling using the baseline parameters described in Section~\ref{subsec:Radio SSA+FFA Model} are shown in Figure~\ref{fig:sedradio}. 
We do not capture the peak of the SED during any of the three epochs. However, if we consider only solutions with a characteristic velocity $v = R/t$ (where $R$ is taken from Equation~\ref{eq3}, and t is time of post-explosion), that is less than 20,000 km $s^{-1}$ (shown in blue in Figure~\ref{fig:sedradio}) we would infer that the peak of the SED was located between $\sim$1--5 GHz between $\sim$1700--3000 days post-explosion\footnote{Formally, this velocity is the speed of the homologously expanding ejecta which has reached the reverse shock at the time, t. In detailed consideration of the hydrodynamics of interacting SN, this is comparable to, but slightly larger, than the instantaneous velocity of the forward shock, $v_s = dR/dt$ (see Appendix~\ref{apsec:x-ray model})}. We view this number for the characteristic velocity as a conservative limit, since the early blackbody radius evolution (\S~\ref{subsec:temprad}) and broad component of the observed H$\alpha$ emission (\S~\ref{sec:spec props}) imply shock velocities of only $\sim$3000--7000 km $s^{-1}$. 

When considering only solutions with $v <$ 20,000 km $s^{-1}$, the inferred radius of the radio-emitting material at each epoch ranges from a few times 10$^{16}$ to a few times 10$^{17}$ cm, and CSM densities from a few times 10$^{-22}$ to a few times 10$^{-19}$ g cm$^{-3}$ (see Figure~\ref{fig:densityplot}). Adopting the expression $\dot{M} = 4\pi R^{2} \rho_{\rm{csm}} v_{\rm{w}}$ and a fiducial wind speed of $v_{\rm{w}} = 100$ km s$^{-1}$, these allowed densities and radii correspond to equivalent pre-explosion mass-loss rates of $\sim2\times10^{-4}$ to $\sim7\times10^{-3}$ M$_\odot$ yr$^{-1}$ for epoch 1 and $\sim2\times10^{-4}$ to $\sim2\times10^{-3}$ M$_\odot$ yr$^{-1}$ for epochs 2 and 3.

 \begin{figure*}
 \centering
\includegraphics[width=0.4\textwidth,height=0.25\textheight]{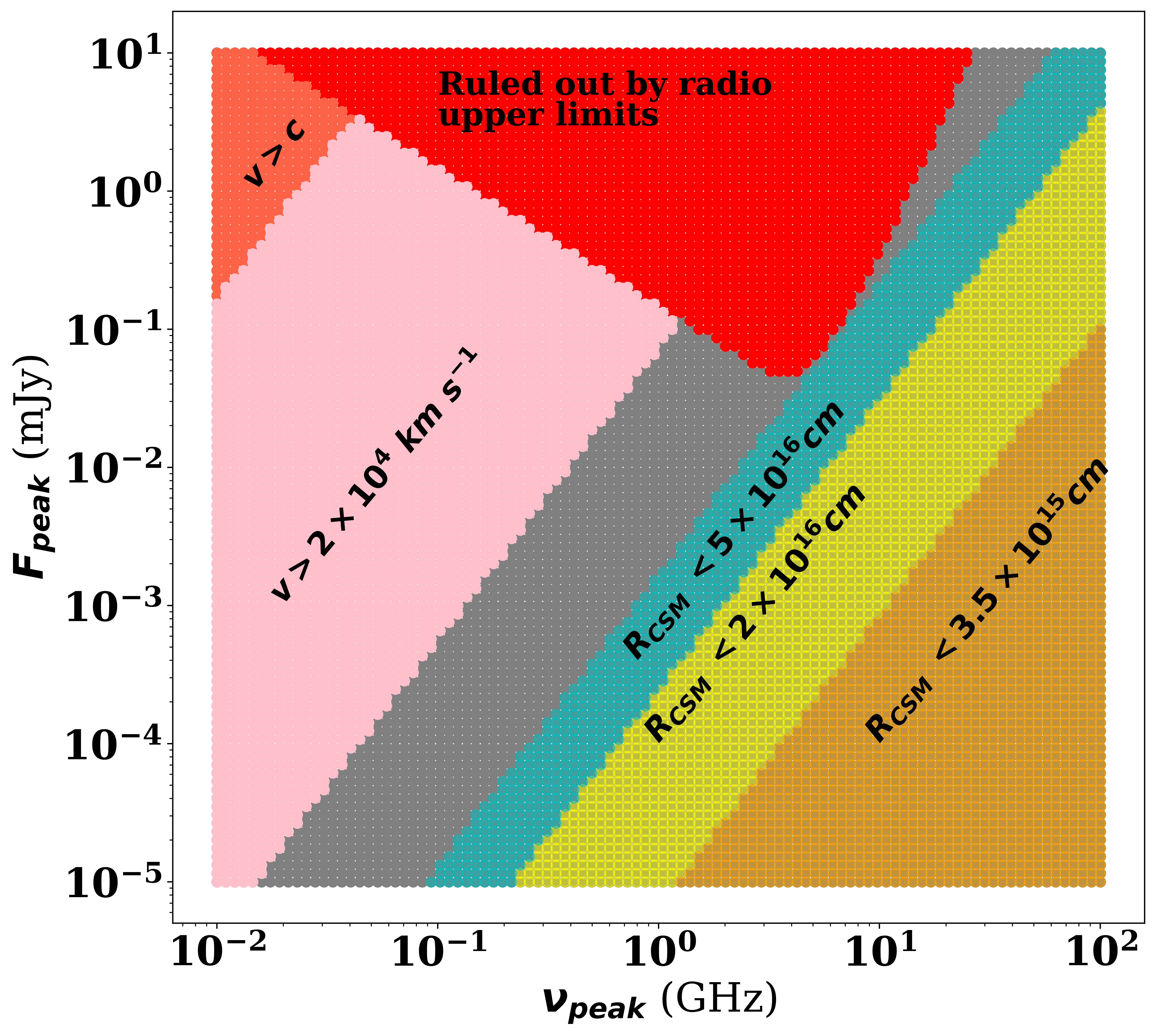}
\includegraphics[width=0.4\textwidth,height=0.25\textheight]{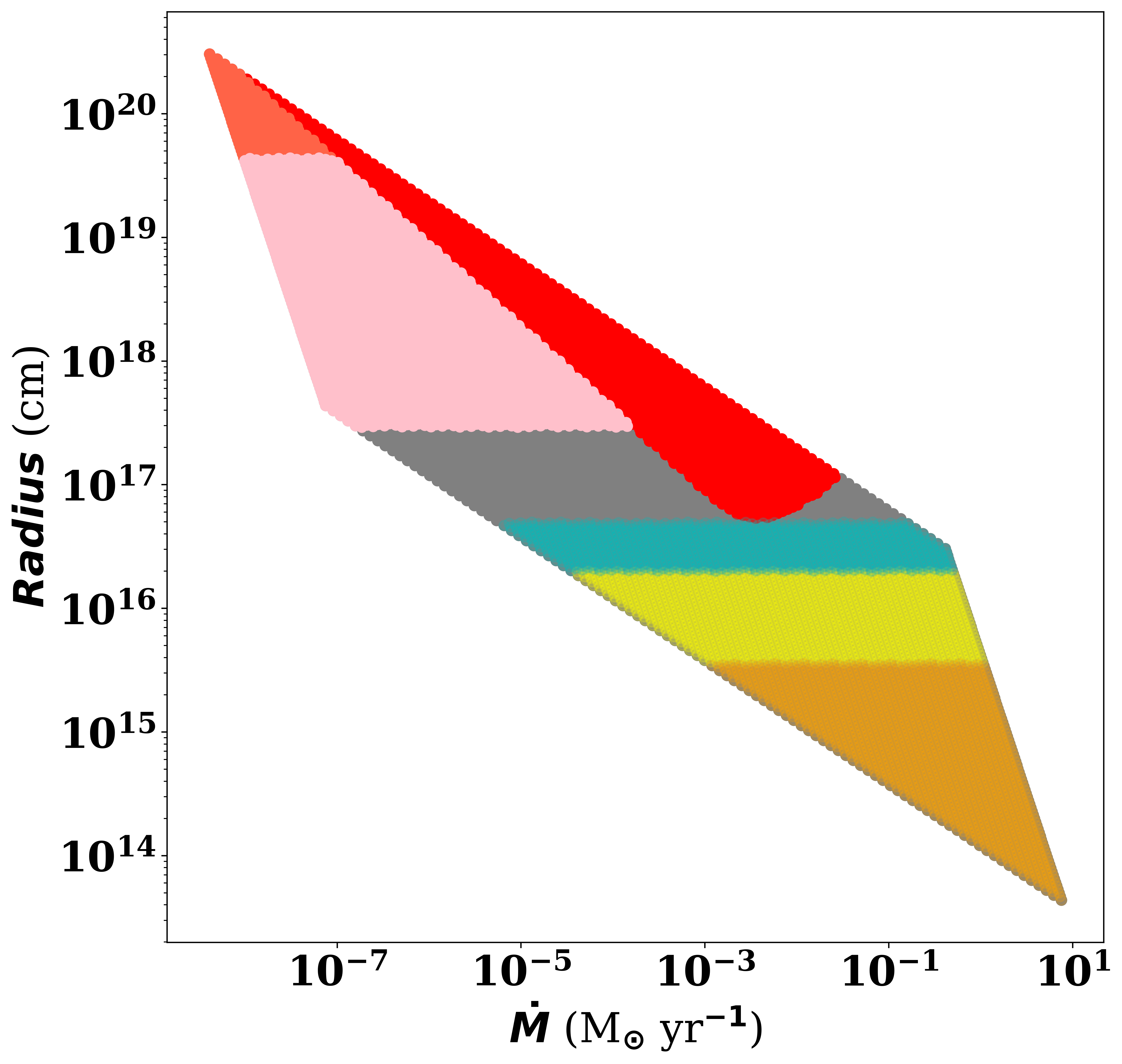}
\caption{Parameter spaces of the radio upper limit modeling assuming that the observed emission is not from the SN. \emph{Left:} The peak fluxes and frequencies used in our model grid. The red region is ruled out by the radio upper limits. We additionally highlight regions whose solutions imply certain values for either the characteristic velocity or radius of the radio-emitting material. Specifically, we highlight where: (i) $v>c$ (peach), (ii) $v>$20,000 km s$^{-1}$ (pink), (iii), $R<3.5\times10^{15}$ cm (orange), (iv) $R<2\times10^{16}$ cm (yellow), and (v) $R<5\times10^{15}$ cm (teal). Each of these is disfavored based on different physical arguments stemming from the optical observations of PS1-11aop (see text in \S \ref{subsec:radio-limits} for details). The remaining solutions are shown in grey. Compared to solutions found from modeling the radio detections are true emissions from the SN, these include (i) solutions with similar peak frequencies but lower peak fluxes and (ii) a small patch of solutions with higher peak frequencies and fluxes. \emph{Right:} The resulting inferred radius of the radio-emitting region and mass loss rates of the model grid with the same color scale. The mass loss rates presented assume a wind speed of 100 km s$^{-1}$.
}
\label{fig:radio-limit}
\end{figure*}

For our adopted explosion epoch (5 days prior to g-band maximum), the values of $v=R/t$ allowed for each epoch range from $\sim$4200$-$5000 km $s^{-1}$ up to our imposed upper limit of 20,000 km $s^{-1}$. In general, the solutions with larger inferred radii/velocities have lower inferred CSM densities and pre-explosion mass-loss rates. If we examine only solutions with characteristic velocities $<$7,000 km $s^{-1}$ (more closely aligned with inferences from the optical data described above, and indicative of a shock that has been decelerated by interaction with a dense CSM) then the inferred pre-explosion mass-loss rates are slightly more restricted, spanning $\sim7\times10^{-4}$ to $\sim7\times10^{-3}$ M$_\odot$ yr$^{-1}$.

Given the range of allowed SEDs for each radio epoch, we do not directly constrain the slope of the CSM density profile. If we assume that the CSM has a power law density profile of the form, $\rho \propto r^{-\alpha}$, then the allowed SEDs constrain $\alpha$ to be in the range of 0.1 $\lesssim$ $\alpha$ $\lesssim$ 3.0. If we restrict ourselves to solutions with $\alpha = 2$, corresponding to wind-like density profiles, between the radii probed by all three epochs then the allowed pre-explosion mass-loss rates still span $\sim2\times10^{-4}$ to $\sim2\times10^{-3}$ M$_\odot$ yr$^{-1}$.

\subsubsection{Case 2 Results: Radio Emission as Upper Limits}\label{subsec:radio-limits}

As described in Section~\ref{sec: argument for sn}, while we observe a fading of roughly a factor of two in the X-band, (which we found were statistically significant with $p=0.025$; Section~\ref{subsec:radio obs}), the star formation rates that would be estimated by our radio detections are only a factor of $\sim$4--6 times higher than those found based on H$_{\alpha}$ emission in the host. Thus, following the same procedure as above, we consider which radio SEDs---and associated physical parameters---are allowed if we treat all of our radio observations as limits on the true flux from the SN. We focused on results from epochs 2 and 3, which are more constraining as they have observations at multiple frequencies.

In Figure~\ref{fig:radio-limit} we show results from this analysis for our second epoch of radio observations. The left panel shows the peak frequency and peak flux of our grid of SSA+FFA radio SEDs, while the right panel shows the inferred mass-loss rate and characteristic radius associated with each of these SEDs. We emphasize that the ``edges'' these inferred physical parameters are dictated by the size of the grid of $F_{\nu_p}$ and $\nu_p$ that we consider. If we had adopted a larger grid even lower/higher radii and densities could be found. Models marked in red are ruled out by our radio upper limits. Broadly speaking, there are two new classes of solutions allowed compared to the results described in Section~\ref{subsec:radio-model-result}: (i) SEDs with similar peak frequencies but lower peak fluxes and (ii) SEDs that have higher peak frequencies (and in some cases higher peak fluxes). The former tend to have lower implied CSM densities than the previous model set-up ($<2\times10^{-4}$ M$_\odot$ yr$^{-1}$). However, the latter, which represent SEDs that are still optically thick even at this late time can have higher implied CSM densities than previously allowed ($\gtrsim 1\times10^{-2}$ M$_\odot$ yr$^{-1}$). 
 
While many SEDs are formally allowed by the data, some imply physical parameters that are inconsistent with inferences from the optical observations of PS1-11aop.  First, as above, we disfavor solutions that imply characteristic velocities $>$20,000 km s$^{-1}$, which we consider a conservative limit (shown as a pink patch in Figure \ref{fig:radio-limit}). In addition, a number of solutions lead to very small inferred radii that may be unphysical for a shock front that has been expanding for $\gtrsim$1695 days, even if it is decelerating in a dense CSM. In particular, in Figure \ref{fig:radio-limit}, we highlight solutions that have radii lower than three critical values: 
\begin{itemize}
\setlength\itemsep{-0.1em}
    \item \emph{$3.5 \times 10^{15}$ cm:} This is the maximum blackbody radius that we measured from the optical emission of PS1-11aop at $\sim$30 days post-maximum (Figure~\ref{fig:radio-limit}). In the canonical picture, this represents the minimum radius of the interaction zone at that epoch.
    \item \emph{$2.0 \times 10^{16}$ cm:} In our optical spectrum obtained at 357 days post-maximum, we measured a FWHM for the H$\alpha$ features of $\gtrsim$7000 km s$^{-1}$. Assuming this is an unshocked ejecta that has yet to reach the reverse shock, it places a lower limit on the radius of the shock of $R = V \times t = 2 \times 10^{16}$ cm at this epoch.
    \item \emph{$5.0 \times 10^{16}$ cm:} If the shock front was located at $\gtrsim2 \times 10^{16}$ at 357 days, then it should have reached a larger radius by the time of the second radio observation (1695 days). While details depend on the deceleration of the shock in the dense CSM, \citet{Fransson1996} find that for a SN ejecta expanding into a CSM with a constant power-law density profile that radius should evolve with time as $R \propto t^{(n-3)/(n-s)}$, where $n$ is the power law index of the SN ejecta and $s$ is the slope of the CSM. Using this relation and testing $ 5 < n < 12$ and $0.5 < s < 2$, we estimate that the shock should be located at a radius of $> 5 \times$10$^{16}$ at the time of the second radio epoch, if the reverse shock has not yet reached the inner ejecta which is characterized by a shallower density profile.
\end{itemize}

Based on these arguments, we disfavor solutions that imply radii below these values (particularly the first two, which are less model-dependent). The remaining solutions are shown in grey in Figure~\ref{fig:radio-limit}. Many fall within the set that has lower implied densities than those found by fitting our radio detections, above. However, a patch of the solutions remains which peak at frequencies $>$15 GHz and fluxes $>$1 mJy, implying mass loss rates $\gtrsim$10$^{-2}$ M$_\odot$ yr$^{-1}$. In particular, we note again that even larger mass-loss rates than shown in Figure~\ref{fig:radio-limit} cannot formally be ruled out, as the edges are dictated by the size of our input grid. Such solutions would imply that the SN would have been easily detected with higher frequency observations. 

Finally, we note that this second set of high-density solutions would be incompatible with a scenario where (i) our first epoch of X-band observations \emph{did} contain true light from the SN, even if epochs 2 and 3 do not and (ii) the CSM density distribution is smooth. This is because these solutions are still optically thick at 9\,GHz, implying flux at those frequencies should be \emph{rising} in time. However, the predicted 9\,GHz flux density of these SEDs ranges from a factor of $\sim$80 to more than 10 orders of magnitude fainter than the flux density in our first epoch detection (at 1261 days). The more extreme discrepancies are found for solutions with higher inferred mass-loss rates. Thus, this discrepancy would be amplified for any extension of our grid. However, we will discuss the implications of both sets of solutions in Section~\ref{sec:Discussion and Conclusion}, below.

\subsubsection{Impact of Parameter Variation}\label{subsec:radio-param-impact}

In Section~\ref{subsec:radio-model-result}, we describe the results from our ``baseline'' radio model. This model contains some free parameters, whose values can impact the values of the physical properties we infer. We describe the impact of varying a number of these assumptions on the density of the CSM surrounding the progenitor of PS1-11aop.

\emph{Epoch of Explosion:} In our baseline model, we assumed the epoch of explosion was shortly before discovery by PS1. If we instead consider an explosion epoch $\sim$100 days prior to g-band maximum (as favored by the MOSFiT modeling described in \S~\ref{subsec:MOSFiT modelling}) we find that the derived parameters (v$_{sh}$, R, and $\rho_{CSM}$) vary by $<$6\%. 

\emph{Free-free absorption:} In our baseline model, we calculate the free-free optical depth assuming the CSM exterior to the radio-emitting region has a wind-like density profile. However, in the case that the radio emission is due to the SN, all of our observations appear to be at frequencies higher than the SED peak. Thus, we find minimal ($<$2\%) modifications to the fluxes due to the inclusion of free-free absorption. Our results are therefore relatively insensitive to the exact shape of the density profile assumed when calculating $\tau_{ff}$. Recalculating our results with values of $1 < s < 3$ in Equation~\ref{eq5}, we find variations in the derived parameters of $<$0.1\%. 

\emph{Equipartition:} In our baseline model, we assume $\alpha = \epsilon_E/\epsilon_B = 1$ (i.e. equipartition of energy between the relativistic electrons and magnetic fields) with values of $\epsilon_E = \epsilon_B = 0.1$. However, these parameters are uncertain. Constraints on $\alpha$ that have been determined from SNe with both radio and X-ray observations typically falling in the range of 10$-$100 \citep[e.g.][]{Horesh2013, Yadav2014, Horesh2020, Ruiz-Carmona2022}. From equation Equation~\ref{eq4}, we see that our inferred CSM densities are inversely proportional to both $\alpha$ and $\epsilon_B$. Thus, to test the cumulative affect of increasing $\alpha$ (and correspondingly decreasing $\epsilon_B$), we rerun our models setting $\alpha = 10$ and $\epsilon_B = 0.01$. We find that our inferred CSM densities/mass loss rates increase by a factor of $\sim$1.5--3.  Thus we find that most free parameters in our model could either have minimal effect or could increase our estimates for the CSM density by a factor of a few.

\section{X-ray Modelling} \label{sec:x-ray modelling}

As discussed in Section~\ref{sec: argument for sn}, while we do not find any direct evidence for an AGN in the host of PS1-11aop, we also only have a single epoch of X-ray observations, which limits our ability to constrain its nature. Thus, our main goal in modeling the X-ray emission detected at the location of PS1-11aop $\sim$1640 post-discovery is to understand the implications if it is due to CSM interaction, and whether or not this is consistent with inferences made from the optical and radio emission, described above. As this does not impact the main conclusions of this manuscript, the bulk of this analysis can be found in Appendix~\ref{apsec:x-ray model} and \ref{apsec:X-ray result}. Here we summarize our modeling framework and main conclusions.

\subsection{X-ray Model Framework} \label{subsec:X-ray Model}

To interpret the X-ray detection from PS1-11aop, we first use a model framework similar to that employed by \citet{Chandra2015}. In brief: for a given SN ejecta and CSM density profile, we (i) numerically integrate the equations of motion to determine the shock radius and velocity as a function of time, (ii) calculate the density and total mass swept up at both the forward and reverse shock at each time step, (iii) determine the cooling time, and thus whether the reverse shock is adiabatic or radiative, (iv) estimate the resulting unabsorbed X-ray luminosity from both the forward and reverse shock, (v) estimate the total column depth of cool gas along the line of sight, and (vi) calculate a resulting observable X-ray flux. 

We make two notes about this methodology. First, while self-similar solutions to the equations of motion and resulting X-ray luminosities have been published (e.g. \citealt{Fransson1996,ChevalierFransson2003}), these assume a single power-law density profile for the CSM---which may not be applicable for the case of PS1-11aop. In the framework of \citet{Fransson1996} the predicted X-ray luminosity is sensitive to the \emph{entire} CSM mass swept up prior to the epoch of observation, thus necessitating more the generalized approach. Second, when estimating the unabsorbed X-ray luminosity, we assume the X-ray emission is dominated by thermal bremsstrahlung emission (Equation~\ref{eq:Lx}). However, in cases where the reverse shock is radiative, line emission can become important. We therefore also (i) calculate an approximate radiated luminosity by multiplying the kinetic luminosity of each shock by a radiative efficiency factor and (ii) perform an independent estimate of the density in the X-ray emitting region using the emission measure calculated in Section~\ref{subsec:x-ray obs}.

\subsection{Summary of Conclusions from X-ray Modelling}\label{sec:xray_summary}

Based on the modeling in Appdendix~\ref{apsec:X-ray result}, we find the luminous X-ray emission at the location of PS1-11aop could possibly be explained by interaction with a dense CSM described by either a single power-law or shell-like density distribution. Under the assumption that the X-rays are due to thermal bremsstrahlung, the high densities in the X-ray emitting region inferred from the emission measure would indicate a significant contribution from the reverse shock. In the broader context of the optical and radio observations of PS1-11aop, we would favor a shell-like CSM environment.  However, in this case, fine-tuning may be required with the density of the inner shell and the location of the transition between the inner and outer CSM regions to prevent the velocity and temperature of the reverse shock from decreasing to the point that X-rays are no longer expected.  In addition, in all cases, we find that the reverse shock is still radiative, and thus further modeling which takes into account various instabilities and line emission is required for a more detailed assessment of the X-ray detection. Alternatively, the luminous X-ray emission could indicate (i) an aspherical CSM with the radio and X-ray emission at 4.5 years post-explosion coming from different locations or (ii) an alternative power source for the observed X-rays.

\section{Discussion} \label{sec:Discussion and Conclusion}

PS1-11aop was a luminous and long-lived transient with coincident radio and X-ray emission detected between $\sim$4 and 10 years post-explosion. Analysis of this multi-wavelength data has allowed us to examine the structure of the CSM out to large distances. Here we discuss the implications of our results for the overall structure of the CSM surrounding PS1-11aop, the timescale of any mass ejections from the progenitor star, and implications for the mass ejection mechanism. Throughout, we rely primarily on the results of the optical and radio analysis presented in Sections~\ref{sec:optical modelling} and \ref{sec:radio modelling}. As described above, the X-ray results are generally compatible with this picture.

\subsection{Overall CSM structure} \label{subsec:Overall CSM structure}

\begin{figure}[t]
\includegraphics[width=.45\textwidth,height = 0.25 \textheight,trim=.05cm 0cm 0cm .10cm,clip]{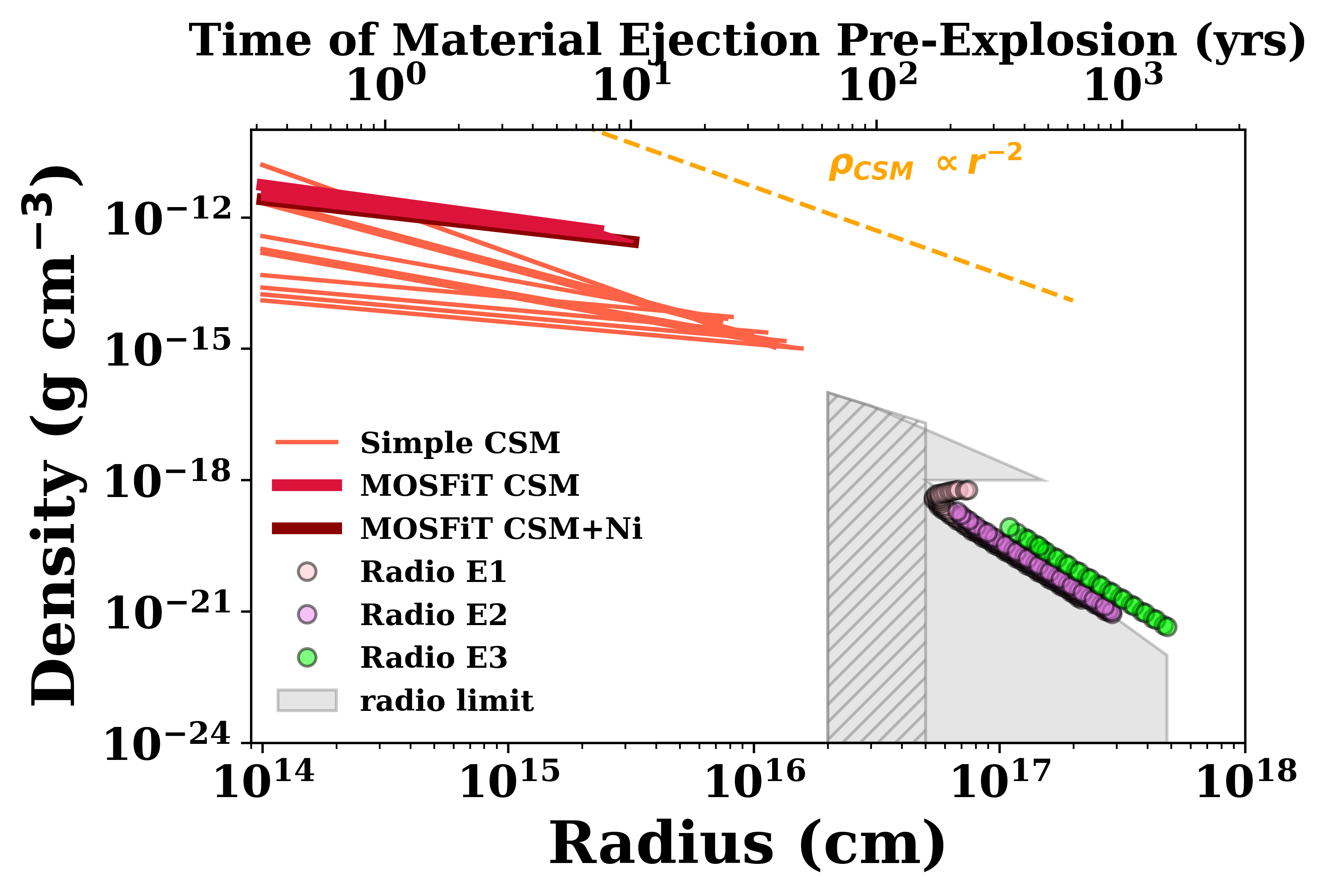}
\caption {Inferred density profiles surrounding PS1-11aop based on optical and radio data. The pink lines represent inferences from fitting the bolometric light curve with a simple CSM model described in S \ref{subsec:CSM model}. Different lines correspond to different assumptions for the epoch of the explosion and the slope of the density profile. Red and dark red bands represent MOSFiT solutions for CSM only and CSM $+$ Ni models, respectively. Pink, magenta, and green circles represent solutions allowed by modeling the three epochs of radio data as true emission from the supernova. (We emphasize that different points of the same color are not separate constraints, but rather different allowed solutions). Finally, the grey-shaded region represents solutions allowed when treating the observed radio emission as upper limits on the true flux from the SN. As described in Section~\ref{subsec:radio-limits}, we estimate that the shock has reached a radius of $\gtrsim5\times10^{16}$ cm by the time of the radio observations, so solutions below the valid distinguished by a hatched region.
Time scales of the mass ejections pre-explosion are shown on the top axis (assuming an ejection velocity of 100\,km\,s$^{-1}$). 
} 
\label{fig:densityplot}
\end{figure}

Figure \ref{fig:densityplot} shows the full extent of the CSM radius (1$\times 10^{14}$\,cm $- 5 \times 10^{17}$\,cm) that was probed in this study. The two red bands represent inferences from the two different MOSFiT models taken from the best-fit inner density, CSM slope, and total CSM mass as presented in Table \ref{tab:MOSFiTresults}. The thickness of the bands represents the uncertainty described by the posterior of each of these parameters. The pink area represents the range of radii and densities inferred from the simple CSM model (\S~\ref{subsec:CSM model}) for shock velocities between 3000 and 7000 km s$^{-1}$. 
The gray area shows the region allowed by the radio upper limits assuming that the emission is not caused by the SN itself. We have plotted all allowed radio solutions with inferred radii greater than 2$\times$10$^{16}$ cm, but, as described in Section~\ref{subsec:radio-limits} we estimate that the shock has likely reached a radius of $\gtrsim$5$\times$10$^{16}$ by the time the radio observations were obtained.
Finally, in colored circles, we plot the range of solutions allowed by all three epochs of radio observations in our baseline model (\S \ref{subsec:radio-model-result}) for an emission caused by SN. We emphasize that different circles of the same color do not represent independent constraints on the density, but rather different solutions that are allowed by the data.

We consider implications in two scenarios: (i) that the observed radio emission is due to the SN and (ii) that the observed radio emission is upper limits on the true flux.
In the former case, it is clear from Figure~\ref{fig:densityplot} that the densities inferred from the optical and the radio emission of PS1-11aop cannot be described by a single power law form. We require at least a two-zone model where the optical and radio are probing different phases in the mass loss history of the progenitor star.
In particular, the optical probes dense material that would have been ejected during the final stages of the star's life, while the radio regime probes the pre-explosion mass loss history of the system \emph{prior} to that final ``eruption''. Features of this CSM profile include:

\begin{itemize}
    \setlength\itemsep{-0.1em}
    \item \emph{Dense Inner Region:} While the density inferred for the inner CSM varies depending on the type of model used, the explosion date, and assumed shock velocity, all models investigated in Section~\ref{sec:optical modelling} favor large total CSM masses ($\sim$10--30 M$_{\odot}$) and high equivalent progenitor mass-loss rates ($\sim$0.02--1 M$_{\odot}$ yr$^{-1}$ for an assumed wind speed of 100 km s$^{-1}$). MOSFiT modeling prefers a density distribution that is shallower than a wind-like environment.
    \item \emph{Sparse Outer Region:} In the case that the radio emission is caused by the SN, our baseline radio results yielded CSM densities of $\sim$10$^{-21}$ to 10$^{-18}$ g cm$^{-3}$ at radii $>$5.4$\times10^{16}$\,cm. These corresponded to mass-loss rates of $\sim$10$^{-4}$ to 10$^{-3}$ M$_{\odot}$ yr$^{-1}$ for an assumed wind speed of 100 km s$^{-1}$. While we do not directly constrain the slope of the CSM density distribution, wind-like environments are allowed.
    \item \emph{Radius of Transition:} Based on the combined optical and radio modeling, the transition between the outer and inner CSM regions likely occurred between $\sim$$3\times10^{15}$\,cm and $5\times10^{16}$\,cm. The lower end of the range is the maximum blackbody radius found from our optical light curve and is the outer CSM radius favored by MOSFit. The upper end of this range is the minimum radius allowed when modeling the first epoch of radio emission. If the break in power-law evolution of the bolometric light curve at late-time (see Figure~\ref{fig:SimpleCSM}) represents the emergence of the shock from the dense shell (as was argued for SN\,2010jl; \citealt{Fransson_2014,Moriya2014}) then we would infer a break radius between $\sim 0.8-2 \times 10^{16}$ cm. 
\end{itemize}

In the event that the radio emission detected is an upper limit on the true emission from the SN, then additional density distributions are possible. As described in Section~\ref{subsec:radio-limits}, most of the allowed radio solutions in this scenario imply CSM densities that are lower than those found by modeling the data as detections. In this case, the same general picture as described above applies, but with a sparser outer environment. 

However, there is also an allowed parameter space with higher densities (corresponding to models that have high peak frequencies and fluxes). In this case, it may be possible to fit the CSM distribution surrounding PS1-11aop with a single zone, characterized by mass loss rates of $\sim$0.1 M$_\odot$ extending out to radii $\gtrsim$5$\times$10$^{16}$ cm. However, in this scenario, the origin of the sharp increase in the decline rate of the bolometric luminosity observed between the first and second seasons of optical observations is less clear. While a decrease in luminosity is expected when a SN enters the momentum-conserving phase \citep{Ofek2014}, \citet{Svirski2012} and \citet{Moriya2014} find that it should decline as $L \propto t^{-1.5}$. This is significantly shallower than the late-time decline we find in Section~\ref{subsec:CSM model}, which ranges from $L \propto t^{-2.5}$ to $L \propto t^{-3.5}$, depending on the assumed explosion epoch. A similarly steep decline ($L \propto t^{-3}$) was observed after 320 days in SN\,2010jl, and has been argued to be due to the shock emerging from the dense CSM shell or a sudden increase in the slope of the CSM density distribution \citep{Fransson_2014, Moriya2014}. Based on these arguments we favor the overall CSM density inferred from the lower-density radio solutions.

\begin{figure}[t]
\centering
\includegraphics[width=.4\textwidth,height = 0.3 \textheight,trim=.4cm 0cm 0.45cm .10cm,clip]{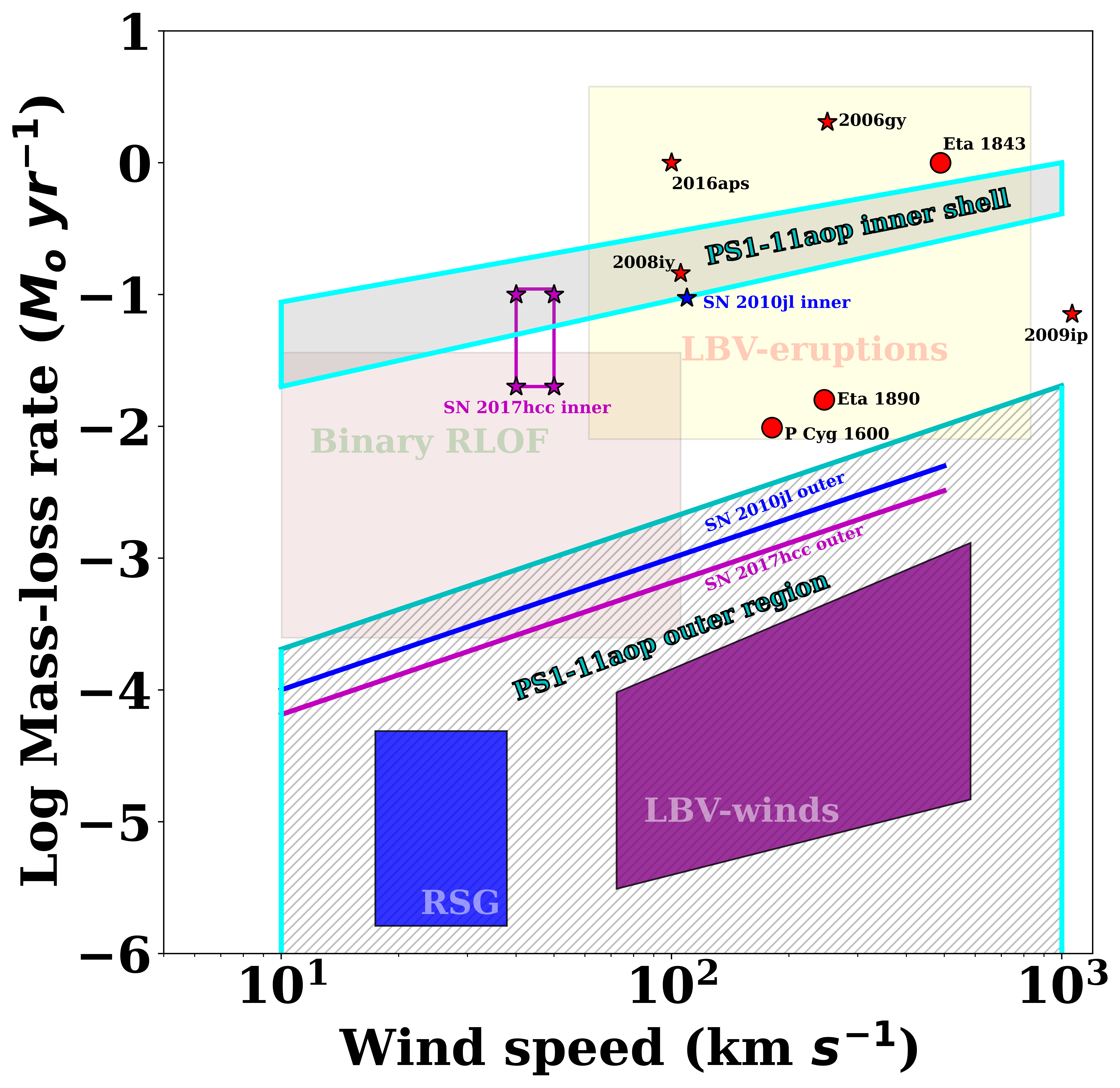}
\caption {A two-zone model for the density distribution surrounding PS1-11aop compared to the phase space of mass-loss rates versus wind velocity adapted from \cite{Smith2014,smith2017RS}. The cyan-enclosed box represents the parameter space allowed for PS1-11aop. The dense inner region is the region probed by the optical analysis while the gray dashed region represents constraints on the density from the radio emission, which probes a sparser outer region. Also shown are boxes that represent the parameter space for winds from red supergiants (RSGs), winds from luminous blue variables (LBVs), eruptions from LBVs, and non-conservative mass transfer from binary Roche lobe overflow (RLOF).  Stars and circles represent constraints on CSM shells from both type IIn supernova and LBVs, respectively. We particularly highlight values SN 2010jl (blue; \citealt{Fransson_2014}) and SN 2017hcc (magenta; \citealt{Chandra2022}), both of which are also consistent with a dense inner and sparse outer CSM profile.} 
\label{fig:nathan-smith}
\end{figure}

A summary of the inferred properties of both the inner and outer CSM regions for PS1-11aop for the two-zone model are shown in Figure~\ref{fig:nathan-smith}, where we plot progenitor mass-loss rate versus wind speed. Constraints for both CSM regions are shown as diagonal lines because we do not directly measure the wind/expansion speed for either CSM component (and $\rho \propto \dot{M}/v_w$). For the outer and inner CSM region, we plot lines that span a wide range of wind speeds (10-1000 km s$^{-1}$) assuming that the radio emission comes from the SN. Although, other luminous Type IIn SNe have expansion velocities spanning 30$-$600 km\,s$^{-1}$ \citep{Miller2010,Fransson_2014,Smith&Andrew2020}. 

The colored regions in Figure~\ref{fig:nathan-smith} correspond to parameter spaces for mass loss from evolved massive stars taken from \cite{Smith2014}. These correspond to eruptions from luminous blue variable (LBV) stars, steady winds from LBV stars, winds from red supergiants (RSGs), and non-conservative binary Roche-lobe overflow (RLOF). The red circles and stars represent examples of well-studied LBVs and Type IIn SNe, respectively, with constraints on both their pre-explosion mass-loss rates and wind velocities \citep[][and references therein]{Smith2014}. From this, we see that the dense inner CSM region observed in PS1-11aop is consistent with those found for Type IIn SNe and LBV giant eruptions. If the radio emission observed is from the SN, then the density in sparser outer region is still relatively high compared to steady mass loss rates observed for evolved stars in nearby galaxies, and more consistent with expectations for non-conservative mass transfer in binary systems. However, if they are a limit on the true radio emission, then the density in the outer region is comparable with the steady winds from either LBVs or RSGs.

This overall density structure resembles that of the luminous Type IIn SN2017hcc \citep{Prieto2017, Smith&Andrew2020,Chandra2022} and SN2010jl \citep{Fransson_2014}, both of which also have constraints on the density of the CSM at multiple physical scales. Modelling of the prompt emission of SN2017hcc ($\sim30-100$ days) revealed mass-loss rates of 0.1$-$10$^{-2}$\,M$_\odot$yr$^{-1}$, while radio emission observed at $\sim$ 1000\,days was consistent with lower densities ($\sim$ 6.5$\times$10$^{-4}$\,M$_\odot$yr$^{-1}$)  \citep{Chandra2022}. The timescale probed by the late-time radio emission corresponds to a CSM radius of $\sim$10$^{16}$\,cm. SN2010jl was followed for over 1100 days in the optical and displayed a two-component light curve as described above. \cite{Fransson_2014} estimated a mass-loss rate of $\sim$0.1\,M$_\odot$yr$^{-1}$ at radii $\lesssim$10$^{16}$\,cm based on observations within the first year of the explosion. In addition, they observed narrow emission lines in the nebular phase ($\sim$1100 days) which they use to estimate a mass-loss rate of $\sim$ 10$^{-3}$\,M$_\odot$yr$^{-1}$ at a radii of $\sim$10$^{17}$\,cm. Constraints on the densities of the inner and outer CSM regions for SN\,2017hcc and SN\,2010jl are also shown in Figure~\ref{fig:nathan-smith}.

\subsection{Implications for Timing of Evolutionary Phases} \label{subsec:implication-for-time}

The two-zone CSM described above implies two distinct eras in the pre-explosion life of the progenitor star. While we do not directly constrain the wind/expansion velocity of either CSM component, we can comment on the timing of these evolutionary phases under different assumptions. On the top axis of Figure~\ref{fig:densityplot}, we plot the implied timing of the mass ejection ($t = R/v_w$) for an assumed wind speed of $v_{w} = $100 km s$^{-1}$. Under this assumption, the dense shell probed by the optical emission would represent material ejected in the final $\lesssim$10-100 years pre-explosion. In contrast, the sparser material probed by the radio---which is located at radii between $\sim$5--50$\times$ 10$^{16}$ cm---would have been ejected sometime between $\sim$150 and 1500\,years pre-explosion. These results scale inversely with the assumed wind/ejection speed. For example, for a faster wind speed of 1000 km s$^{-1}$, we would infer that the dense shell probes material ejected in the final $\lesssim$1-10 years pre-explosion and the sparser outer region probes material ejected between 15 and 150 years pre-explosion. 

Finally, we note that if it is possible to explain the multiwavelength observations of PS1-11aop with a single-zone CSM (e.g. the radio emission provides upper limits, the high-density solutions are accurate, and there is a different explanation for the break in power-law behavior of the bolometric light curve between 300-400 days) this would require that the progenitor star underwent sustained mass loss of $\sim$0.1 M$_\odot$ yr$^{-1}$ for $\gtrsim$150 years prior to explosion (for an assumed wind speed of 100 km s$^{-1}$.

\subsection{Implications for Mechanism of Final Mass Ejection } \label{subsec:Overall implication}

As described above, given the full multiwavelength dataset for PS1-11aop, we favor a CSM density profile surrounding PS1-11aop described by a dense inner shell with multiple solar masses of material, surrounded by a sparser outer region. Here we discuss this result in the context of multiple mass loss mechanisms that have been proposed for luminous Type IIn SNe.

\subsubsection{Pulsational Pair Instability}\label{subsec:PPI}

Stars with initial masses between $\sim$70--140 M$_{\odot}$ may undergo a pulsational pair-instability which can trigger multiple ejections of shells of material in the final years and centuries before explosion \citep{Woosley2017}. Collisions between these shells can subsequently lead to bright transients, and this mechanism has been proposed as an origin for luminous interacting supernova. While constraints on the mass in the dense shell for PS1-11aop are $\sim$7--30 M$_\odot$, if there were multiple previous shell ejections it is possible that the progenitor was in the mass range where PPI may occur. 

While the light curves predicted for PPI are very diverse, in many cases multiple bumps are present over a time period of $\sim$300 days post-explosion due to the interactions of multiple shells \citep{Woosley2017,Chen2023}. We do not observe this in PS1-11aop, whose light curve showed a relatively smooth decline. It is possible that the timing of both the pre-explosion upper limits and longer timescale monitoring of PS1-11aop was not right to detect another shell (e.g. such as the models shown in Figures 18 and 19 of \cite{Woosley2017} where a faint and short-lived transient precedes a longer and more luminous transient by $\sim$18 months). However, the densities we infer from the radio observations of PS1-11aop are 3 orders of magnitude lower than predicted for the inter-shell regions ($\gtrsim$10$^{-15}$\,g\,cm$^{-3}$) in \cite{Woosley2017} and \cite{Chen2023}. 

In addition, our host galaxy analysis (Section~\ref{sec:Host Galaxy}) found that PS1-11aop exploded in a super-solar metallicity environment. This is contrary to typical inferences for PISN, which tend to require stars with metallicity less than 1/3 solar, such that the star maintains enough mass late in its evolution to undergo pair instability \citep{Woosley2017}. A super-solar PISN would likely require significant modifications to current mass loss prescriptions.

\subsubsection{Pulsation-Driven Superwinds in RSGs Stars}\label{subsec:RSG-winds}

As described by \cite{Yoon2010}, some RSGs are unstable to radial pulsations that may drive strong ``super-winds'' when the pulsation growth rate is high.  This mechanism can potentially remove almost the entire hydrogen envelope of the star and higher-mass RSGs are expected to be more susceptible. However, while the CSM probed by the radio emission is consistent with expectations for red supergiant stars (for low assumed wind speeds; Figure~\ref{fig:nathan-smith}) this model has several challenges in explaining the dense inner CSM observed in PS1-11aop. 

First, \cite{Yoon2010} find that massive RSGs are first susceptible to pulsation-drive superwinds prior to core-helium depletion. This timescale ($\sim$10$^{5}$ years prior to core-collapse) is incompatible with the inferred timeline of the ejection of the dense shell in PS1-11aop. 
In addition, the energy available from the growth of pulsations can only drive mass-loss rates up to $\sim$0.01 M$_\odot$ year$^{-1}$ \citep{Yoon2010}, which is on the very low end of the range estimated for the dense shell in PS1-11aop. It is therefore unclear that pulsation-driven superwinds could produce a confined dense shell of $\sim$15-30 M$_\odot$.

\subsubsection{Late-stage nuclear burning}\label{subsec:late-stage-nuclear-burning}

Alternatively, the late phase of the final mass ejection in PS1-11aop may indicate a connection to the late nuclear burning stages. It has been proposed that energy from these stages may be transported to the surface of the star either through turbulent mixing \citep{Smith&Arnett2014} or convectively-driven gravity waves \citep{Quataert2012,Shiode2014,Fuller2017}.  For the range of mass ejection velocities considered in Section~\ref{subsec:implication-for-time}, above, the timing of the final mass ejection in PS1-11aop would correspond to either the neon or oxygen-burning phases \citep{Shiode2014}. 

However, in the models of \cite{Fuller2017} for hydrogen-rich SN progenitors, the wave heating mechanism leads primarily to inflation in the stellar envelope as opposed to an ejection of a large amount of material as the energy is deposited in the base of the H-rich envelope. This model may therefore struggle to eject the large amounts of material observed in PS1-11aop, unless (i) this mechanism behaves differently in higher mass stars (\citealt{Fuller2017} modeled a 15\,M$_\odot$ star) or (ii) the inflation of the envelope via wave heating triggers another mechanism, such as late-stage binary interaction.

\subsubsection{Binary Interaction}\label{subsec:binary-interaction}

It is also possible that the dense shell observed in PS1-11aop is the result of non-conservative mass transfer or late-stage interactions in a binary system. For low assumed ejection speeds, the densities inferred for this inner shell are consistent with current prescriptions for binary Roche lobe overflow (Figure~\ref{fig:nathan-smith}; \citealt{Smith2014}). However, the large CSM masses found ($\sim$7--30 M$_\odot$) are more similar to expectations for massive common envelope ejections or stellar mergers \citep{Chevalier_2012}. In this context, if the radio detections are upper limits on the true emission, the density in the sparser outer density region is compatible with expectations for the steady winds from multiple classes of evolved stars with convective envelopes (LBVs, RSGs). Such stars are predicted to undergo unstable or runaway mass transfer if they fill their Roche lobes. If the dense shell observed in PS1-11aop was the result of a common envelope ejection, it would have required the star to undergo late-stage (case-C) mass transfer. This type of interaction is expected to be relatively rare \citep{Podsiadlowski1992,Margutti2017}, but may be more common if coupled with envelope expansion due to wave heating, as described above. Alternatively, binary models have also been presented for some luminous Type IIn SN, where a thermonuclear explosion is preceded by the ejection of a common envelope \citep[e.g.][]{Jerkstrand2020}.

\section{Summary and Conclusions} \label{sec:conclusion}
We have presented a detailed multi-wavelength study of the luminous and slowly evolving Type IIn SN PS1-11aop. Here we summarize our main conclusions. 

\emph{Optical Light Curve:} PS1-11aop reached a peak absolute magnitude of M$\lesssim$-20.5 mag and was still brighter than -18.5 mag more than one-year post-explosion, translating to a peak bolometric luminosity of L$\sim 5 \times 10^{43}$ erg s$^{-1}$ and a total radiated energy E$> 8 \times 10^{50}$ ergs. Its light curve showed a power-law decline that steepened between $\sim$250 and 450 days (depending on the epoch of explosion). This morphology is similar to the luminous Type IIn SN\,2010jl.

\emph{Optical Spectroscopy:} Two optical spectra obtained during the explosion show intermediate/broad H$\alpha$ components that are not present in spectra of the host galaxy. The FWHM of this feature increase from $\sim$3500 km s$^{-1}$ at 90 days to $\sim$7500 km s$^{-1}$ at 357 days, indicating there may be contributions from unshocked ejecta at late times.

\emph{Optical Modelling:} Modelling the optical light curve in the context of interaction between the SN ejecta and CSM material, we infer a dense CSM that can be described by pre-SN mass-loss rates of $\sim$0.05--1 M$_\odot$ yr$^{-1}$ (for an assumed wind speed of 100 km s$^{01}$) and an outer radius between $\sim 0.3-2 \times 10^{16}$ cm. Integrating these density distributions, we find total CSM masses between $\sim$7 and 27 M$_\odot$, depending on both the model and efficiency of energy conversion.  

\emph{Host Galaxy:} PS1-11aop exploded near the center of a star-forming galaxy with super-solar metallicity and no evidence of AGN activity. 

\emph{Radio Emission and Modelling:} Luminous radio emission detected at the location of PS1-11aop on three epochs between $\sim$3.5 and 8 years post-explosion. We observe fading by a factor of $\sim$2 at 9 GHz over this time period and the observed emission is a factor of $\sim$4--7 times higher than expectations for star formation from the host galaxy. However, we take an agnostic approach and model the emission in two cases: (i) assuming it originates from the SN blastwave interacting with CSM and (ii) that it is an upper limit on the true radio emission associated with the explosion.  

In the former case, we infer pre-SN mass loss rates of $\sim$0.02--2$\times$10$^{-3}$ M$_\odot$ year$^{-1}$ at radii of $\sim$0.5-5$\times$10$^{17}$ cm. In the latter case, we primarily identify solutions with similar radii but even lower densities/pre-explosion mass-loss rates. While a small set of higher-density solutions is formally allowed in the case that the observed radio emission is not due to the SN, we disfavor these solutions based on a combination of (i) the observed radio evolution, (ii) the expected radius of the blastwave at late times and (iii) the observed break in the optical light curve which can be interpreted as blastwave exiting the dense CSM region. 

\emph{X-ray Emission and Modelling:} Luminous X-ray emission was also detected from the location of PS1-11aop in one epoch of observations $\sim$4.5 years post-explosion. This emission can be modeled in the context of CSM interaction consistently results from optical and radio. However, we note that (i) the inferred density in the X-ray emitting region is large, implying a significant contribution from the reverse shock and (ii) the shock is expected to be radiative, even at late times, indicating more sophisticated models may be necessary. 

\emph{Overall CSM Structure:} Together, the multiwavelength data for PS1-11aop is consistent with a two-zone CSM structure with a dense inner region and sparer outer region (probed by the optical and radio emission, respectively). This overall CSM structure is similar to what has been inferred for the luminous Type IIn SN\,2010jl and SN\,2017hcc \cite{Fransson_2014,Chandra2022}. If the detected radio emission is due to the SN explosion, then the inferred CSM densities even in the sparer outer region are higher than found for most evolved stars in nearby galaxies. Overall, this density structure may be consistent with a progenitor that underwent a late-stage binary interaction or LBV-like eruption.

\emph{Software:} \texttt{photpipe} \citep{Rest2014}, \texttt{IRAF} \citep{Blondin2012}, \texttt{PyRAF}  \citep{pyraf2012}, \texttt{CASA} \citep{casa2007}, \texttt{pwkit} \citep{peterwilliams2017}, \texttt{CIAO} \citep{Fruscione2006}, \texttt{Source Extractor} \citep{Bertin1996}, \texttt{MOSFiT} \citep{Guillochon_2018}, 
\texttt{Prospector} \citep{prospector2021}, \texttt{Dynesty} \citep{Speagle2020}, \texttt{python-fsps} \citep{Conroy_2009},  Matplotlib \citep{matplotlib}, NumPy \citep{numpy},
Astropy \citep{Astropy2013, Astropy2018, astropy2022}, SAOImage DS9\citep{ds92003}  
\\

\acknowledgements

We thank Wen-fai Fong, Nathan Sanders, and Josh Speagle for useful discussions.
We also acknowledge the PanSTARRS1 collaboration for their time and support during the Medium Deep Survey. The National Radio Astronomy Observatory (NRAO) is a facility of the National Science Foundation operated under a cooperative agreement by Associated Universities, Inc. We acknowledge NRAO for telescope time awarded through the Karl G. Jansky Very Large Array (VLA) interferometer for program numbers 15B-237, 17A-226, and 21A-317. We also thank the staff for their help in the preparation of observations. The authors acknowledge support for this work provided by the National Aeronautics and Space Administration through Chandra Award Number GO7-18045A issued by the Chandra X-ray Center, which is operated by the Smithsonian Astrophysical Observatory for and on behalf of the National Aeronautics Space Administration under contract NAS8-03060. This paper employs a list of Chandra datasets, obtained by the Chandra X-ray Observatory, contained in the Chandra Data Collection (CDC) ~\dataset[doi:10.25574/cdc.298]{https://doi.org/10.25574/cdc.298}.
The Dunlap Institute is funded through an endowment established by the David Dunlap family and the University of Toronto.

M.R.D. acknowledges support from the NSERC through grant RGPIN-2019-06186, the Canada Research Chairs Program, the Canadian Institute for Advanced Research (CIFAR), and the Dunlap Institute at the University of Toronto. T.E. is supported by NASA through the NASA Hubble Fellowship grant HST-HF2-51504.001-A awarded by the Space Telescope Science Institute, which is operated by the Association of Universities for Research in Astronomy, Inc., for NASA, under contract NAS5-26555. 
%%%
R.M.\ acknowledges support from the National Science Foundation under Award No. AST-2221789
and AST-2224255.  The Margutti team at UC Berkeley is partially funded by the Heising-Simons Foundation under grants \# 2018-0911 and \#2021-3248 (PI: Margutti).
The Berger Time-Domain Group at Harvard is supported by NSF and NASA grants. R.L. acknowledges support from the European Research Council (ERC) under the European Unionâs Horizon Europe research and innovation program (grant agreement No. 1010422). D.M. acknowledges NSF support from grants PHY-2209451 and AST-2206532. V.A.V. acknowledges support by the NSF through grant AST-2108676.
 
\clearpage 

\appendix

\section{Detailed Description of X-ray Model Framework}\label{apsec:x-ray model}

For the density structure of the ejecta, we adopt the standard assumption of a broken power law, with a steep outer and flat inner region. Specifically, we take:
\begin{equation}\label{eq:rho_ej}
    \rho_{\rm{ej}} = \rho_o \left(\frac{t}{t_o}\right)^{-3} \left(\frac{v_ot}{r}\right)^n \begin{cases}
    n = n_0 & \text{for $r/t > v_o$} \\
    n = n_1 & \text{for $r/t < v_o$}
    \end{cases}
\end{equation}
where $\rho_o$, $t_o$, and $v_o$ are scaling factors (the ejecta with a velocity of $v_o$ would have a density of $\rho_o$ at time $t_o$). Written in this form, $v_o$ is also the transition velocity between the outer and inner portions of the density profile. We take $n_0 = 12$ and $n_1 = 2$. To calculate the transition velocity and scaling factors we use the formalism from \citet{Chevalier1994} which depends on the ejecta mass and explosion energy of the supernova, as well as $n_0$ and $n_1$.

For the structure of the CSM, we consider two broad cases. First, we consider a single power law structure ($\rho_{\rm{CSM}} \propto r^{-s}$). In this case, our solutions for the hydrodynamics of the shock converge to those described by the self-similar solutions in \cite{Fransson1996} at early times, but can deviate at late times because we explicitly include the transition to the shallow inner ejecta density profile. Second, we allow for a dense inner shell and sparser outer medium, both described by power laws and connected by a steep transition region. Explicitly, we parameterize this as:

\begin{equation}\label{eq:rho_csm}
    \rho_{\rm{CSM}} = \frac{1}{4\pi r_0^2} \begin{cases}
    \left(\frac{\dot{M}_1}{v_w}\right) \left( \frac{r_0}{r}\right)^{s_1} & \text{for $r < r_0$} \\
    \left(\frac{\dot{M}_1}{v_w}\right) \left( \frac{r_0}{r}\right)^{s_2} & \text{for $r_0 < r < r_0+\delta r$} \\
    \left(\frac{\dot{M}_2}{v_w}\right) \left( \frac{r_0}{r}\right)^{s_3} & \text{for $r > r_0+\delta r$}
    \end{cases}
\end{equation}
where $s_1$, $s_2$, and $s_3$ are the power-law indices for each of the three CSM segments (denser inner, steep transition, and sparser outer regions, respectively). The parameters $\dot{M}_1/v_w$ and $\dot{M}_2/v_w$ set the normalizations for the inner dense and sparser outer CSM regions. We note that if either $s_1$ or $s_3$ is set to 2 (appropriate for a wind-like medium) then these parameters describe the implied mass-loss rate of the progenitor star ($\dot{M}$) for a given wind speed ($v_w$); otherwise, they simply describe the equivalent normalization for the density of that CSM segment at the reference radius, $r_0$. As written, $r_0$ also specifies the radius where the dense inner region ends and the sharp transition begins. To specify a given CSM profile, we set values for $s_1$, $s_2$, $s_3$, $r_0$, $\dot{M}_1/v_w$, and $\dot{M}_2/v_w$. $\delta r$ can then be calculated based on the distance required to smoothly transition from the inner to the outer density profiles with the specified value of $s_2$. In general, we set $s_2 =8$ and consider a range of values for $s_1$ and $s_3$ between 0 and 3.

Hydrodynamic evolution of the shock is determined by numerical integration of the equation of motion as detailed in \citet{ChevalierFransson2003}. This equation assumes the thin-shell approximation and can be written as:
\begin{equation}\label{eq:eom}
    \left(M_{\rm{ej,s}} + M_{\rm{CSM,s}}\right) \frac{dv_s}{dt} = 4 \pi R_s \left[ \rho_{\rm{ej}} \left(\frac{R_s}{t} - v_s \right)^2 - \rho_{\rm{CSM}} v_s^2 \right]
\end{equation}
where $R_s$ is the shock radius, $v_s = dR_s/dt$ is the shock velocity, $M_{\rm{CSM,s}}$ is the total CSM mass swept up by the forward shock,  $M_{\rm{ej,s}}$ is the total amount of ejecta swept up by the reverse shock,
and $\rho_{\rm{ej}}$ and $\rho_{\rm{CSM}}$ are the current density of the ejecta and CSM at the shock radius, respectively. The sum $M_{\rm{ej,s}} + M_{\rm{CSM,s}}$ is the total amount of shocked material. These can be calculated by integrating the ejecta and CSM density profiles above for a given radius, $R_s$, and time, $t$.

The numerical integration was performed using a 4th-order Runge-Kutta scheme. This provides $R_s$ and $v_s$ (and thus $\rho_{\rm{ej}}$, $\rho_{\rm{CSM}}$, $M_{\rm{CSM,s}}$ and $M_{\rm{ej,s}}$) as a function of time. The initial conditions required are a shock radius ($R_s$) and shock velocity ($v_s$) at a given time ($t$). Since the inner portion of our CSM profile is described by a power law, we utilize the solutions of \citet{Fransson1996} evaluated at 100 days. 

We calculate the cooling time for the forward and reverse shocks: as
\begin{equation}\label{eq:tcool}
    t_{\rm{cool}} = \frac{3kT}{2(1-\mu/\mu_A)n_i\Lambda}
\end{equation}
where $T$ is the shock temperature, $\mu$ is the mean mass per particle, $\mu_A$ is the mean atomic weight, $n_i$ is the number density of ions, and $\Lambda$ is the cooling function. We take the values for $\mu$, $\mu_A$, and $\Lambda$ provided in \citet{Nymark2006} for a solar composition gas. These results assume ionization equilibrium. We calculate the temperature for each shock in the strong shock limit:
\begin{equation}\label{eq:t_shock}
    T = 2\frac{(\gamma -1)}{(\gamma+1)^2} \frac{\mu m_H}{k} v^2
\end{equation}
where $v$ is the shock velocity. $v = v_s$ for the forward shock and $v = (R_s/t) - v_s$ for the reverse shock. We take $\gamma = 5/3$ and assume solar composition. 

From the cooling times, we can determine if the shocks are adiabatic ($t_{\rm{cool}} > t$) or radiative ($t_{\rm{cool}} < t$). In general, the forward shock is always adiabatic, while the reverse shock can be either adiabatic or radiative at the time of the X-ray observation of PS1-11aop, depending on the assumed CSM density distribution.

We estimate the unabsorbed X-ray luminosity that results from the shock properties found above in two ways. First, we follow \citet{Fransson1996} and calculate the free-free luminosity from each shock as: 
\begin{equation}\label{eq:Lx}
    L_k = 4 \pi \int j_{ff} (\rho/m_H)^2 r^2 dr \approx j_{ff}M_{s,k} \rho_k/m_H^2
\end{equation}
where k denotes either the forward or reverse shock and $M_{s,k}$ and $\rho_k$ are the total mass swept up by and current density at the shock. $j_{ff}$ is the emissivity, which is a function of energy and depends on the electron temperature at the shock as:
\begin{equation}\label{eq:jff}
    j_{ff} = 1.64 \times 10^{-20} \zeta g_{ff} T_e^{-0.5} e^{-E/kT_e} \text{ergs cm$^2$ s$^{-1}$ keV$^{-1}$}
\end{equation}
where $\zeta$ is a composition factor and $g_{ff}$ is the free-free Gaunt factor. We take $\zeta=0.86$ for solar composition. For the Gaunt factor, we adopt the approximations from \citet{Nymark2006} and \citet{Fransson1996} for the forward shock and reverse shock, respectively, and consider energies between 0.3 and 10.0 keV. For the electron temperature, we adopt the shock temperature described above, which assumes the equipartition between electrons and ions. This assumption will be discussed below. 

When the reverse shock is radiative line emission (which we do not treat) can become important. Thus, for comparison, we also follow \citet{Chandra2015} and calculate an approximate radiative luminosity as the kinetic luminosity ($L_{\rm{kinetic}}$) times a radiation efficiency ($\eta$):
\begin{equation}\label{eq:Lrad}
    L_{\rm{rad}} = \eta L_{\rm{kinetic}} = \left(\frac{t}{t+t_{\rm{cool}}} \right) \left( 2 \pi R_s^2 \rho v^3\right)
\end{equation}
where $\rho$ is the pre-shock density and $v$ is the shock velocity (listed above for both the forward and reverse shock). We will discuss the implications of a comparison between our two luminosity calculations for PS1-11aop below. In both cases, we divide the above luminosities by 2 as only half of the luminosity escapes outward \citep{Fransson1996}.

For high CSM densities, absorption due to neutral hydrogen along the line of sight will be non-negligible. This includes contributions from both the CSM exterior to the forward shock (applicable to emission from both the forward and reverse shock) and the cold dense shell (applicable to emission from the reverse shock when it is radiative). For the former, we estimate a column depth by integrating the CSM density profile from $R_s$ out to a radius of 1e18 cm. Since some of the CSM can be ionized this is an upper limit. For the latter, we approximate the column depth of the cold dense shell as:
\begin{equation}\label{eq:Ncds}
    N_{\rm{cds}} = \frac{M_{\rm{ej,s}}}{4 \pi R_s^2 \mu_A m_H}
\end{equation}
following \citet{Nymark2006}. We use these column densities to estimate a final, observable X-ray flux from both the forward and reverse shocks, using the Wisconsin cross sections from \cite{Morrison1983} for energies between 0.3 and 10.0 keV.

\section{Application of X-ray Modelling to PS1-11aop} \label{apsec:X-ray result}

\subsection{Constant Power Law Density Profile}\label{subsec:x-ray-constant}

 \begin{figure*}[ht]
\includegraphics[width=\textwidth]{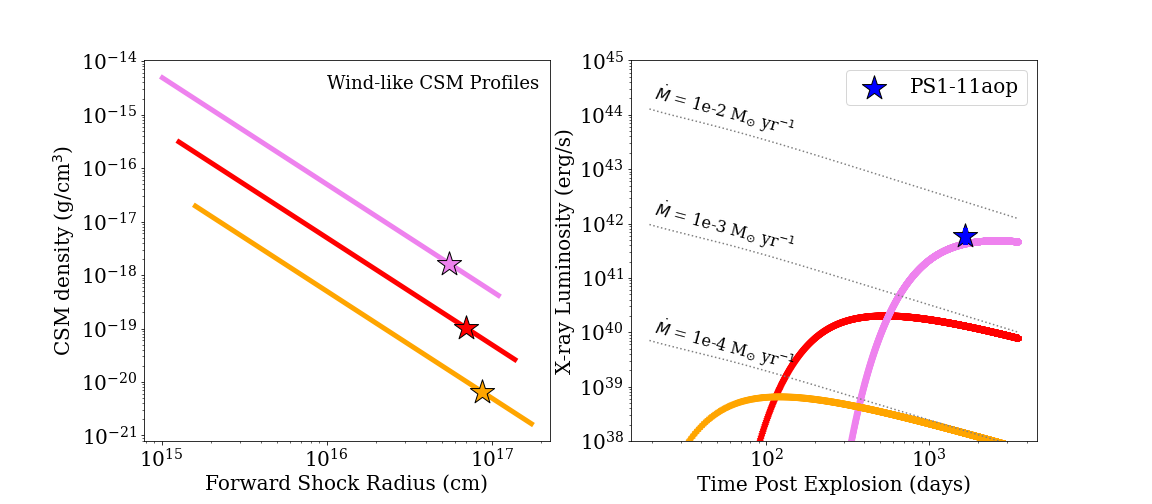}
\caption{Comparison of the X-ray observation of PS1-11aop to predictions for interaction with wind-like CSM density profiles. \emph{Left:} The CSM density profiles (density versus radius) for constant progenitor mass-loss rates of $10^{-4}$, $10^{-3}$, and $10^{-2}$ M$_\odot$ yr$^{-1}$. Stars mark the location of the shock at the time of the X-ray observation of PS1-11aop. \emph{Right:} Predictions for the unabsorbed (dotted lines) and absorbed (solid colored lines) thermal bremsstrahlung X-ray luminosity that results from interaction with the CSM density profiles shown in the left panel. Luminosities are calculated in the 0.3-10 keV range. The X-ray detection of PS1-11aop is shown as a blue star, where the errorbar covers the full range for thermal bremsstrahlung emission with temperatures between 0.8 keV $<$ kT $<$ 80 keV (Section~\ref{subsec:x-ray obs}).} 
\label{fig:xray_wind}
\end{figure*}

We first consider whether the late-time X-ray detection can be explained by interaction due to the CSM structure with a single power-law-like structure. Figure~\ref{fig:xray_wind} illustrates a comparison to predictions for a simple wind-like CSM with $\rho \propto r^{-2}$. The left panel shows three density distributions, which correspond to progenitor mass-loss rates of $10^{-4}$, $10^{-3}$, and $10^{-2}$ M$_\odot$ yr$^{-1}$ (for an assumed wind speed of 100 km $s^{-1}$). Stars show the inferred location of the shock front at the time of the X-ray observation of PS1-11aop. 

The right panel shows the resulting predictions for both the unabsorbed (dotted) and absorbed (solid) X-ray light curves along with the detection of PS1-11aop (blue star). In all cases, the predicted free-free emission from the reverse shock dominates over the forward shock at the epoch when PS1-11aop was observed, with the forward shock contributing $\lesssim10^{40}$ erg s$^{-1}$ (unabsorbed) at 1640 days for the range of mass-loss rates shown. In addition, we find that for mass-loss rates $\gtrsim$10$^{-3}$ M$_\odot$ yr$^{-1}$, the reverse shock is still radiative even at these late times. When absorption due to both the CSM and the ``cold dense shell'' (located between the forward and reverse shock) is taken into account, we find that we require progenitor pre-explosion mass-loss rates of $\sim$10$^{-2}$ M$_\odot$ yr$^{-1}$ to reach the luminosity and timescale of the X-ray detection of PS1-11aop. In this case, the cold dense shell dominates the local absorption with an estimated $N_{\rm{cds}} \approx 1\times 10^{23}$ cm$^{-2}$.

For a wind-like environment with  $\dot{M} = $10$^{-2}$ M$_\odot$ yr$^{-1}$, the forward shock is located at a radius of $\sim 5 \times 10^{16}$cm and has decelerated to a velocity of $\sim 3500$ km s$^{-1}$ at the time of the X-ray observation of PS1-11aop. The temperature of the reverse shock has also declined to kT$\lesssim$1 keV at these late times. While our shock temperature is calculated assuming equipartition between the electrons and the ions, we note that for the high densities at the reverse shock ($\sim$10$^{-16}$ g cm$^{-3}$), the timescale for equipartition would be $\lesssim$100 days \citep{Fransson1996}.

While these results show that a constant power-law density profile could potentially explain the X-ray detection of PS1-11aop, we offer a note of caution. As described above, we find that the reverse shock is radiative at late times. In this case, Equation~\ref{eq:Lx} is not expected to fully capture the behavior of the X-ray emission. If we instead calculate an approximate radiated luminosity based on the kinetic luminosities of both shocks (Equation~\ref{eq:Lrad}) we find slightly lower values of only a few $\times 10^{41}$ erg s$^{-1}$ at the time of the observation of PS1-11aop.

\subsection{Shell-like CSM Profiles}\label{subsec:x-ray-shell}

\begin{figure*}[ht]
\includegraphics[width=\textwidth]{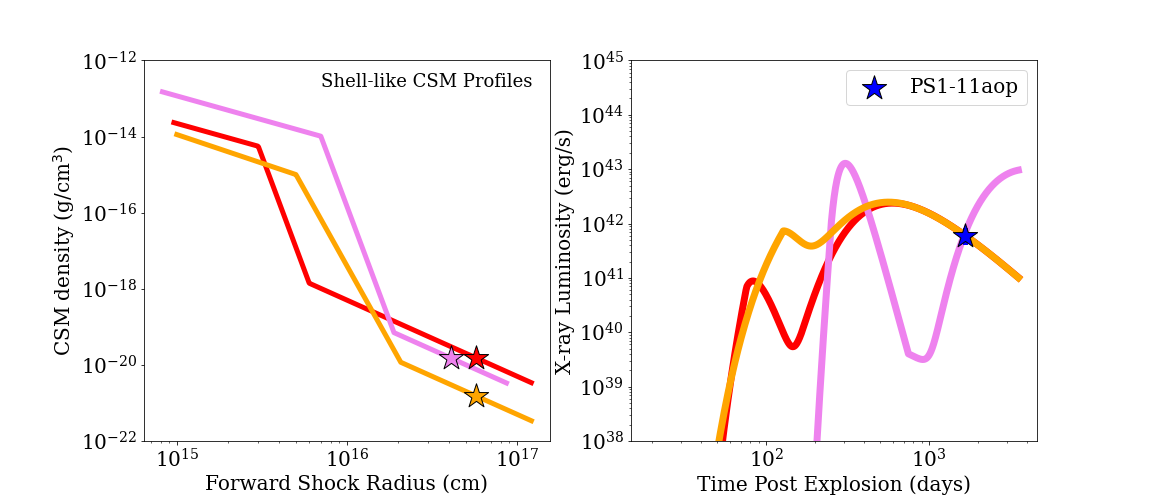}
\caption{Comparison of the X-ray observation of PS1-11aop to predictions for interaction with shell-like CSM density profiles. \emph{Left:} Three fiducial shell-like CSM density profiles. All three have (i) a dense inner region that is consistent with expectations for progenitor mass-loss rates of 0.05--1 M$_\odot$ yr$^{-1}$ and a shallow density profile ($\rho \propto r^{-s}$ where 1.0 $<$ s $<$ 1.5), (ii) a transition radius between 3--7$\times$10$^{15}$ cm, and (iii) a sparser outer region that is consistent with expectations for progenitor mass-loss rates of 10$^{-5}$--10$^{-4}$ M$_\odot$ yr$^{-1}$ and a wind-like density profile ($\rho \propto r^{-2}$). Stars mark the location of the shock at the time of the X-ray observation of PS1-11aop \emph{Right:} Predictions for the absorbed thermal bremsstrahlung X-ray luminosity that results from interaction with the CSM density profiles shown in the left panel. Luminosities are calculated in the 0.3-10 keV range. The X-ray detection of PS1-11aop is shown as a blue star. At the time of the observation of PS1-11aop, the predicted X-ray emission is dominated by the reverse shock. Earlier predicted emission peaks between $\sim$50-200 days post-explosion are dominated by the forward shock.} 
\label{fig:xray_shell}
\end{figure*}

Motivated by the results from the optical and radio, we also consider simple shell-like density structures. Given the large number of free parameters, we do not attempt a full exploration of CSM profiles that could potentially explain our observed X-ray luminosity. Rather, we simply consider whether the X-ray detection is compatible with the general density structure implied by our baseline radio and optical modeling. To investigate this, we generate a series of CSM profiles that have (i) an inner region with a shallow (s $<$ 2 for $\rho \propto r^{-s}$) density profile, (ii) a transition radius between 3$-$10$\times$10$^{15}$ cm, and (iii) an outer wind-like density profile consistent with a progenitor mass-loss rate between 10$^{-5}$ and 10$^{-4}$ M$_\odot$ yr$^{-1}$. From these, we search for models that reproduce the observed X-ray flux. 

In Figure~\ref{fig:xray_shell}, we plot three representative models that fulfill these criteria which span a range of densities for the inner shell. These models have shock radii, shock velocities, and CSM densities at the time of the PS1-11aop observation of 4$-$6$\times 10^{16}$ cm, 2800$-$4000 km s$^{-1}$, and $\sim$0.1--2$\times 10^{-20}$ g cm$^{-3}$, respectively---consistent with inferences from the radio observations described above. However, due to the shell at smaller radii, the density at the reverse shock at these late times ($\sim$10$^{-17}-$10$^{-15}$ g cm$^{-3}$) are approximately 2$-$4 orders of magnitude higher than that found for a shock expanding into a medium described by a single power-law with a mass-loss rate of 10$^{-4}$ M$_\odot$ yr$^{-1}$. 

Other general properties of these solutions are similar to what was found for the single power-law profile in \S \ref{subsec:x-ray-constant}. Namely, we find that at the time of the X-ray observation: (i) the predicted free-free luminosity is dominated by the reverse shock\footnote{Although some models show a 'double peak' feature where the first maximum is dominated by emission from the forward shock.}, (ii) the temperature at the reverse shock has cooled to $\lesssim$1 keV, (iii) the reverse shock is still radiative, and (iv) estimates for extra absorption due to the cold dense shell are on the order of $N_{\rm{cds}} \approx 10^{23}$ cm$^{-2}$. 

As above, we caution that the fact that the reverse shock is radiative means that our simple model likely does not capture the full behavior of the predicted X-ray emission. In addition, we find that the velocity of the reverse shock ($v_{rs}$), and hence its inferred temperature and kinetic luminosity (see Equations~\ref{eq:t_shock} and \ref{eq:Lrad}), are quite sensitive to the structure of the inner CSM shell. This is because $v_{rs} = (R_s/t) - v_s$, where  $R_s$ is the shock radius, $t$ is the time since the explosion, and $v_s = dR_s/dt$ is the forward shock velocity. Once the forward shock exits the dense inner shell, $v_s$ decelerates at a much slower pace, causing $v_{rs}$ to decrease. For some shell-like density profiles, this decrease in $v_{rs}$ is dramatic enough that no X-rays are predicted at late times. In general, shell-like profiles with some combination of (i) a lower inner density, (ii) a shallower transition region, or (iii) a larger transition radius, maintain the highest kinetic luminosities.

\subsection{Constraints on Density from Emission Measure} \label{x-ray-EM}

In section \S \ref{apsec:X-ray result}, we found that the predicted free-free emission at the time of the observation of PS1-11aop was dominated by the reverse shock using the framework of \citet{Fransson1996}. However, given that (i) the absorption due to the cold dense shell is uncertain and (ii) the total reservoir of kinetic luminosity available in the forward shock tends to exceed that of the reverse shock due to its higher velocity, here we perform an independent estimate of the density in the X-ray emitting region. Assuming that the X-rays are due to thermal bremsstrahlung, the normalization of the spectrum is proportional to the emission measure: $EM = \int n_e n_I dV$ where $n_e$ and $n_I$ are the number density of electrons and ions, respectively. With an assumption of the volume of the X-ray emitting region, it is therefore possible to estimate the density.

We take $\mu = 0.61$ and $\mu_I = 1.29$ based on \citet{Nymark2006} for a solar composition gas and assume a volume for the X-ray emitting shell of the form $V_{\rm{shell}} = 4 \pi R_s^2 \Delta R f$ where $R_s$ is the radius of the X-ray emitting shell, $\Delta R$ is the width of the shell, and $f$ is a volume filling factor. We initially consider radii in the range of (3$-$10)$\times$10$^{16}$ cm based on radio observations of PS1-11aop at similar epochs, $\Delta R   = 0.08 R_s$ based on the models of \citet{Chevalier1989} for the width of the shocked region, and $f=$1. 

Using the emission measures listed in Table~\ref{tab:xrayflux} (which assume no additional absorption beyond that of the Milky Way), this yields estimated densities of $\sim$10$^{-17}-10^{-16}$ g cm$^{-3}$, with larger values associated with cooler bremsstrahlung emission. If we adopt an additional local absorption of $N_H = 10^{22}$ or $10^{23}$ cm$^{-2}$, the emission measure increases by a factor of 3--5, which leads to a factor of 2--3 increase in the estimated density. In addition, for only partial filling of the volume, these results would increase as $f^{-1/2}$. Overall, these densities are similar to those found at the reverse shock in the hydrodynamic modeling above, but over an order of magnitude above those found in the CSM \emph{even} for the case of a constant 10$^{-2}$ M$_{\odot}$ yr$^{-1}$ density distribution. 

%\bibliographystyle{aasjournal}
%\bibliography{PS11aop-bibliography}{}

\begin{thebibliography}{}
\expandafter\ifx\csname natexlab\endcsname\relax\def\natexlab#1{#1}\fi
\providecommand{\url}[1]{\href{#1}{#1}}
\providecommand{\dodoi}[1]{doi:~\href{http://doi.org/#1}{\nolinkurl{#1}}}
\providecommand{\doeprint}[1]{\href{http://ascl.net/#1}{\nolinkurl{http://ascl.net/#1}}}
\providecommand{\doarXiv}[1]{\href{https://arxiv.org/abs/#1}{\nolinkurl{https://arxiv.org/abs/#1}}}

\bibitem[{{Ahn} {et~al.}(2012){Ahn}, {Alexandroff}, {Allende Prieto},
  {Anderson}, {Anderton}, {Andrews}, {Aubourg}, {Bailey}, {Balbinot}, {Barnes},
  {Bautista}, {Beers}, {Beifiori}, {Berlind}, {Bhardwaj}, {Bizyaev}, {Blake},
  {Blanton}, {Blomqvist}, {Bochanski}, {Bolton}, {Borde}, {Bovy}, {Brandt},
  {Brinkmann}, {Brown}, {Brownstein}, {Bundy}, {Busca}, {Carithers}, {Carnero},
  {Carr}, {Casetti-Dinescu}, {Chen}, {Chiappini}, {Comparat}, {Connolly},
  {Crepp}, {Cristiani}, {Croft}, {Cuesta}, {da Costa}, {Davenport}, {Dawson},
  {de Putter}, {De Lee}, {Delubac}, {Dhital}, {Ealet}, {Ebelke}, {Edmondson},
  {Eisenstein}, {Escoffier}, {Esposito}, {Evans}, {Fan}, {Femen{\'\i}a
  Castell{\'a}}, {Fern{\'a}ndez Alvar}, {Ferreira}, {Filiz Ak}, {Finley},
  {Fleming}, {Font-Ribera}, {Frinchaboy}, {Garc{\'\i}a-Hern{\'a}ndez},
  {Garc{\'\i}a P{\'e}rez}, {Ge}, {G{\'e}nova-Santos}, {Gillespie}, {Girardi},
  {Gonz{\'a}lez Hern{\'a}ndez}, {Grebel}, {Gunn}, {Guo}, {Haggard}, {Hamilton},
  {Harris}, {Hawley}, {Hearty}, {Ho}, {Hogg}, {Holtzman}, {Honscheid},
  {Huehnerhoff}, {Ivans}, {Ivezi{\'c}}, {Jacobson}, {Jiang}, {Johansson},
  {Johnson}, {Kauffmann}, {Kirkby}, {Kirkpatrick}, {Klaene}, {Knapp}, {Kneib},
  {Le Goff}, {Leauthaud}, {Lee}, {Lee}, {Long}, {Loomis}, {Lucatello},
  {Lundgren}, {Lupton}, {Ma}, {Ma}, {MacDonald}, {Mack}, {Mahadevan}, {Maia},
  {Majewski}, {Makler}, {Malanushenko}, {Malanushenko}, {Manchado},
  {Mandelbaum}, {Manera}, {Maraston}, {Margala}, {Martell}, {McBride},
  {McGreer}, {McMahon}, {M{\'e}nard}, {Meszaros}, {Miralda-Escud{\'e}},
  {Montero-Dorta}, {Montesano}, {Morrison}, {Muna}, {Munn}, {Murayama},
  {Myers}, {Neto}, {Nguyen}, {Nichol}, {Nidever}, {Noterdaeme}, {Nuza}, {Ogand
  o}, {Olmstead}, {Oravetz}, {Owen}, {Padmanabhan}, {Palanque-Delabrouille},
  {Pan}, {Parejko}, {Parihar}, {P{\^a}ris}, {Pattarakijwanich}, {Pepper},
  {Percival}, {P{\'e}rez-Fournon}, {P{\'e}rez-R{\`a}fols}, {Petitjean},
  {Pforr}, {Pieri}, {Pinsonneault}, {Porto de Mello}, {Prada}, {Price-Whelan},
  {Raddick}, {Rebolo}, {Rich}, {Richards}, {Robin}, {Rocha-Pinto}, {Rockosi},
  {Roe}, {Ross}, {Ross}, {Rossi}, {Rubi{\~n}o-Martin}, {Samushia}, {Sanchez
  Almeida}, {S{\'a}nchez}, {Santiago}, {Sayres}, {Schlegel}, {Schlesinger},
  {Schmidt}, {Schneider}, {Schultheis}, {Schwope}, {Sc{\'o}ccola}, {Seljak},
  {Sheldon}, {Shen}, {Shu}, {Simmerer}, {Simmons}, {Skibba}, {Skrutskie},
  {Slosar}, {Sobreira}, {Sobeck}, {Stassun}, {Steele}, {Steinmetz}, {Strauss},
  {Streblyanska}, {Suzuki}, {Swanson}, {Tal}, {Thakar}, {Thomas}, {Thompson},
  {Tinker}, {Tojeiro}, {Tremonti}, {Vargas Maga{\~n}a}, {Verde}, {Viel},
  {Vikas}, {Vogt}, {Wake}, {Wang}, {Weaver}, {Weinberg}, {Weiner}, {West},
  {White}, {Wilson}, {Wisniewski}, {Wood-Vasey}, {Yanny}, {Y{\`e}che}, {York},
  {Zamora}, {Zasowski}, {Zehavi}, {Zhao}, {Zheng}, {Zhu}, \& {Zinn}}]{Ahn2012}
{Ahn}, C.~P., {Alexandroff}, R., {Allende Prieto}, C., {et~al.} 2012, \apjs,
  203, 21, \dodoi{10.1088/0067-0049/203/2/21}

\bibitem[{{Aihara} {et~al.}(2011){Aihara}, {Allende Prieto}, {An}, {Anderson},
  {Aubourg}, {Balbinot}, {Beers}, {Berlind}, {Bickerton}, {Bizyaev}, {Blanton},
  {Bochanski}, {Bolton}, {Bovy}, {Brandt}, {Brinkmann}, {Brown}, {Brownstein},
  {Busca}, {Campbell}, {Carr}, {Chen}, {Chiappini}, {Comparat}, {Connolly},
  {Cortes}, {Croft}, {Cuesta}, {da Costa}, {Davenport}, {Dawson}, {Dhital},
  {Ealet}, {Ebelke}, {Edmondson}, {Eisenstein}, {Escoffier}, {Esposito},
  {Evans}, {Fan}, {Femen{\'\i}a Castell{\'a}}, {Font-Ribera}, {Frinchaboy},
  {Ge}, {Gillespie}, {Gilmore}, {Gonz{\'a}lez Hern{\'a}ndez}, {Gott}, {Gould},
  {Grebel}, {Gunn}, {Hamilton}, {Harding}, {Harris}, {Hawley}, {Hearty}, {Ho},
  {Hogg}, {Holtzman}, {Honscheid}, {Inada}, {Ivans}, {Jiang}, {Johnson},
  {Jordan}, {Jordan}, {Kazin}, {Kirkby}, {Klaene}, {Knapp}, {Kneib},
  {Kochanek}, {Koesterke}, {Kollmeier}, {Kron}, {Lampeitl}, {Lang}, {Le Goff},
  {Lee}, {Lin}, {Long}, {Loomis}, {Lucatello}, {Lundgren}, {Lupton}, {Ma},
  {MacDonald}, {Mahadevan}, {Maia}, {Makler}, {Malanushenko}, {Malanushenko},
  {Mandelbaum}, {Maraston}, {Margala}, {Masters}, {McBride}, {McGehee},
  {McGreer}, {M{\'e}nard}, {Miralda-Escud{\'e}}, {Morrison}, {Mullally},
  {Muna}, {Munn}, {Murayama}, {Myers}, {Naugle}, {Neto}, {Nguyen}, {Nichol},
  {O'Connell}, {Ogando}, {Olmstead}, {Oravetz}, {Padmanabhan},
  {Palanque-Delabrouille}, {Pan}, {Pandey}, {P{\^a}ris}, {Percival},
  {Petitjean}, {Pfaffenberger}, {Pforr}, {Phleps}, {Pichon}, {Pieri}, {Prada},
  {Price-Whelan}, {Raddick}, {Ramos}, {Reyl{\'e}}, {Rich}, {Richards}, {Rix},
  {Robin}, {Rocha-Pinto}, {Rockosi}, {Roe}, {Rollinde}, {Ross}, {Ross},
  {Rossetto}, {S{\'a}nchez}, {Sayres}, {Schlegel}, {Schlesinger}, {Schmidt},
  {Schneider}, {Sheldon}, {Shu}, {Simmerer}, {Simmons}, {Sivarani}, {Snedden},
  {Sobeck}, {Steinmetz}, {Strauss}, {Szalay}, {Tanaka}, {Thakar}, {Thomas},
  {Tinker}, {Tofflemire}, {Tojeiro}, {Tremonti}, {Vandenberg}, {Vargas
  Maga{\~n}a}, {Verde}, {Vogt}, {Wake}, {Wang}, {Weaver}, {Weinberg}, {White},
  {White}, {Yanny}, {Yasuda}, {Yeche}, \& {Zehavi}}]{Aihara2011}
{Aihara}, H., {Allende Prieto}, C., {An}, D., {et~al.} 2011, \apjs, 193, 29,
  \dodoi{10.1088/0067-0049/193/2/29}

\bibitem[{{Alexander} {et~al.}(2015){Alexander}, {Soderberg}, \&
  {Chomiuk}}]{Alexander2015}
{Alexander}, K.~D., {Soderberg}, A.~M., \& {Chomiuk}, L.~B. 2015, \apj, 806,
  106, \dodoi{10.1088/0004-637X/806/1/106}

\bibitem[{{Allington-Smith} {et~al.}(1994){Allington-Smith}, {Breare}, {Ellis},
  {Gellatly}, {Glazebrook}, {Jorden}, {Maclean}, {Oates}, {Shaw}, {Tanvir},
  {Taylor}, {Taylor}, {Webster}, \& {Worswick}}]{Allington-Smith1994}
{Allington-Smith}, J., {Breare}, M., {Ellis}, R., {et~al.} 1994, \pasp, 106,
  983, \dodoi{10.1086/133471}

\bibitem[{{Arcavi} {et~al.}(2014){Arcavi}, {Gal-Yam}, {Sullivan}, {Pan},
  {Cenko}, {Horesh}, {Ofek}, {De Cia}, {Yan}, {Yang}, {Howell}, {Tal},
  {Kulkarni}, {Tendulkar}, {Tang}, {Xu}, {Sternberg}, {Cohen}, {Bloom},
  {Nugent}, {Kasliwal}, {Perley}, {Quimby}, {Miller}, {Theissen}, \&
  {Laher}}]{Arcavi2014}
{Arcavi}, I., {Gal-Yam}, A., {Sullivan}, M., {et~al.} 2014, \apj, 793, 38,
  \dodoi{10.1088/0004-637X/793/1/38}

\bibitem[{{Arnett}(1982)}]{Arnett1982}
{Arnett}, W.~D. 1982, \apj, 253, 785, \dodoi{10.1086/159681}

\bibitem[{{Arnett} \& {Meakin}(2011)}]{Arnett2011}
{Arnett}, W.~D., \& {Meakin}, C. 2011, \apj, 741, 33,
  \dodoi{10.1088/0004-637X/741/1/33}

\bibitem[{{Astropy Collaboration} {et~al.}(2013){Astropy Collaboration},
  {Robitaille}, {Tollerud}, {Greenfield}, {Droettboom}, {Bray}, {Aldcroft},
  {Davis}, {Ginsburg}, {Price-Whelan}, {Kerzendorf}, {Conley}, {Crighton},
  {Barbary}, {Muna}, {Ferguson}, {Grollier}, {Parikh}, {Nair}, {Unther},
  {Deil}, {Woillez}, {Conseil}, {Kramer}, {Turner}, {Singer}, {Fox}, {Weaver},
  {Zabalza}, {Edwards}, {Azalee Bostroem}, {Burke}, {Casey}, {Crawford},
  {Dencheva}, {Ely}, {Jenness}, {Labrie}, {Lim}, {Pierfederici}, {Pontzen},
  {Ptak}, {Refsdal}, {Servillat}, \& {Streicher}}]{Astropy2013}
{Astropy Collaboration}, {Robitaille}, T.~P., {Tollerud}, E.~J., {et~al.} 2013,
  \aap, 558, A33, \dodoi{10.1051/0004-6361/201322068}

\bibitem[{{Astropy Collaboration} {et~al.}(2018){Astropy Collaboration},
  {Price-Whelan}, {Sip{\H{o}}cz}, {G{\"u}nther}, {Lim}, {Crawford}, {Conseil},
  {Shupe}, {Craig}, {Dencheva}, {Ginsburg}, {VanderPlas}, {Bradley},
  {P{\'e}rez-Su{\'a}rez}, {de Val-Borro}, {Aldcroft}, {Cruz}, {Robitaille},
  {Tollerud}, {Ardelean}, {Babej}, {Bach}, {Bachetti}, {Bakanov}, {Bamford},
  {Barentsen}, {Barmby}, {Baumbach}, {Berry}, {Biscani}, {Boquien}, {Bostroem},
  {Bouma}, {Brammer}, {Bray}, {Breytenbach}, {Buddelmeijer}, {Burke},
  {Calderone}, {Cano Rodr{\'\i}guez}, {Cara}, {Cardoso}, {Cheedella}, {Copin},
  {Corrales}, {Crichton}, {D'Avella}, {Deil}, {Depagne}, {Dietrich}, {Donath},
  {Droettboom}, {Earl}, {Erben}, {Fabbro}, {Ferreira}, {Finethy}, {Fox},
  {Garrison}, {Gibbons}, {Goldstein}, {Gommers}, {Greco}, {Greenfield},
  {Groener}, {Grollier}, {Hagen}, {Hirst}, {Homeier}, {Horton}, {Hosseinzadeh},
  {Hu}, {Hunkeler}, {Ivezi{\'c}}, {Jain}, {Jenness}, {Kanarek}, {Kendrew},
  {Kern}, {Kerzendorf}, {Khvalko}, {King}, {Kirkby}, {Kulkarni}, {Kumar},
  {Lee}, {Lenz}, {Littlefair}, {Ma}, {Macleod}, {Mastropietro}, {McCully},
  {Montagnac}, {Morris}, {Mueller}, {Mumford}, {Muna}, {Murphy}, {Nelson},
  {Nguyen}, {Ninan}, {N{\"o}the}, {Ogaz}, {Oh}, {Parejko}, {Parley}, {Pascual},
  {Patil}, {Patil}, {Plunkett}, {Prochaska}, {Rastogi}, {Reddy Janga},
  {Sabater}, {Sakurikar}, {Seifert}, {Sherbert}, {Sherwood-Taylor}, {Shih},
  {Sick}, {Silbiger}, {Singanamalla}, {Singer}, {Sladen}, {Sooley},
  {Sornarajah}, {Streicher}, {Teuben}, {Thomas}, {Tremblay}, {Turner},
  {Terr{\'o}n}, {van Kerkwijk}, {de la Vega}, {Watkins}, {Weaver}, {Whitmore},
  {Woillez}, {Zabalza}, \& {Astropy Contributors}}]{Astropy2018}
{Astropy Collaboration}, {Price-Whelan}, A.~M., {Sip{\H{o}}cz}, B.~M., {et~al.}
  2018, \aj, 156, 123, \dodoi{10.3847/1538-3881/aabc4f}

\bibitem[{{Astropy Collaboration} {et~al.}(2022){Astropy Collaboration},
  {Price-Whelan}, {Lim}, {Earl}, {Starkman}, {Bradley}, {Shupe}, {Patil},
  {Corrales}, {Brasseur}, {N{\"o}the}, {Donath}, {Tollerud}, {Morris},
  {Ginsburg}, {Vaher}, {Weaver}, {Tocknell}, {Jamieson}, {van Kerkwijk},
  {Robitaille}, {Merry}, {Bachetti}, {G{\"u}nther}, {Aldcroft},
  {Alvarado-Montes}, {Archibald}, {B{\'o}di}, {Bapat}, {Barentsen},
  {Baz{\'a}n}, {Biswas}, {Boquien}, {Burke}, {Cara}, {Cara}, {Conroy},
  {Conseil}, {Craig}, {Cross}, {Cruz}, {D'Eugenio}, {Dencheva}, {Devillepoix},
  {Dietrich}, {Eigenbrot}, {Erben}, {Ferreira}, {Foreman-Mackey}, {Fox},
  {Freij}, {Garg}, {Geda}, {Glattly}, {Gondhalekar}, {Gordon}, {Grant},
  {Greenfield}, {Groener}, {Guest}, {Gurovich}, {Handberg}, {Hart},
  {Hatfield-Dodds}, {Homeier}, {Hosseinzadeh}, {Jenness}, {Jones}, {Joseph},
  {Kalmbach}, {Karamehmetoglu}, {Ka{\l}uszy{\'n}ski}, {Kelley}, {Kern},
  {Kerzendorf}, {Koch}, {Kulumani}, {Lee}, {Ly}, {Ma}, {MacBride}, {Maljaars},
  {Muna}, {Murphy}, {Norman}, {O'Steen}, {Oman}, {Pacifici}, {Pascual},
  {Pascual-Granado}, {Patil}, {Perren}, {Pickering}, {Rastogi}, {Roulston},
  {Ryan}, {Rykoff}, {Sabater}, {Sakurikar}, {Salgado}, {Sanghi}, {Saunders},
  {Savchenko}, {Schwardt}, {Seifert-Eckert}, {Shih}, {Jain}, {Shukla}, {Sick},
  {Simpson}, {Singanamalla}, {Singer}, {Singhal}, {Sinha}, {Sip{\H{o}}cz},
  {Spitler}, {Stansby}, {Streicher}, {{\v{S}}umak}, {Swinbank}, {Taranu},
  {Tewary}, {Tremblay}, {de Val-Borro}, {Van Kooten}, {Vasovi{\'c}}, {Verma},
  {de Miranda Cardoso}, {Williams}, {Wilson}, {Winkel}, {Wood-Vasey}, {Xue},
  {Yoachim}, {Zhang}, {Zonca}, \& {Astropy Project Contributors}}]{astropy2022}
{Astropy Collaboration}, {Price-Whelan}, A.~M., {Lim}, P.~L., {et~al.} 2022,
  \apj, 935, 167, \dodoi{10.3847/1538-4357/ac7c74}

\bibitem[{{Baldwin} {et~al.}(1981){Baldwin}, {Phillips}, \&
  {Terlevich}}]{BPT1-1981}
{Baldwin}, J.~A., {Phillips}, M.~M., \& {Terlevich}, R. 1981, \pasp, 93, 5,
  \dodoi{10.1086/130766}

\bibitem[{{Bauer} {et~al.}(2008){Bauer}, {Dwarkadas}, {Brandt}, {Immler},
  {Smartt}, {Bartel}, \& {Bietenholz}}]{Bauer2008ApJ}
{Bauer}, F.~E., {Dwarkadas}, V.~V., {Brandt}, W.~N., {et~al.} 2008, \apj, 688,
  1210, \dodoi{10.1086/589761}

\bibitem[{Benetti {et~al.}(2013)Benetti, Nicholl, Cappellaro, Pastorello,
  Smartt, Elias-Rosa, Drake, Tomasella, Turatto, Harutyunyan, Taubenberger,
  Hachinger, Morales-Garoffolo, Chen, Djorgovski, Fraser, Gal-Yam, Inserra,
  Mazzali, \& Young}]{Benetti2013}
Benetti, S., Nicholl, M., Cappellaro, E., {et~al.} 2013, Monthly Notices of the
  Royal Astronomical Society, 441, \dodoi{10.1093/mnras/stu538}

\bibitem[{{Bertin} \& {Arnouts}(1996)}]{Bertin1996}
{Bertin}, E., \& {Arnouts}, S. 1996, \aaps, 117, 393,
  \dodoi{10.1051/aas:1996164}

\bibitem[{Blondin {et~al.}(2012)Blondin, Matheson, Kirshner, Mandel, Berlind,
  Calkins, Challis, Garnavich, Jha, Modjaz, Riess, \& Schmidt}]{Blondin2012}
Blondin, S., Matheson, T., Kirshner, R.~P., {et~al.} 2012, AJ, 143, 126,
  \dodoi{10.1088/0004-6256/143/5/126}

\bibitem[{{Brennan} {et~al.}(2023){Brennan}, {Schulze}, {Lunnan}, {Sollerman},
  {Yan}, {Fransson}, {Irani}, {Melinder}, {Chen}, {De}, {Fremling}, {Kim},
  {Perley}, {Pessi}, {Drake}, {Graham}, {Laher}, {Masci}, {Purdum}, \&
  {Rodriguez}}]{Brennan2023}
{Brennan}, S.~J., {Schulze}, S., {Lunnan}, R., {et~al.} 2023, arXiv e-prints,
  arXiv:2312.13280, \dodoi{10.48550/arXiv.2312.13280}

\bibitem[{Brooks \& Gelman(1998)}]{Gelman1998}
Brooks, S., \& Gelman, A. 1998, Journal of Computational and Graphical
  Statistics, 7, 434

\bibitem[{{Cardelli} {et~al.}(1989){Cardelli}, {Clayton}, \&
  {Mathis}}]{Cardelli1989}
{Cardelli}, J.~A., {Clayton}, G.~C., \& {Mathis}, J.~S. 1989, \apj, 345, 245,
  \dodoi{10.1086/167900}

\bibitem[{Carnall {et~al.}(2019)Carnall, Leja, Johnson, McLure, Dunlop, \&
  Conroy}]{Carnall_2019}
Carnall, A.~C., Leja, J., Johnson, B.~D., {et~al.} 2019, The Astrophysical
  Journal, 873, 44, \dodoi{10.3847/1538-4357/ab04a2}

\bibitem[{{Chabrier}(2003)}]{Chabrier2003}
{Chabrier}, G. 2003, \pasp, 115, 763, \dodoi{10.1086/376392}

\bibitem[{Chandra {et~al.}(2012)Chandra, Chevalier, Chugai, Fransson, Irwin,
  Soderberg, Chakraborti, \& Immler}]{Chandra_2012}
Chandra, P., Chevalier, R.~A., Chugai, N., {et~al.} 2012, The Astrophysical
  Journal, 755, 110, \dodoi{10.1088/0004-637x/755/2/110}

\bibitem[{{Chandra} {et~al.}(2015){Chandra}, {Chevalier}, {Chugai}, {Fransson},
  \& {Soderberg}}]{Chandra2015}
{Chandra}, P., {Chevalier}, R.~A., {Chugai}, N., {Fransson}, C., \&
  {Soderberg}, A.~M. 2015, \apj, 810, 32, \dodoi{10.1088/0004-637X/810/1/32}

\bibitem[{{Chandra} {et~al.}(2012){Chandra}, {Chevalier}, {Irwin}, {Chugai},
  {Fransson}, \& {Soderberg}}]{Chandra2012}
{Chandra}, P., {Chevalier}, R.~A., {Irwin}, C.~M., {et~al.} 2012, \apjl, 750,
  L2, \dodoi{10.1088/2041-8205/750/1/L2}

\bibitem[{{Chandra} {et~al.}(2022){Chandra}, {Chevalier}, {James}, \&
  {Fox}}]{Chandra2022}
{Chandra}, P., {Chevalier}, R.~A., {James}, N. J.~H., \& {Fox}, O.~D. 2022,
  \mnras, 517, 4151, \dodoi{10.1093/mnras/stac2915}

\bibitem[{{Chandra} \& {Soderberg}(2007)}]{PonaamandAlicia2007}
{Chandra}, P., \& {Soderberg}, A. 2007, The Astronomer's Telegram, 1182, 1

\bibitem[{{Chandra} {et~al.}(2009){Chandra}, {Stockdale}, {Chevalier}, {Van
  Dyk}, {Ray}, {Kelley}, {Weiler}, {Panagia}, \& {Sramek}}]{Chandra2009}
{Chandra}, P., {Stockdale}, C.~J., {Chevalier}, R.~A., {et~al.} 2009, \apj,
  690, 1839, \dodoi{10.1088/0004-637X/690/2/1839}

\bibitem[{{Charalampopoulos} {et~al.}(2022){Charalampopoulos}, {Leloudas},
  {Malesani}, {Wevers}, {Arcavi}, {Nicholl}, {Pursiainen}, {Lawrence},
  {Anderson}, {Benetti}, {Cannizzaro}, {Chen}, {Galbany}, {Gromadzki},
  {Guti{\'e}rrez}, {Inserra}, {Jonker}, {M{\"u}ller-Bravo}, {Onori}, {Short},
  {Sollerman}, \& {Young}}]{Charalampopoulos2022}
{Charalampopoulos}, P., {Leloudas}, G., {Malesani}, D.~B., {et~al.} 2022, \aap,
  659, A34, \dodoi{10.1051/0004-6361/202142122}

\bibitem[{{Chatzopoulos} \& {Wheeler}(2012)}]{Chatz2012}
{Chatzopoulos}, E., \& {Wheeler}, J.~C. 2012, \apj, 748, 42,
  \dodoi{10.1088/0004-637X/748/1/42}

\bibitem[{{Chatzopoulos} {et~al.}(2013){Chatzopoulos}, {Wheeler}, {Vinko},
  {Horvath}, \& {Nagy}}]{Chatzopoulos2013}
{Chatzopoulos}, E., {Wheeler}, J.~C., {Vinko}, J., {Horvath}, Z.~L., \& {Nagy},
  A. 2013, \apj, 773, 76, \dodoi{10.1088/0004-637X/773/1/76}

\bibitem[{{Chen} {et~al.}(2023){Chen}, {Whalen}, {Woosley}, \&
  {Zhang}}]{Chen2023}
{Chen}, K.-J., {Whalen}, D.~J., {Woosley}, S.~E., \& {Zhang}, W. 2023, \apj,
  955, 39, \dodoi{10.3847/1538-4357/ace968}

\bibitem[{{Chevalier}(1982)}]{Chevalier1982}
{Chevalier}, R.~A. 1982, \apj, 258, 790, \dodoi{10.1086/160126}

\bibitem[{{Chevalier}(1987)}]{Chevaliar1987Natur}
---. 1987, \nat, 329, 611, \dodoi{10.1038/329611a0}

\bibitem[{{Chevalier}(1998)}]{Chevalier1998ApJ}
---. 1998, \apj, 499, 810, \dodoi{10.1086/305676}

\bibitem[{Chevalier(2012)}]{Chevalier_2012}
Chevalier, R.~A. 2012, The Astrophysical Journal, 752, L2,
  \dodoi{10.1088/2041-8205/752/1/l2}

\bibitem[{{Chevalier} \& {Fransson}(1994)}]{Chevalier1994}
{Chevalier}, R.~A., \& {Fransson}, C. 1994, \apj, 420, 268,
  \dodoi{10.1086/173557}

\bibitem[{{Chevalier} \& {Fransson}(2003)}]{ChevalierFransson2003}
---. 2003, in Supernovae and Gamma-Ray Bursters, ed. K.~{Weiler}, Vol. 598,
  171--194, \dodoi{10.1007/3-540-45863-8_10}

\bibitem[{Chevalier \& Fransson(2006)}]{Chevalier_2006}
Chevalier, R.~A., \& Fransson, C. 2006, The Astrophysical Journal, 651,
  381–391, \dodoi{10.1086/507606}

\bibitem[{{Chevalier} \& {Fransson}(2017)}]{ChevalierFransson2017}
{Chevalier}, R.~A., \& {Fransson}, C. 2017, in Handbook of Supernovae, ed.
  A.~W. {Alsabti} \& P.~{Murdin}, 875, \dodoi{10.1007/978-3-319-21846-5_34}

\bibitem[{{Chevalier} \& {Liang}(1989)}]{Chevalier1989}
{Chevalier}, R.~A., \& {Liang}, E.~P. 1989, \apj, 344, 332,
  \dodoi{10.1086/167802}

\bibitem[{{Chornock} {et~al.}(2014){Chornock}, {Berger}, {Gezari}, {Zauderer},
  {Rest}, {Chomiuk}, {Kamble}, {Soderberg}, {Czekala}, {Dittmann}, {Drout},
  {Foley}, {Fong}, {Huber}, {Kirshner}, {Lawrence}, {Lunnan}, {Marion},
  {Narayan}, {Riess}, {Roth}, {Sanders}, {Scolnic}, {Smartt}, {Smith},
  {Stubbs}, {Tonry}, {Burgett}, {Chambers}, {Flewelling}, {Hodapp}, {Kaiser},
  {Magnier}, {Martin}, {Neill}, {Price}, \& {Wainscoat}}]{Chornock2014}
{Chornock}, R., {Berger}, E., {Gezari}, S., {et~al.} 2014, \apj, 780, 44,
  \dodoi{10.1088/0004-637X/780/1/44}

\bibitem[{{Chugai} \& {Danziger}(1994)}]{Chugai1994}
{Chugai}, N.~N., \& {Danziger}, I.~J. 1994, \mnras, 268, 173,
  \dodoi{10.1093/mnras/268.1.173}

\bibitem[{{Cohen} \& {Soker}(2024)}]{Cohen2024}
{Cohen}, T., \& {Soker}, N. 2024, \mnras, 527, 10025,
  \dodoi{10.1093/mnras/stad3745}

\bibitem[{Conroy {et~al.}(2009)Conroy, Gunn, \& White}]{Conroy_2009}
Conroy, C., Gunn, J.~E., \& White, M. 2009, The Astrophysical Journal, 699,
  486, \dodoi{10.1088/0004-637X/699/1/486}

\bibitem[{{Crowther}(2007)}]{Crowther2007}
{Crowther}, P.~A. 2007, \araa, 45, 177,
  \dodoi{10.1146/annurev.astro.45.051806.110615}

\bibitem[{{Dessart} {et~al.}(2015){Dessart}, {Audit}, \&
  {Hillier}}]{Dessart2015}
{Dessart}, L., {Audit}, E., \& {Hillier}, D.~J. 2015, \mnras, 449, 4304,
  \dodoi{10.1093/mnras/stv609}

\bibitem[{{Dickinson} {et~al.}(2024){Dickinson}, {Smith}, {Andrews}, {Milne},
  {Kilpatrick}, \& {Milisavljevic}}]{Dickinson2024}
{Dickinson}, D., {Smith}, N., {Andrews}, J.~E., {et~al.} 2024, \mnras, 527,
  7767, \dodoi{10.1093/mnras/stad3631}

\bibitem[{{Dwarkadas} \& {Gruszko}(2012)}]{Dwarkadas2012}
{Dwarkadas}, V.~V., \& {Gruszko}, J. 2012, \mnras, 419, 1515,
  \dodoi{10.1111/j.1365-2966.2011.19808.x}

\bibitem[{{Fabricant} {et~al.}(2005){Fabricant}, {Fata}, {Roll}, {Hertz},
  {Caldwell}, {Gauron}, {Geary}, {McLeod}, {Szentgyorgyi}, {Zajac}, {Kurtz},
  {Barberis}, {Bergner}, {Brown}, {Conroy}, {Eng}, {Geller}, {Goddard},
  {Honsa}, {Mueller}, {Mink}, {Ordway}, {Tokarz}, {Woods}, {Wyatt}, {Epps}, \&
  {Dell'Antonio}}]{Fabricant2005}
{Fabricant}, D., {Fata}, R., {Roll}, J., {et~al.} 2005, \pasp, 117, 1411,
  \dodoi{10.1086/497385}

\bibitem[{{Filippenko} {et~al.}(1992){Filippenko}, {Richmond}, {Matheson},
  {Shields}, {Burbidge}, {Cohen}, {Dickinson}, {Malkan}, {Nelson}, {Pietz},
  {Schlegel}, {Schmeer}, {Spinrad}, {Steidel}, {Tran}, \&
  {Wren}}]{Filippenko1992}
{Filippenko}, A.~V., {Richmond}, M.~W., {Matheson}, T., {et~al.} 1992, \apjl,
  384, L15, \dodoi{10.1086/186252}

\bibitem[{{Fransson} {et~al.}(1996){Fransson}, {Lundqvist}, \&
  {Chevalier}}]{Fransson1996}
{Fransson}, C., {Lundqvist}, P., \& {Chevalier}, R.~A. 1996, \apj, 461, 993,
  \dodoi{10.1086/177119}

\bibitem[{Fransson {et~al.}(2014)Fransson, Ergon, Challis, Chevalier, France,
  Kirshner, Marion, Milisavljevic, Smith, Bufano, \& et~al.}]{Fransson_2014}
Fransson, C., Ergon, M., Challis, P.~J., {et~al.} 2014, The Astrophysical
  Journal, 797, 118, \dodoi{10.1088/0004-637x/797/2/118}

\bibitem[{{Fruscione} {et~al.}(2006){Fruscione}, {McDowell}, {Allen},
  {Brickhouse}, {Burke}, {Davis}, {Durham}, {Elvis}, {Galle}, {Harris},
  {Huenemoerder}, {Houck}, {Ishibashi}, {Karovska}, {Nicastro}, {Noble},
  {Nowak}, {Primini}, {Siemiginowska}, {Smith}, \& {Wise}}]{Fruscione2006}
{Fruscione}, A., {McDowell}, J.~C., {Allen}, G.~E., {et~al.} 2006, in Society
  of Photo-Optical Instrumentation Engineers (SPIE) Conference Series, Vol.
  6270, Society of Photo-Optical Instrumentation Engineers (SPIE) Conference
  Series, ed. D.~R. {Silva} \& R.~E. {Doxsey}, 62701V,
  \dodoi{10.1117/12.671760}

\bibitem[{{Fuller}(2017)}]{Fuller2017}
{Fuller}, J. 2017, \mnras, 470, 1642, \dodoi{10.1093/mnras/stx1314}

\bibitem[{Gal-Yam(2012)}]{Gal_Yam_2012}
Gal-Yam, A. 2012, Science, 337, 927–932, \dodoi{10.1126/science.1203601}

\bibitem[{Guillochon {et~al.}(2018)Guillochon, Nicholl, Villar, Mockler,
  Narayan, Mandel, Berger, \& Williams}]{Guillochon_2018}
Guillochon, J., Nicholl, M., Villar, V.~A., {et~al.} 2018, The Astrophysical
  Journal Supplement Series, 236, 6, \dodoi{10.3847/1538-4365/aab761}

\bibitem[{Harris {et~al.}(2020)Harris, Millman, van~der Walt, Gommers,
  Virtanen, Cournapeau, Wieser, Taylor, Berg, Smith, Kern, Picus, Hoyer, van
  Kerkwijk, Brett, Haldane, del R{'{\i}}o, Wiebe, Peterson,
  G{'{e}}rard-Marchant, Sheppard, Reddy, Weckesser, Abbasi, Gohlke, \&
  Oliphant}]{numpy}
Harris, C.~R., Millman, K.~J., van~der Walt, S.~J., {et~al.} 2020, Nature, 585,
  357, \dodoi{10.1038/s41586-020-2649-2}

\bibitem[{{Hatsukade} {et~al.}(2021){Hatsukade}, {Tominaga}, {Morokuma},
  {Morokuma-Matsui}, {Matsuda}, {Tamura}, {Niinuma}, \&
  {Motogi}}]{Hatsukade2021}
{Hatsukade}, B., {Tominaga}, N., {Morokuma}, T., {et~al.} 2021, \apj, 922, 17,
  \dodoi{10.3847/1538-4357/ac20d5}

\bibitem[{{HI4PI Collaboration} {et~al.}(2016){HI4PI Collaboration}, {Ben
  Bekhti}, {Fl{\"o}er}, {Keller}, {Kerp}, {Lenz}, {Winkel}, {Bailin},
  {Calabretta}, {Dedes}, {Ford}, {Gibson}, {Haud}, {Janowiecki}, {Kalberla},
  {Lockman}, {McClure-Griffiths}, {Murphy}, {Nakanishi}, {Pisano}, \&
  {Staveley-Smith}}]{HI4PI2016}
{HI4PI Collaboration}, {Ben Bekhti}, N., {Fl{\"o}er}, L., {et~al.} 2016, \aap,
  594, A116, \dodoi{10.1051/0004-6361/201629178}

\bibitem[{Hoffman {et~al.}(2008)Hoffman, Leonard, Chornock, Filippenko, Barth,
  \& Matheson}]{Hoffman_2008}
Hoffman, J.~L., Leonard, D.~C., Chornock, R., {et~al.} 2008, The Astrophysical
  Journal, 688, 1186, \dodoi{10.1086/592261}

\bibitem[{{H{\"o}fner} \& {Olofsson}(2018)}]{Hofner2018}
{H{\"o}fner}, S., \& {Olofsson}, H. 2018, \aapr, 26, 1,
  \dodoi{10.1007/s00159-017-0106-5}

\bibitem[{{Horesh} {et~al.}(2013){Horesh}, {Stockdale}, {Fox}, {Frail},
  {Carpenter}, {Kulkarni}, {Ofek}, {Gal-Yam}, {Kasliwal}, {Arcavi}, {Quimby},
  {Cenko}, {Nugent}, {Bloom}, {Law}, {Poznanski}, {Gorbikov}, {Polishook},
  {Yaron}, {Ryder}, {Weiler}, {Bauer}, {Van Dyk}, {Immler}, {Panagia},
  {Pooley}, \& {Kassim}}]{Horesh2013}
{Horesh}, A., {Stockdale}, C., {Fox}, D.~B., {et~al.} 2013, \mnras, 436, 1258,
  \dodoi{10.1093/mnras/stt1645}

\bibitem[{{Horesh} {et~al.}(2020){Horesh}, {Sfaradi}, {Ergon}, {Barbarino},
  {Sollerman}, {Moldon}, {Dobie}, {Schulze}, {P{\'e}rez-Torres}, {Williams},
  {Fremling}, {Gal-Yam}, {Kulkarni}, {O'Brien}, {Lundqvist}, {Murphy},
  {Fender}, {Anand}, {Belicki}, {Bellm}, {Coughlin}, {De}, {Golkhou}, {Graham},
  {Green}, {Hankins}, {Kasliwal}, {Kupfer}, {Laher}, {Masci}, {Miller},
  {Neill}, {Ofek}, {Perrott}, {Porter}, {Reiley}, {Rigault}, {Rodriguez},
  {Rusholme}, {Shupe}, \& {Titterington}}]{Horesh2020}
{Horesh}, A., {Sfaradi}, I., {Ergon}, M., {et~al.} 2020, \apj, 903, 132,
  \dodoi{10.3847/1538-4357/abbd38}

\bibitem[{Hunter(2007)}]{matplotlib}
Hunter, J.~D. 2007, Computing in Science \& Engineering, 9, 90,
  \dodoi{10.1109/MCSE.2007.55}

\bibitem[{Inserra {et~al.}(2013)Inserra, Smartt, Scalzo, Fraser, Pastorello,
  Childress, Pignata, Jerkstrand, Kotak, Benetti, Della~Valle, Gal-Yam,
  Mazzali, Smith, Sullivan, Valenti, Yaron, Young, \& Reichart}]{Inserra2013}
Inserra, C., Smartt, S., Scalzo, R., {et~al.} 2013, Monthly Notices of the
  Royal Astronomical Society, 437, \dodoi{10.1093/mnrasl/slt138}

\bibitem[{{Jerkstrand} {et~al.}(2020){Jerkstrand}, {Maeda}, \&
  {Kawabata}}]{Jerkstrand2020}
{Jerkstrand}, A., {Maeda}, K., \& {Kawabata}, K.~S. 2020, Science, 367, 415,
  \dodoi{10.1126/science.aaw1469}

\bibitem[{{Johnson} {et~al.}(2021{\natexlab{a}}){Johnson}, {Leja}, {Conroy}, \&
  {Speagle}}]{Johnson2021}
{Johnson}, B.~D., {Leja}, J., {Conroy}, C., \& {Speagle}, J.~S.
  2021{\natexlab{a}}, \apjs, 254, 22, \dodoi{10.3847/1538-4365/abef67}

\bibitem[{{Johnson} {et~al.}(2021{\natexlab{b}}){Johnson}, {Leja}, {Conroy}, \&
  {Speagle}}]{prospector2021}
---. 2021{\natexlab{b}}, \apjs, 254, 22, \dodoi{10.3847/1538-4365/abef67}

\bibitem[{{Joye} \& {Mandel}(2003)}]{ds92003}
{Joye}, W.~A., \& {Mandel}, E. 2003, in Astronomical Society of the Pacific
  Conference Series, Vol. 295, Astronomical Data Analysis Software and Systems
  XII, ed. H.~E. {Payne}, R.~I. {Jedrzejewski}, \& R.~N. {Hook}, 489

\bibitem[{{Kaiser} {et~al.}(2010){Kaiser}, {Burgett}, {Chambers}, {Denneau},
  {Heasley}, {Jedicke}, {Magnier}, {Morgan}, {Onaka}, \& {Tonry}}]{Kaiser2010}
{Kaiser}, N., {Burgett}, W., {Chambers}, K., {et~al.} 2010, in Society of
  Photo-Optical Instrumentation Engineers (SPIE) Conference Series, Vol. 7733,
  Ground-based and Airborne Telescopes III, ed. L.~M. {Stepp}, R.~{Gilmozzi},
  \& H.~J. {Hall}, 77330E, \dodoi{10.1117/12.859188}

\bibitem[{{Kankare} {et~al.}(2017){Kankare}, {Kotak}, {Mattila}, {Lundqvist},
  {Ward}, {Fraser}, {Lawrence}, {Smartt}, {Meikle}, {Bruce}, {Harmanen},
  {Hutton}, {Inserra}, {Kangas}, {Pastorello}, {Reynolds},
  {Romero-Ca{\~n}izales}, {Smith}, {Valenti}, {Chambers}, {Hodapp}, {Huber},
  {Kaiser}, {Kudritzki}, {Magnier}, {Tonry}, {Wainscoat}, \&
  {Waters}}]{Kankare2017}
{Kankare}, E., {Kotak}, R., {Mattila}, S., {et~al.} 2017, Nature Astronomy, 1,
  865, \dodoi{10.1038/s41550-017-0290-2}

\bibitem[{{Kauffmann} {et~al.}(2003){Kauffmann}, {Heckman}, {Tremonti},
  {Brinchmann}, {Charlot}, {White}, {Ridgway}, {Brinkmann}, {Fukugita}, {Hall},
  {Ivezi{\'c}}, {Richards}, \& {Schneider}}]{Kauffmann2003}
{Kauffmann}, G., {Heckman}, T.~M., {Tremonti}, C., {et~al.} 2003, \mnras, 346,
  1055, \dodoi{10.1111/j.1365-2966.2003.07154.x}

\bibitem[{{Kennicutt}(1998)}]{kennicutt1998}
{Kennicutt}, Robert~C., J. 1998, \araa, 36, 189,
  \dodoi{10.1146/annurev.astro.36.1.189}

\bibitem[{{Kewley} {et~al.}(2001){Kewley}, {Dopita}, {Sutherland}, {Heisler},
  \& {Trevena}}]{Kewley2001}
{Kewley}, L.~J., {Dopita}, M.~A., {Sutherland}, R.~S., {Heisler}, C.~A., \&
  {Trevena}, J. 2001, \apj, 556, 121, \dodoi{10.1086/321545}

\bibitem[{{Kiewe} {et~al.}(2012){Kiewe}, {Gal-Yam}, {Arcavi}, {Leonard},
  {Emilio Enriquez}, {Cenko}, {Fox}, {Moon}, {Sand}, {Soderberg}, \&
  {CCCP}}]{Kiewe2012}
{Kiewe}, M., {Gal-Yam}, A., {Arcavi}, I., {et~al.} 2012, \apj, 744, 10,
  \dodoi{10.1088/0004-637X/744/1/10}

\bibitem[{{Komatsu} {et~al.}(2011){Komatsu}, {Smith}, {Dunkley}, {Bennett},
  {Gold}, {Hinshaw}, {Jarosik}, {Larson}, {Nolta}, {Page}, {Spergel},
  {Halpern}, {Hill}, {Kogut}, {Limon}, {Meyer}, {Odegard}, {Tucker}, {Weiland},
  {Wollack}, \& {Wright}}]{Komatsu2011}
{Komatsu}, E., {Smith}, K.~M., {Dunkley}, J., {et~al.} 2011, \apjs, 192, 18,
  \dodoi{10.1088/0067-0049/192/2/18}

\bibitem[{{Lehmer} {et~al.}(2016){Lehmer}, {Basu-Zych}, {Mineo}, {Brandt},
  {Eufrasio}, {Fragos}, {Hornschemeier}, {Luo}, {Xue}, {Bauer}, {Gilfanov},
  {Ranalli}, {Schneider}, {Shemmer}, {Tozzi}, {Trump}, {Vignali}, {Wang},
  {Yukita}, \& {Zezas}}]{Lehmer2016}
{Lehmer}, B.~D., {Basu-Zych}, A.~R., {Mineo}, S., {et~al.} 2016, \apj, 825, 7,
  \dodoi{10.3847/0004-637X/825/1/7}

\bibitem[{{Magnier} {et~al.}(2008){Magnier}, {Liu}, {Monet}, \&
  {Chambers}}]{Maigner2008}
{Magnier}, E.~A., {Liu}, M., {Monet}, D.~G., \& {Chambers}, K.~C. 2008, in A
  Giant Step: from Milli- to Micro-arcsecond Astrometry, ed. W.~J. {Jin},
  I.~{Platais}, \& M.~A.~C. {Perryman}, Vol. 248, 553--559,
  \dodoi{10.1017/S1743921308020139}

\bibitem[{{Margalit} {et~al.}(2022){Margalit}, {Quataert}, \&
  {Ho}}]{Margalit2022}
{Margalit}, B., {Quataert}, E., \& {Ho}, A. Y.~Q. 2022, \apj, 928, 122,
  \dodoi{10.3847/1538-4357/ac53b0}

\bibitem[{{Margutti} {et~al.}(2014){Margutti}, {Milisavljevic}, {Soderberg},
  {Chornock}, {Zauderer}, {Murase}, {Guidorzi}, {Sanders}, {Kuin}, {Fransson},
  {Levesque}, {Chandra}, {Berger}, {Bianco}, {Brown}, {Challis},
  {Chatzopoulos}, {Cheung}, {Choi}, {Chomiuk}, {Chugai}, {Contreras}, {Drout},
  {Fesen}, {Foley}, {Fong}, {Friedman}, {Gall}, {Gehrels}, {Hjorth}, {Hsiao},
  {Kirshner}, {Im}, {Leloudas}, {Lunnan}, {Marion}, {Martin}, {Morrell},
  {Neugent}, {Omodei}, {Phillips}, {Rest}, {Silverman}, {Strader},
  {Stritzinger}, {Szalai}, {Utterback}, {Vinko}, {Wheeler}, {Arnett},
  {Campana}, {Chevalier}, {Ginsburg}, {Kamble}, {Roming}, {Pritchard}, \&
  {Stringfellow}}]{Margutti2014}
{Margutti}, R., {Milisavljevic}, D., {Soderberg}, A.~M., {et~al.} 2014, \apj,
  780, 21, \dodoi{10.1088/0004-637X/780/1/21}

\bibitem[{{Margutti} {et~al.}(2017){Margutti}, {Kamble}, {Milisavljevic},
  {Zapartas}, {de Mink}, {Drout}, {Chornock}, {Risaliti}, {Zauderer},
  {Bietenholz}, {Cantiello}, {Chakraborti}, {Chomiuk}, {Fong}, {Grefenstette},
  {Guidorzi}, {Kirshner}, {Parrent}, {Patnaude}, {Soderberg}, {Gehrels}, \&
  {Harrison}}]{Margutti2017}
{Margutti}, R., {Kamble}, A., {Milisavljevic}, D., {et~al.} 2017, \apj, 835,
  140, \dodoi{10.3847/1538-4357/835/2/140}

\bibitem[{{Margutti} {et~al.}(2023){Margutti}, {Bright}, {Matthews},
  {Coppejans}, {Alexander}, {Berger}, {Bietenholz}, {Chornock}, {DeMarchi},
  {Drout}, {Eftekhari}, {Jacobson-Gal{\'a}n}, {Laskar}, {Milisavljevic},
  {Murase}, {Nicholl}, {Omand}, {Stroh}, {Terreran}, \&
  {VanderLey}}]{Margutti2023}
{Margutti}, R., {Bright}, J.~S., {Matthews}, D.~J., {et~al.} 2023, \apjl, 954,
  L45, \dodoi{10.3847/2041-8213/acf1fd}

\bibitem[{{McMullin} {et~al.}(2007){McMullin}, {Waters}, {Schiebel}, {Young},
  \& {Golap}}]{casa2007}
{McMullin}, J.~P., {Waters}, B., {Schiebel}, D., {Young}, W., \& {Golap}, K.
  2007, Astronomical Society of the Pacific Conference Series, Vol. 376, {CASA
  Architecture and Applications}, ed. R.~A. {Shaw}, F.~{Hill}, \& D.~J. {Bell},
  127

\bibitem[{{Miller} {et~al.}(2010){Miller}, {Silverman}, {Butler}, {Bloom},
  {Chornock}, {Filippenko}, {Ganeshalingam}, {Klein}, {Li}, {Nugent}, {Smith},
  \& {Steele}}]{Miller2010}
{Miller}, A.~A., {Silverman}, J.~M., {Butler}, N.~R., {et~al.} 2010, \mnras,
  404, 305, \dodoi{10.1111/j.1365-2966.2010.16280.x}

\bibitem[{{Mineo} {et~al.}(2012{\natexlab{a}}){Mineo}, {Gilfanov}, \&
  {Sunyaev}}]{Mineo2012a}
{Mineo}, S., {Gilfanov}, M., \& {Sunyaev}, R. 2012{\natexlab{a}}, \mnras, 419,
  2095, \dodoi{10.1111/j.1365-2966.2011.19862.x}

\bibitem[{{Mineo} {et~al.}(2012{\natexlab{b}}){Mineo}, {Gilfanov}, \&
  {Sunyaev}}]{Mineo2012b}
---. 2012{\natexlab{b}}, \mnras, 426, 1870,
  \dodoi{10.1111/j.1365-2966.2012.21831.x}

\bibitem[{{Moran} {et~al.}(2023){Moran}, {Fraser}, {Kotak}, {Pastorello},
  {Benetti}, {Brennan}, {Guti{\'e}rrez}, {Kankare}, {Kuncarayakti}, {Mattila},
  {Reynolds}, {Anderson}, {Brown}, {Campana}, {Chambers}, {Chen}, {Della
  Valle}, {Dennefeld}, {Elias-Rosa}, {Galbany}, {Galindo-Guil}, {Gromadzki},
  {Hiramatsu}, {Inserra}, {Leloudas}, {M{\"u}ller-Bravo}, {Nicholl},
  {Reguitti}, {Shahbandeh}, {Smartt}, {Tartaglia}, \& {Young}}]{Moran2023}
{Moran}, S., {Fraser}, M., {Kotak}, R., {et~al.} 2023, \aap, 669, A51,
  \dodoi{10.1051/0004-6361/202244565}

\bibitem[{{Moriya}(2014)}]{Moriya2014}
{Moriya}, T.~J. 2014, arXiv e-prints, arXiv:1402.2519,
  \dodoi{10.48550/arXiv.1402.2519}

\bibitem[{{Morrison} \& {McCammon}(1983)}]{Morrison1983}
{Morrison}, R., \& {McCammon}, D. 1983, \apj, 270, 119, \dodoi{10.1086/161102}

\bibitem[{{Nicholl} {et~al.}(2020){Nicholl}, {Blanchard}, {Berger}, {Chornock},
  {Margutti}, {Gomez}, {Lunnan}, {Miller}, {Fong}, {Terreran},
  {Vigna-G{\'o}mez}, {Bhirombhakdi}, {Bieryla}, {Challis}, {Laher}, {Masci}, \&
  {Paterson}}]{Nicholl2020}
{Nicholl}, M., {Blanchard}, P.~K., {Berger}, E., {et~al.} 2020, Nature
  Astronomy, 4, 893, \dodoi{10.1038/s41550-020-1066-7}

\bibitem[{{Nyholm} {et~al.}(2020){Nyholm}, {Sollerman}, {Tartaglia}, {Taddia},
  {Fremling}, {Blagorodnova}, {Filippenko}, {Gal-Yam}, {Howell},
  {Karamehmetoglu}, {Kulkarni}, {Laher}, {Leloudas}, {Masci}, {Kasliwal},
  {Mor{\r{a}}}, {Moriya}, {Ofek}, {Papadogiannakis}, {Quimby}, {Rebbapragada},
  \& {Schulze}}]{Nyholm2020}
{Nyholm}, A., {Sollerman}, J., {Tartaglia}, L., {et~al.} 2020, \aap, 637, A73,
  \dodoi{10.1051/0004-6361/201936097}

\bibitem[{{Nymark} {et~al.}(2009){Nymark}, {Chandra}, \&
  {Fransson}}]{Nymark2009}
{Nymark}, T.~K., {Chandra}, P., \& {Fransson}, C. 2009, \aap, 494, 179,
  \dodoi{10.1051/0004-6361:200810884}

\bibitem[{{Nymark} {et~al.}(2006){Nymark}, {Fransson}, \& {Kozma}}]{Nymark2006}
{Nymark}, T.~K., {Fransson}, C., \& {Kozma}, C. 2006, \aap, 449, 171,
  \dodoi{10.1051/0004-6361:20054169}

\bibitem[{{Ofek} {et~al.}(2014{\natexlab{a}}){Ofek}, {Arcavi}, {Tal},
  {Sullivan}, {Gal-Yam}, {Kulkarni}, {Nugent}, {Ben-Ami}, {Bersier}, {Cao},
  {Cenko}, {De Cia}, {Filippenko}, {Fransson}, {Kasliwal}, {Laher}, {Surace},
  {Quimby}, \& {Yaron}}]{Ofek2014b}
{Ofek}, E.~O., {Arcavi}, I., {Tal}, D., {et~al.} 2014{\natexlab{a}}, \apj, 788,
  154, \dodoi{10.1088/0004-637X/788/2/154}

\bibitem[{{Ofek} {et~al.}(2014{\natexlab{b}}){Ofek}, {Zoglauer}, {Boggs},
  {Barri{\'e}re}, {Reynolds}, {Fryer}, {Harrison}, {Cenko}, {Kulkarni},
  {Gal-Yam}, {Arcavi}, {Bellm}, {Bloom}, {Christensen}, {Craig}, {Even},
  {Filippenko}, {Grefenstette}, {Hailey}, {Laher}, {Madsen}, {Nakar}, {Nugent},
  {Stern}, {Sullivan}, {Surace}, \& {Zhang}}]{Ofek2014}
{Ofek}, E.~O., {Zoglauer}, A., {Boggs}, S.~E., {et~al.} 2014{\natexlab{b}},
  \apj, 781, 42, \dodoi{10.1088/0004-637X/781/1/42}

\bibitem[{{Osterbrock} \& {Pogge}(1985)}]{Osterbrock1985}
{Osterbrock}, D.~E., \& {Pogge}, R.~W. 1985, \apj, 297, 166,
  \dodoi{10.1086/163513}

\bibitem[{{Pettini} \& {Pagel}(2004)}]{Pettini2004}
{Pettini}, M., \& {Pagel}, B. E.~J. 2004, \mnras, 348, L59,
  \dodoi{10.1111/j.1365-2966.2004.07591.x}

\bibitem[{{Podsiadlowski} {et~al.}(1992){Podsiadlowski}, {Joss}, \&
  {Hsu}}]{Podsiadlowski1992}
{Podsiadlowski}, P., {Joss}, P.~C., \& {Hsu}, J.~J.~L. 1992, \apj, 391, 246,
  \dodoi{10.1086/171341}

\bibitem[{{Pooley} {et~al.}(2002){Pooley}, {Lewin}, {Fox}, {Miller}, {Lacey},
  {Van Dyk}, {Weiler}, {Sramek}, {Filippenko}, {Leonard}, {Immler},
  {Chevalier}, {Fabian}, {Fransson}, \& {Nomoto}}]{pooley2002ApJ}
{Pooley}, D., {Lewin}, W. H.~G., {Fox}, D.~W., {et~al.} 2002, \apj, 572, 932,
  \dodoi{10.1086/340346}

\bibitem[{{Prieto} {et~al.}(2017){Prieto}, {Chen}, {Dong}, {Shappee},
  {Seibert}, {Bersier}, {Holoien}, {Kochanek}, {Stanek}, \&
  {Thompson}}]{Prieto2017}
{Prieto}, J.~L., {Chen}, P., {Dong}, S., {et~al.} 2017, Research Notes of the
  American Astronomical Society, 1, 28, \dodoi{10.3847/2515-5172/aa9c46}

\bibitem[{Pérez-Torres {et~al.}(2009)Pérez-Torres, Alberdi, Colina,
  Torrelles, Panagia, Wilson, Kankare, \& Mattila}]{Prez_Torres_2009}
Pérez-Torres, M.~A., Alberdi, A., Colina, L., {et~al.} 2009, Monthly Notices
  of the Royal Astronomical Society, 399, 1641–1649,
  \dodoi{10.1111/j.1365-2966.2009.15389.x}

\bibitem[{Quataert \& Shiode(2012)}]{Quataert2012}
Quataert, E., \& Shiode, J. 2012, Monthly Notices of the Royal Astronomical
  Society: Letters, 423, L92–L96, \dodoi{10.1111/j.1745-3933.2012.01264.x}

\bibitem[{{Ranalli} {et~al.}(2012){Ranalli}, {Comastri}, {Zamorani},
  {Cappelluti}, {Civano}, {Georgantopoulos}, {Gilli}, {Schinnerer},
  {Smol{\v{c}}i{\'c}}, \& {Vignali}}]{Ranalli2012}
{Ranalli}, P., {Comastri}, A., {Zamorani}, G., {et~al.} 2012, \aap, 542, A16,
  \dodoi{10.1051/0004-6361/201118723}

\bibitem[{{Ransome} {et~al.}(2021){Ransome}, {Habergham-Mawson}, {Darnley},
  {James}, {Filippenko}, \& {Schlegel}}]{Ransome2021}
{Ransome}, C.~L., {Habergham-Mawson}, S.~M., {Darnley}, M.~J., {et~al.} 2021,
  \mnras, 506, 4715, \dodoi{10.1093/mnras/stab1938}

\bibitem[{{Renzo} {et~al.}(2020){Renzo}, {Farmer}, {Justham}, {G{\"o}tberg},
  {de Mink}, {Zapartas}, {Marchant}, \& {Smith}}]{Renzo2020}
{Renzo}, M., {Farmer}, R., {Justham}, S., {et~al.} 2020, \aap, 640, A56,
  \dodoi{10.1051/0004-6361/202037710}

\bibitem[{{Rest} {et~al.}(2011){Rest}, {Foley}, {Gezari}, {Narayan}, {Draine},
  {Olsen}, {Huber}, {Matheson}, {Garg}, {Welch}, {Becker}, {Challis},
  {Clocchiatti}, {Cook}, {Damke}, {Meixner}, {Miknaitis}, {Minniti}, {Morelli},
  {Nikolaev}, {Pignata}, {Prieto}, {Smith}, {Stubbs}, {Suntzeff}, {Walker},
  {Wood-Vasey}, {Zenteno}, {Wyrzykowski}, {Udalski}, {Szyma{\'n}ski}, {Kubiak},
  {Pietrzy{\'n}ski}, {Soszy{\'n}ski}, {Szewczyk}, {Ulaczyk}, \&
  {Poleski}}]{Rest2011}
{Rest}, A., {Foley}, R.~J., {Gezari}, S., {et~al.} 2011, \apj, 729, 88,
  \dodoi{10.1088/0004-637X/729/2/88}

\bibitem[{{Rest} {et~al.}(2014){Rest}, {Scolnic}, {Foley}, {Huber}, {Chornock},
  {Narayan}, {Tonry}, {Berger}, {Soderberg}, {Stubbs}, {Riess}, {Kirshner},
  {Smartt}, {Schlafly}, {Rodney}, {Botticella}, {Brout}, {Challis}, {Czekala},
  {Drout}, {Hudson}, {Kotak}, {Leibler}, {Lunnan}, {Marion}, {McCrum},
  {Milisavljevic}, {Pastorello}, {Sanders}, {Smith}, {Stafford}, {Thilker},
  {Valenti}, {Wood-Vasey}, {Zheng}, {Burgett}, {Chambers}, {Denneau}, {Draper},
  {Flewelling}, {Hodapp}, {Kaiser}, {Kudritzki}, {Magnier}, {Metcalfe},
  {Price}, {Sweeney}, {Wainscoat}, \& {Waters}}]{Rest2014}
{Rest}, A., {Scolnic}, D., {Foley}, R.~J., {et~al.} 2014, \apj, 795, 44,
  \dodoi{10.1088/0004-637X/795/1/44}

\bibitem[{{Ruan} {et~al.}(2016){Ruan}, {Anderson}, {Cales}, {Eracleous},
  {Green}, {Morganson}, {Runnoe}, {Shen}, {Wilkinson}, {Blanton}, {Dwelly},
  {Georgakakis}, {Greene}, {LaMassa}, {Merloni}, \& {Schneider}}]{Ruan2016}
{Ruan}, J.~J., {Anderson}, S.~F., {Cales}, S.~L., {et~al.} 2016, \apj, 826,
  188, \dodoi{10.3847/0004-637X/826/2/188}

\bibitem[{{Ruiz-Carmona} {et~al.}(2022){Ruiz-Carmona}, {Sfaradi}, \&
  {Horesh}}]{Ruiz-Carmona2022}
{Ruiz-Carmona}, R., {Sfaradi}, I., \& {Horesh}, A. 2022, \aap, 666, A82,
  \dodoi{10.1051/0004-6361/202142024}

\bibitem[{{Sana} {et~al.}(2012){Sana}, {de Mink}, {de Koter}, {Langer},
  {Evans}, {Gieles}, {Gosset}, {Izzard}, {Le Bouquin}, \&
  {Schneider}}]{Sana2012}
{Sana}, H., {de Mink}, S.~E., {de Koter}, A., {et~al.} 2012, Science, 337, 444,
  \dodoi{10.1126/science.1223344}

\bibitem[{{Schinzel} {et~al.}(2009){Schinzel}, {Taylor}, {Stockdale}, {Granot},
  \& {Ramirez-Ruiz}}]{schinelz2009ApJ}
{Schinzel}, F.~K., {Taylor}, G.~B., {Stockdale}, C.~J., {Granot}, J., \&
  {Ramirez-Ruiz}, E. 2009, \apj, 691, 1380,
  \dodoi{10.1088/0004-637X/691/2/1380}

\bibitem[{{Schlafly} \& {Finkbeiner}(2011)}]{Schlafly2011}
{Schlafly}, E.~F., \& {Finkbeiner}, D.~P. 2011, \apj, 737, 103,
  \dodoi{10.1088/0004-637X/737/2/103}

\bibitem[{{Schlegel}(1990)}]{Schlegel1990}
{Schlegel}, E.~M. 1990, \mnras, 244, 269

\bibitem[{{Schlegel} \& {Petre}(2006)}]{Schlegel2006}
{Schlegel}, E.~M., \& {Petre}, R. 2006, \apj, 646, 378, \dodoi{10.1086/504890}

\bibitem[{{Schlegel} {et~al.}(1999){Schlegel}, {Ryder}, {Staveley-Smith},
  {Petre}, {Colbert}, {Dopita}, \& {Campbell-Wilson}}]{Shle1999AJ}
{Schlegel}, E.~M., {Ryder}, S., {Staveley-Smith}, L., {et~al.} 1999, \aj, 118,
  2689, \dodoi{10.1086/301145}

\bibitem[{{Schmidt} {et~al.}(1989){Schmidt}, {Weymann}, \&
  {Foltz}}]{Schmidt1989}
{Schmidt}, G.~D., {Weymann}, R.~J., \& {Foltz}, C.~B. 1989, \pasp, 101, 713,
  \dodoi{10.1086/132495}

\bibitem[{{Schr{\o}der} {et~al.}(2020){Schr{\o}der}, {MacLeod}, {Loeb},
  {Vigna-G{\'o}mez}, \& {Mandel}}]{Schroder2020}
{Schr{\o}der}, S.~L., {MacLeod}, M., {Loeb}, A., {Vigna-G{\'o}mez}, A., \&
  {Mandel}, I. 2020, \apj, 892, 13, \dodoi{10.3847/1538-4357/ab7014}

\bibitem[{{Science Software Branch at STScI}(2012)}]{pyraf2012}
{Science Software Branch at STScI}. 2012, {PyRAF: Python alternative for IRAF},
  Astrophysics Source Code Library, record ascl:1207.011.
\newblock \doeprint{1207.011}

\bibitem[{{Shiode} \& {Quataert}(2014)}]{Shiode2014}
{Shiode}, J.~H., \& {Quataert}, E. 2014, \apj, 780, 96,
  \dodoi{10.1088/0004-637X/780/1/96}

\bibitem[{{Smith} {et~al.}(2007{\natexlab{a}}){Smith}, {Ryder}, {B{\"o}ttcher},
  {Tingay}, {Stacy}, {Pakull}, \& {Liang}}]{SmithI2007}
{Smith}, I.~A., {Ryder}, S.~D., {B{\"o}ttcher}, M., {et~al.}
  2007{\natexlab{a}}, \apj, 669, 1130, \dodoi{10.1086/521698}

\bibitem[{{Smith}(2014)}]{Smith2014}
{Smith}, N. 2014, \araa, 52, 487, \dodoi{10.1146/annurev-astro-081913-040025}

\bibitem[{{Smith}(2017{\natexlab{a}})}]{Smith2017hsn}
---. 2017{\natexlab{a}}, in Handbook of Supernovae, ed. A.~W. {Alsabti} \&
  P.~{Murdin}, 403, \dodoi{10.1007/978-3-319-21846-5_38}

\bibitem[{{Smith}(2017{\natexlab{b}})}]{smith2017RS}
---. 2017{\natexlab{b}}, Philosophical Transactions of the Royal Society of
  London Series A, 375, 20160268, \dodoi{10.1098/rsta.2016.0268}

\bibitem[{{Smith} \& {Andrews}(2020)}]{Smith&Andrew2020}
{Smith}, N., \& {Andrews}, J.~E. 2020, \mnras, 499, 3544,
  \dodoi{10.1093/mnras/staa3047}

\bibitem[{{Smith} {et~al.}(2023){Smith}, {Andrews}, {Milne}, {Filippenko},
  {Brink}, {Kelly}, {Yuk}, \& {Jencson}}]{Smith2023}
{Smith}, N., {Andrews}, J.~E., {Milne}, P., {et~al.} 2023, arXiv e-prints,
  arXiv:2312.00253, \dodoi{10.48550/arXiv.2312.00253}

\bibitem[{Smith \& Arnett(2014)}]{Smith&Arnett2014}
Smith, N., \& Arnett, W.~D. 2014, The Astrophysical Journal, 785, 82,
  \dodoi{10.1088/0004-637x/785/2/82}

\bibitem[{Smith {et~al.}(2008)Smith, Chornock, Li, Ganeshalingam, Silverman,
  Foley, Filippenko, \& Barth}]{Smith_2008}
Smith, N., Chornock, R., Li, W., {et~al.} 2008, The Astrophysical Journal, 686,
  467–484, \dodoi{10.1086/591021}

\bibitem[{{Smith} {et~al.}(2008{\natexlab{a}}){Smith}, {Chornock}, {Li},
  {Ganeshalingam}, {Silverman}, {Foley}, {Filippenko}, \& {Barth}}]{Smith2008}
{Smith}, N., {Chornock}, R., {Li}, W., {et~al.} 2008{\natexlab{a}}, \apj, 686,
  467, \dodoi{10.1086/591021}

\bibitem[{{Smith} {et~al.}(2010){Smith}, {Chornock}, {Silverman}, {Filippenko},
  \& {Foley}}]{SmithRyan2010}
{Smith}, N., {Chornock}, R., {Silverman}, J.~M., {Filippenko}, A.~V., \&
  {Foley}, R.~J. 2010, \apj, 709, 856, \dodoi{10.1088/0004-637X/709/2/856}

\bibitem[{{Smith} {et~al.}(2011){Smith}, {Li}, {Filippenko}, \&
  {Chornock}}]{Smith2011}
{Smith}, N., {Li}, W., {Filippenko}, A.~V., \& {Chornock}, R. 2011, \mnras,
  412, 1522, \dodoi{10.1111/j.1365-2966.2011.17229.x}

\bibitem[{{Smith} \& {McCray}(2007)}]{Smith2007B}
{Smith}, N., \& {McCray}, R. 2007, \apjl, 671, L17, \dodoi{10.1086/524681}

\bibitem[{{Smith} {et~al.}(2007{\natexlab{b}}){Smith}, {Li}, {Foley},
  {Wheeler}, {Pooley}, {Chornock}, {Filippenko}, {Silverman}, {Quimby},
  {Bloom}, \& {Hansen}}]{Smith2007}
{Smith}, N., {Li}, W., {Foley}, R.~J., {et~al.} 2007{\natexlab{b}}, \apj, 666,
  1116, \dodoi{10.1086/519949}

\bibitem[{{Smith} {et~al.}(2008{\natexlab{b}}){Smith}, {Foley}, {Bloom}, {Li},
  {Filippenko}, {Gavazzi}, {Ghez}, {Konopacky}, {Malkan}, {Marshall}, {Pooley},
  {Treu}, \& {Woo}}]{Smith2008A}
{Smith}, N., {Foley}, R.~J., {Bloom}, J.~S., {et~al.} 2008{\natexlab{b}}, \apj,
  686, 485, \dodoi{10.1086/590141}

\bibitem[{{Smith} {et~al.}(2017){Smith}, {Kilpatrick}, {Mauerhan}, {Andrews},
  {Margutti}, {Fong}, {Graham}, {Zheng}, {Kelly}, {Filippenko}, \&
  {Fox}}]{Smith2017}
{Smith}, N., {Kilpatrick}, C.~D., {Mauerhan}, J.~C., {et~al.} 2017, \mnras,
  466, 3021, \dodoi{10.1093/mnras/stw3204}

\bibitem[{{Speagle}(2020)}]{Speagle2020}
{Speagle}, J.~S. 2020, \mnras, 493, 3132, \dodoi{10.1093/mnras/staa278}

\bibitem[{{Stroh} {et~al.}(2021){Stroh}, {Terreran}, {Coppejans}, {Bright},
  {Margutti}, {Bietenholz}, {De Colle}, {DeMarchi}, {Duran}, {Milisavljevic},
  {Murase}, {Paterson}, \& {Williams}}]{Stroh2021}
{Stroh}, M.~C., {Terreran}, G., {Coppejans}, D.~L., {et~al.} 2021, \apjl, 923,
  L24, \dodoi{10.3847/2041-8213/ac375e}

\bibitem[{{Stubbs} {et~al.}(2010){Stubbs}, {Doherty}, {Cramer}, {Narayan},
  {Brown}, {Lykke}, {Woodward}, \& {Tonry}}]{Stubbe2010}
{Stubbs}, C.~W., {Doherty}, P., {Cramer}, C., {et~al.} 2010, \apjs, 191, 376,
  \dodoi{10.1088/0067-0049/191/2/376}

\bibitem[{{Suzuki} {et~al.}(2020){Suzuki}, {Moriya}, \&
  {Takiwaki}}]{Suzuki2020}
{Suzuki}, A., {Moriya}, T.~J., \& {Takiwaki}, T. 2020, \apj, 899, 56,
  \dodoi{10.3847/1538-4357/aba0ba}

\bibitem[{{Suzuki} {et~al.}(2021){Suzuki}, {Nicholl}, {Moriya}, \&
  {Takiwaki}}]{Suzuki2021}
{Suzuki}, A., {Nicholl}, M., {Moriya}, T.~J., \& {Takiwaki}, T. 2021, \apj,
  908, 99, \dodoi{10.3847/1538-4357/abd6ce}

\bibitem[{{Svirski} {et~al.}(2012){Svirski}, {Nakar}, \& {Sari}}]{Svirski2012}
{Svirski}, G., {Nakar}, E., \& {Sari}, R. 2012, \apj, 759, 108,
  \dodoi{10.1088/0004-637X/759/2/108}

\bibitem[{{Tabatabaei} {et~al.}(2017){Tabatabaei}, {Schinnerer}, {Krause},
  {Dumas}, {Meidt}, {Damas-Segovia}, {Beck}, {Murphy}, {Mulcahy}, {Groves},
  {Bolatto}, {Dale}, {Galametz}, {Sandstrom}, {Boquien}, {Calzetti},
  {Kennicutt}, {Hunt}, {De Looze}, \& {Pellegrini}}]{Tabatabaei2017}
{Tabatabaei}, F.~S., {Schinnerer}, E., {Krause}, M., {et~al.} 2017, \apj, 836,
  185, \dodoi{10.3847/1538-4357/836/2/185}

\bibitem[{{Taddia} {et~al.}(2013){Taddia}, {Stritzinger}, {Sollerman},
  {Phillips}, {Anderson}, {Boldt}, {Campillay}, {Castell{\'o}n}, {Contreras},
  {Folatelli}, {Hamuy}, {Heinrich-Josties}, {Krzeminski}, {Morrell}, {Burns},
  {Freedman}, {Madore}, {Persson}, \& {Suntzeff}}]{Taddia2013}
{Taddia}, F., {Stritzinger}, M.~D., {Sollerman}, J., {et~al.} 2013, \aap, 555,
  A10, \dodoi{10.1051/0004-6361/201321180}

\bibitem[{{Tartaglia} {et~al.}(2020){Tartaglia}, {Pastorello}, {Sollerman},
  {Fransson}, {Mattila}, {Fraser}, {Taddia}, {Tomasella}, {Turatto},
  {Morales-Garoffolo}, {Elias-Rosa}, {Lundqvist}, {Harmanen}, {Reynolds},
  {Cappellaro}, {Barbarino}, {Nyholm}, {Kool}, {Ofek}, {Gao}, {Jin}, {Tan},
  {Sand}, {Ciabattari}, {Wang}, {Zhang}, {Huang}, {Li}, {Mo}, {Rui}, {Xiang},
  {Zhang}, {Hosseinzadeh}, {Howell}, {McCully}, {Valenti}, {Benetti}, {Callis},
  {Carracedo}, {Fremling}, {Kangas}, {Rubin}, {Somero}, \&
  {Terreran}}]{Tartaglia2020}
{Tartaglia}, L., {Pastorello}, A., {Sollerman}, J., {et~al.} 2020, \aap, 635,
  A39, \dodoi{10.1051/0004-6361/201936553}

\bibitem[{{Temple} {et~al.}(2005){Temple}, {Raychaudhury}, \&
  {Stevens}}]{Temple2005}
{Temple}, R.~F., {Raychaudhury}, S., \& {Stevens}, I.~R. 2005, \mnras, 362,
  581, \dodoi{10.1111/j.1365-2966.2005.09336.x}

\bibitem[{{Tonry} {et~al.}(2012){Tonry}, {Stubbs}, {Lykke}, {Doherty},
  {Shivvers}, {Burgett}, {Chambers}, {Hodapp}, {Kaiser}, {Kudritzki},
  {Magnier}, {Morgan}, {Price}, \& {Wainscoat}}]{Tonry2012}
{Tonry}, J.~L., {Stubbs}, C.~W., {Lykke}, K.~R., {et~al.} 2012, \apj, 750, 99,
  \dodoi{10.1088/0004-637X/750/2/99}

\bibitem[{{Tremonti} {et~al.}(2004){Tremonti}, {Heckman}, {Kauffmann},
  {Brinchmann}, {Charlot}, {White}, {Seibert}, {Peng}, {Schlegel}, {Uomoto},
  {Fukugita}, \& {Brinkmann}}]{Tremonti2004}
{Tremonti}, C.~A., {Heckman}, T.~M., {Kauffmann}, G., {et~al.} 2004, \apj, 613,
  898, \dodoi{10.1086/423264}

\bibitem[{{Trouille} {et~al.}(2011){Trouille}, {Barger}, \&
  {Tremonti}}]{BPT22011}
{Trouille}, L., {Barger}, A.~J., \& {Tremonti}, C. 2011, \apj, 742, 46,
  \dodoi{10.1088/0004-637X/742/1/46}

\bibitem[{{Valenti} {et~al.}(2008){Valenti}, {Benetti}, {Cappellaro}, {Patat},
  {Mazzali}, {Turatto}, {Hurley}, {Maeda}, {Gal-Yam}, {Foley}, {Filippenko},
  {Pastorello}, {Challis}, {Frontera}, {Harutyunyan}, {Iye}, {Kawabata},
  {Kirshner}, {Li}, {Lipkin}, {Matheson}, {Nomoto}, {Ofek}, {Ohyama}, {Pian},
  {Poznanski}, {Salvo}, {Sauer}, {Schmidt}, {Soderberg}, \&
  {Zampieri}}]{Valenti2008}
{Valenti}, S., {Benetti}, S., {Cappellaro}, E., {et~al.} 2008, \mnras, 383,
  1485, \dodoi{10.1111/j.1365-2966.2007.12647.x}

\bibitem[{{van Dyk} {et~al.}(1993){van Dyk}, {Weiler}, {Sramek}, \&
  {Panagia}}]{vanDyke1993ApJ}
{van Dyk}, S.~D., {Weiler}, K.~W., {Sramek}, R.~A., \& {Panagia}, N. 1993,
  \apjl, 419, L69, \dodoi{10.1086/187139}

\bibitem[{{van Dyk} {et~al.}(1996{\natexlab{a}}){van Dyk}, {Weiler}, {Sramek},
  {Schlegel}, {Filippenko}, {Panagia}, \& {Leibundgut}}]{vanDyk1996}
{van Dyk}, S.~D., {Weiler}, K.~W., {Sramek}, R.~A., {et~al.}
  1996{\natexlab{a}}, \aj, 111, 1271, \dodoi{10.1086/117872}

\bibitem[{{van Dyk} {et~al.}(1996{\natexlab{b}}){van Dyk}, {Weiler}, {Sramek},
  {Schlegel}, {Filippenko}, {Panagia}, \& {Leibundgut}}]{vandyk1996A}
---. 1996{\natexlab{b}}, \aj, 111, 1271, \dodoi{10.1086/117872}

\bibitem[{{van Marle} {et~al.}(2010){van Marle}, {Smith}, {Owocki}, \& {van
  Veelen}}]{vanMarle2010}
{van Marle}, A.~J., {Smith}, N., {Owocki}, S.~P., \& {van Veelen}, B. 2010,
  \mnras, 407, 2305, \dodoi{10.1111/j.1365-2966.2010.16851.x}

\bibitem[{{van Velzen} {et~al.}(2021){van Velzen}, {Gezari}, {Hammerstein},
  {Roth}, {Frederick}, {Ward}, {Hung}, {Cenko}, {Stein}, {Perley}, {Taggart},
  {Foley}, {Sollerman}, {Blagorodnova}, {Andreoni}, {Bellm}, {Brinnel}, {De},
  {Dekany}, {Feeney}, {Fremling}, {Giomi}, {Golkhou}, {Graham}, {Ho},
  {Kasliwal}, {Kilpatrick}, {Kulkarni}, {Kupfer}, {Laher}, {Mahabal}, {Masci},
  {Miller}, {Nordin}, {Riddle}, {Rusholme}, {van Santen}, {Sharma}, {Shupe}, \&
  {Soumagnac}}]{vanVelzen2021}
{van Velzen}, S., {Gezari}, S., {Hammerstein}, E., {et~al.} 2021, \apj, 908, 4,
  \dodoi{10.3847/1538-4357/abc258}

\bibitem[{{Villar} {et~al.}(2017){Villar}, {Berger}, {Metzger}, \&
  {Guillochon}}]{Villar2017}
{Villar}, V.~A., {Berger}, E., {Metzger}, B.~D., \& {Guillochon}, J. 2017,
  \apj, 849, 70, \dodoi{10.3847/1538-4357/aa8fcb}

\bibitem[{{Villar} {et~al.}(2020){Villar}, {Hosseinzadeh}, {Berger},
  {Ntampaka}, {Jones}, {Challis}, {Chornock}, {Drout}, {Foley}, {Kirshner},
  {Lunnan}, {Margutti}, {Milisavljevic}, {Sanders}, {Pan}, {Rest}, {Scolnic},
  {Magnier}, {Metcalfe}, {Wainscoat}, \& {Waters}}]{Villar2020}
{Villar}, V.~A., {Hosseinzadeh}, G., {Berger}, E., {et~al.} 2020, \apj, 905,
  94, \dodoi{10.3847/1538-4357/abc6fd}

\bibitem[{{Vink} {et~al.}(2011){Vink}, {Muijres}, {Anthonisse}, {de Koter},
  {Gr{\"a}fener}, \& {Langer}}]{Vink2011}
{Vink}, J.~S., {Muijres}, L.~E., {Anthonisse}, B., {et~al.} 2011, \aap, 531,
  A132, \dodoi{10.1051/0004-6361/201116614}

\bibitem[{{Weiler} {et~al.}(1990){Weiler}, {Panagia}, \& {Sramek}}]{Weiler1990}
{Weiler}, K.~W., {Panagia}, N., \& {Sramek}, R.~A. 1990, \apj, 364, 611,
  \dodoi{10.1086/169444}

\bibitem[{{Williams} {et~al.}(2017){Williams}, {Clavel}, {Newton}, \&
  {Ryzhkov}}]{peterwilliams2017}
{Williams}, P. K.~G., {Clavel}, M., {Newton}, E., \& {Ryzhkov}, D. 2017,
  {pwkit: Astronomical utilities in Python}.
\newblock \doeprint{1704.001}

\bibitem[{{Woosley}(2017)}]{Woosley2017}
{Woosley}, S.~E. 2017, \apj, 836, 244, \dodoi{10.3847/1538-4357/836/2/244}

\bibitem[{Woosley {et~al.}(2007)Woosley, Blinnikov, \& Heger}]{Woosley2007}
Woosley, S.~E., Blinnikov, S., \& Heger, A. 2007, Nature, 450, 390–392,
  \dodoi{10.1038/nature06333}

\bibitem[{{Wright} {et~al.}(2010){Wright}, {Eisenhardt}, {Mainzer}, {Ressler},
  {Cutri}, {Jarrett}, {Kirkpatrick}, {Padgett}, {McMillan}, {Skrutskie},
  {Stanford}, {Cohen}, {Walker}, {Mather}, {Leisawitz}, {Gautier}, {McLean},
  {Benford}, {Lonsdale}, {Blain}, {Mendez}, {Irace}, {Duval}, {Liu}, {Royer},
  {Heinrichsen}, {Howard}, {Shannon}, {Kendall}, {Walsh}, {Larsen}, {Cardon},
  {Schick}, {Schwalm}, {Abid}, {Fabinsky}, {Naes}, \& {Tsai}}]{Wright2010}
{Wright}, E.~L., {Eisenhardt}, P. R.~M., {Mainzer}, A.~K., {et~al.} 2010, \aj,
  140, 1868, \dodoi{10.1088/0004-6256/140/6/1868}

\bibitem[{{Wu} \& {Fuller}(2021)}]{Wu2021}
{Wu}, S., \& {Fuller}, J. 2021, \apj, 906, 3, \dodoi{10.3847/1538-4357/abc87c}

\bibitem[{{Yadav} {et~al.}(2014){Yadav}, {Ray}, {Chakraborti}, {Stockdale},
  {Chandra}, {Smith}, {Roy}, {Bose}, {Dwarkadas}, {Sutaria}, \&
  {Pooley}}]{Yadav2014}
{Yadav}, N., {Ray}, A., {Chakraborti}, S., {et~al.} 2014, \apj, 782, 30,
  \dodoi{10.1088/0004-637X/782/1/30}

\bibitem[{{Yoon} \& {Cantiello}(2010)}]{Yoon2010}
{Yoon}, S.-C., \& {Cantiello}, M. 2010, \apjl, 717, L62,
  \dodoi{10.1088/2041-8205/717/1/L62}

\bibitem[{{Zapartas} {et~al.}(2021){Zapartas}, {de Mink}, {Justham}, {Smith},
  {Renzo}, \& {de Koter}}]{Zapartas2021}
{Zapartas}, E., {de Mink}, S.~E., {Justham}, S., {et~al.} 2021, \aap, 645, A6,
  \dodoi{10.1051/0004-6361/202037744}

\end{thebibliography}

%\begin{comment}

%\end{comment}

\end{document}